\documentclass[fleqn,usenatbib]{mnras}

\usepackage{newtxtext,newtxmath}

\usepackage[T1]{fontenc}

\DeclareRobustCommand{\VAN}[3]{#2}
\let\VANthebibliography\thebibliography
\def\thebibliography{\DeclareRobustCommand{\VAN}[3]{##3}\VANthebibliography}

\usepackage{graphicx}	
\usepackage{amsmath}	
\usepackage{float}
\usepackage{natbib}
\usepackage{caption}
\usepackage{gensymb}
\usepackage{bm}
\usepackage[output-decimal-marker={,},exponent-product=\cdot]{siunitx}
\usepackage[utf8]{inputenc}
\usepackage{newtxtext,newtxmath}
\usepackage{hyperref}
\usepackage{mathtools}
\usepackage{pifont}
\usepackage{multirow}
\usepackage{hhline}

\newcommand{\mr}{\mathrm}

\title[Hybrid AGN feedback]{A hybrid active galactic nucleus feedback model with spinning black holes, winds and jets}

\author[F. Huško et al.]{Filip Hu\v{s}ko$^{1}$\thanks{E-mail: husko@strw.leidenuniv.nl},
Cedric G. Lacey$^{2}$,
Joop Schaye$^{1}$,
Matthieu Schaller$^{1,3}$,
Evgenii Chaikin$^{1}$,\newauthor
Sylvia Ploeckinger$^{4}$,
Alejandro Benítez Llambay$^{5}$,
Alexander J. Richings$^{6,7}$,
James W. Trayford$^{8}$\\
$^{1}$Leiden Observatory, Leiden University, PO Box 9513, 2300 RA Leiden, the Netherlands\\
$^{2}$Institute for Computational Cosmology, Department of Physics, University of Durham, South Road, Durham, DH1 3LE, UK\\
$^{3}$Lorentz Institute for Theoretical Physics, Leiden University, PO box 9506, 2300 RA Leiden, the Netherlands\\
$^{4}$Department of Astrophysics, University of Vienna, T\"urkenschanzstrasse 17, 1180 Vienna, Austria\\
$^{5}$Dipartimento di Fisica G. Occhialini, Università Degli Studi di Milano Bicocca, Piazza della Scienza, 3 I-20126 Milano MI, Italy\\
$^{6}$Centre for Data Science, Artificial Intelligence and Modelling, University of Hull, Cottingham Road, Hull, HU6 7RX, UK\\
$^{7}$E. A. Milne Centre for Astrophysics, University of Hull, Cottingham Road, Hull, HU6 7RX, UK\\
$^{8}$Institute of Cosmology and Gravitation, University of Portsmouth, Dennis Sciama Building, Burnaby Road, Portsmouth PO1 3FX, UK\\
}

\date{Accepted XXX. Received YYY; in original form ZZZ}

\pubyear{2025}

\begin{document}
\label{firstpage}
\pagerange{\pageref{firstpage}--\pageref{lastpage}}
\maketitle

\begin{abstract}
We present a hybrid active galactic nucleus (AGN) feedback model that features three accretion disc states (the thick, thin, and slim discs at low, moderate, and super-Eddington accretion rates, respectively), and two feedback modes: thermal isotropic and kinetic jets. The model includes black hole (BH) spin evolution due to gas accretion, BH mergers, jet spindown, and Lense-Thirring torques. The BH spin determines the jet directions and affects the feedback efficiencies. 
The model is implemented in the \textsc{swift} code and coupled with the COLIBRE galaxy formation model. We present the first results from hybrid AGN feedback simulations run as part of the COLIBRE suite, focusing on the impact of new parameters and calibration efforts. 
Using the new hybrid AGN feedback model, we find that AGN feedback affects not just massive galaxies, but all galaxies down to $M_*\approx10^8$ M$_\odot$. BH spins are predicted to be near-maximal for intermediate-mass BHs ($M_\mr{BH}\in[10^6,10^8]$ M$_\odot$), and lower for other BH masses, which is in good agreement with observations. The intergalactic medium is hotter and impacted on larger scales in the hybrid AGN feedback simulations compared to those using purely thermal feedback. In the hybrid AGN simulations, we predict that half of the cumulative injected AGN energy is in thermal and the other half in jet form, broadly independent of BH mass and redshift. Jet feedback is important at all redshifts and dominates over thermal feedback at $z<0.5$ and $z>1.5$, but only mildly.  
\end{abstract}



\renewcommand{\arraystretch}{1.18}
\newcommand{\will}[1]{{\color{red}#1}}
\renewcommand*\arraystretch{1.35}




\section{Introduction}

Observations reveal a diversity in the star-formation (SF) activity of galaxies, with lower-mass galaxies typically having higher specific star formation rates (sSFRs), i.e.~SFRs per unit stellar mass, than high-mass galaxies (e.g.~\citealt{Bauer2013}, \citealt{Chang2015}). High-mass ellipticals are typically gas-poor (e.g.~\citealt{Wiklin1995}, \citealt{Young2011}, \citealt{Davis2019}) and have very low or undetectable levels of SF (e.g.~\citealt{Salim2007}, \citealt{Whitaker2012}). Brightest-cluster galaxies (BCGs), the most massive elliptical galaxies in the Universe, are almost all devoid of SF (\citealt{McKelvie2014}). The quiescent nature of these galaxies poses a problem. They are located in the centers of galaxy clusters and are surrounded by hot, dense, pressure-supported gas, the intracluster medium (ICM), which emits X-rays (e.g.~\citealt{Sarazin1986}, \citealt{Stanek2006}). The typical cooling time of the ICM close to the BCG is shorter than a Hubble time. Without mitigation, we would expect massive cooling flows to develop, causing large molecular hydrogen reservoirs and starbursts within the BCG, with SFRs $\approx1000$ M$_\odot$ $\mathrm{yr}^{-1}$ (\citealt{Fabian1994}, \citealt{Hudson2010}). With the exception of a few cases (e.g.~\citealt{Edge2001}, \citealt{McKelvie2014}, \citealt{McDonald2015}), this is not observed.

The lack of SF in massive galaxies can be attributed to energy injection by supermassive black holes (hereafter BHs), i.e.~active galactic nucleus (AGN) feedback (see review by \citealt{Fabian2012}). X-ray observations of galaxy groups and clusters reveal cavities in the ICM (e.g.~\citealt{Gull1973}, \citealt{Boehringer1993}, \citealt{Birzan2004}, \citealt{Eckert2021}). These cavities are cospatial with lobes inflated by jets of relativistic particles launched by the central BHs, which can mainly be observed at radio wavelengths (e.g.~\citealt{Blandford1979}, \citealt{Biermann1987}, \citealt{Urry1995}, \citealt{Markoff2001}). The jet powers, inferred from X-ray observations of cavities, are sufficient to offset the cooling luminosities of the ICM (e.g.~\citealt{Rafferty2006}, \citealt{Hlavacek-Larrondo2012}, \citealt{Russell2013}), solving the aforementioned cooling flow problem. AGN feedback in this situation, as well as in other, lower-mass quenched galaxies, is known as radio jet, `maintenance mode', or mechanical feedback (e.g.~\citealt{Fabian2012}, \citealt{Eckert2021}).

In addition to radio jets, AGN have been observed at other wavelengths (e.g.~optical, X-ray, UV, mid-infrared, etc.; see compilation by \citealt{Shen2020}). The brightest cases are known as quasars (e.g.~\citealt{Schmidt1968}). These observations extend to the highest redshifts (e.g.~\citealt{Fan2001}), with the peak in radiative AGN activity occurring around $z=2$ (e.g.~\citealt{Delvecchio2014}, \citealt{Aird2015}, \citealt{DSilva2023}). AGN are observed to drive outflows (e.g.~\citealt{Crenshaw2003}, \citealt{Tombesi2010}, \citealt{Feruglio2010}). Their origin is unknown, with the radiation possibly driving a wind on the scales of the accretion disc surrounding the BH (e.g.~\citealt{Murray1995}, \citealt{Nomura2016}, \citealt{Fiore2017}, \citealt{QueraBofarull2023}), or on larger scales (e.g.~\citealt{Fabian1999}, \citealt{Costa2018}). What fraction of the luminosity couples to the gas is also uncertain (e.g.~\citealt{Harrison2018}). This mode of AGN feedback, often referred to as `quasar' feedback (\citealt{BestHeckman}), should be more relevant at higher redshifts, when radiative AGN activity is stronger. 

In the standard picture, radiative AGN eject gas from their galaxies/host dark matter (DM) haloes and thus quench them (e.g.~\citealt{DiMatteo2005}), while maintenance-mode (jet) feedback, which is more often observed in the local Universe, is thought to keep the galaxies quenched (e.g.~\citealt{Bower2006}), with almost all massive quenched galaxies having some radio emission, even if at a low level (\citealt{Sabater2019}). In this picture, BHs occupy one of the two different feedback modes depending on their accretion rate expressed as a fraction of the Eddington rate (the Eddington ratio). They are thought to be in the radiatively-efficient quasar mode if the Eddington ratio is above 1 per cent, and in the maintenance mode if it is below that value (\citealt{BestHeckman}).

The above picture of two-mode feedback, in separate Eddington ratio regimes, is very likely an oversimplification. Of order 10 per cent of AGN in the radiatively-efficient, quasar regime, are observed to be radio-loud (\citealt{BestHeckman}), i.e.~they have strong jets. In fact, some of the earliest observed quasars have these strong jets (e.g.~\citealt{Greenstein1964}, \citealt{Kino2005}). While these radio sources are relatively rare and constitute only a small fraction of all radio sources, they are very powerful. These radio sources are found to be hosted by active rather than quenched galaxies (\citealt{Saikia2022}), unlike the majority of low-power radio sources (\citealt{Sabater2019}). The difference in host galaxy properties also implies that BHs in quasars have higher accretion rates, which in turn means that their radio jets occur in a fundamentally different accretion and feedback state to the one that causes `maintenance mode' feedback in quenched galaxies. Conceivably, jets from quasars could be as important for galaxy evolution as the lower-power jets in the standard `maintenance mode' (\citealt{Merloni2008}). Conversely, low-power winds are observed not just in the high-Eddington ratio, quasar regime, but also at low Eddington ratios, in which case, these objects are referred to as `red geysers' (\citealt{Cheung2016}, \citealt{Shi2025}).

Furthermore, the properties of AGN may change at super-Eddington rates, with both jets and winds possibly becoming relevant, and the radiative efficiency dropping (e.g.~\citealt{Abramowicz1988}, \citealt{Sadowski2014}, also see review by \citealt{Marziani2025}). The super-Eddington regime is very rare, especially in the local Universe, but the accretion rates can be so high that this regime may be important for BH and galaxy evolution, despite its short-lived and infrequent nature (\citealt{Husko_2025_SE}). Powerful jets, possibly launched in the super-Eddington state, have been found not only at redshifts up to $z=3$ (\citealt{Smolcic2017}, \citealt{Slaus2024}), but also up to $z=6$ (\citealt{Ghisellini2014} , \citealt{Sbarrato2021}, \citealt{Sbarrato2022}, \citealt{Roy2024}). 

Accretion theory has had great success in explaining many aspects of the AGN observations described above. Standard accretion disc theory (\citealt{ShakuraSunyaev1973}, \citealt{NovikovThorne1973}) has been used to explain observed AGN in the moderate Eddington-ratio regime, including the details of their spectra (e.g.~\citealt{Sun1989}). In this accretion model, which we will refer to as the `thin disc' in this study, the accretion disc is relatively cool, geometrically thin, optically thick and radiatively-efficient. Of order 10 per cent (the precise value depends on BH spin and is between $\approx4$ and $\approx40$ per cent; \citealt{NovikovThorne1973}) of the mass-energy accreting through the disc is released by means of viscous heating and then radiated away. 

The theory of the advection-dominated accretion flow (ADAF; \citealt{NarayanYi1994}), which we will refer to as the `thick disc', describes a hot, geometrically thick, optically thin and radiatively-inefficient accretion flow. This model corresponds to the low-Eddington ratio regime, in which the AGN are not bright in the optical or X-rays, but instead produce energetically-dominant jets. In all accretion discs, the magnetic dynamo effect can lead to an increase of the magnetic field strength. However, in the ADAF, the field is easily advected inwards (\citealt{YuanNarayan2014}). This results in the launching of a jet through the \cite{Blandford1977} (BZ) mechanism. The BZ jet efficiency depends strongly on BH spin, as it is the rotational energy of the BH that powers jet launching. The ADAF model predicts that the transition between the thick and thin discs should occur at $\approx1$ per cent Eddington (\citealt{Narayan1995}), which is consistent with observations of X-ray binaries and AGN (\citealt{Maccarone2023}, \citealt{Done2007}, \citealt{Noda2018}, \citealt{Russell2013}).

In addition to the thin and thick discs, other models have also been developed. The advection-dominated, inflow-outflow solution (ADIOS; \citealt{Blandford1999}) predicts a similar accretion flow as the thick disc (ADAF), but it features strong outflows on account of thermal pressure of the gas. This can explain observations of red geysers (\citealt{Cheung2016}). One of the main features of this model is that the winds take away most of the accreting mass, so that the horizon-scale mass accretion rate at the inner edge of the accretion disc and into the BH is much smaller than that at large radii (onto the disc). In other words, the accretion efficiency is much less than unity. The ADIOS model is preferred over the ADAF one in observations of individual objects such as M87$^*$ and SgrA$^*$ (\citealt{Yuan2003}, \citealt{Kuo2014}), cavity power observations in galaxy groups and clusters (\citealt{Nemmen2015}), and numerical simulations (\citealt{Yuan2012}). 

The slim disc model (e.g.~\citealt{Abramowicz1988}, \citealt{Wang1999}, \citealt{Sadowski2009}) has been used to describe the super-Eddington regime. In the slim disc, the pressure is dominated by radiation rather than gas, but the disc is otherwise quite similar to the thick disc. Most notably, it is highly advective, which should result (in combination with the dynamo effect) in similar jet efficiencies as in the thick disc.

Recent developments in numerical simulation methods have allowed direct tests of accretion disc models. The exception to this is of very thin discs corresponding to the radiatively-efficient thin disc model (\citealt{ShakuraSunyaev1973}), which cannot yet be simulated due to their very thin nature. Recent simulations of accretion disc formation from cosmological initial conditions find a hyper-magnetized disc state at moderate Eddington ratios, which may supplant the original \cite{ShakuraSunyaev1973} solution (\citealt{Hopkins2024}, \citealt{Hopkins2025}). General-relativistic magneto-hydrodynamical (GRMHD) simulations of thick discs have confirmed most of the properties predicted by the ADAF/ADIOS models (e.g.~\citealt{Narayan2003}, \citealt{Tchekhovskoy2010}, \citealt{McKinney2012}, \citealt{Sadowski2013}). These simulations have also confirmed the BZ mechanism of jet launching, and they have converged on a fitting formula for the jet efficiency that is more accurate than the BZ prediction (\citealt{Narayan2021}). The jet efficiency in this formula depends strongly on BH spin and on the magnetization of the disc. It can reach values of up to several hundred per cent. This may appear to violate energy conservation. However, the efficiencies are defined as the power output relative to the mass-energy input into the BH. BHs can tap their own rotational energy through jet launching and lose more energy in the process than they gain mass-energy through accretion. In addition to confirming the operation of the BZ process, GRMHD simulations show that very mass-loaded winds are operating in the thick disc (\citealt{Yuan2012}, \citealt{Guo2023}), resulting in accretion efficiencies of a per cent or less.

GRMHD simulations reveal two classes of objects that differ based on the strength of the magnetic field near the event horizon. The first are so-called `standard and normal evolution' (SANE) discs (e.g.~\citealt{Narayan2012}), while the second are magnetically-arrested (MAD) discs (e.g.~\citealt{Narayan2003}). In the SANE state, the event horizon magnetization is a free parameter, while in the MAD state, it has saturated and is determined by the accretion rate and BH spin. We expect most thick discs to be in the MAD state on theoretical grounds, due to strong advection (\citealt{YuanNarayan2014}), and this is supported by observations (\citealt{EHT2021}, \citealt{EHTSagA5}, \citealt{Yuan2022}). GRMHD simulations of super-Eddington, slim discs show that jet efficiencies can be similarly high as in the thick disc. We also expect them to be MAD, as they are advection-dominated. GRMHD simulations of thinner, sub-Eddington accretion discs (\citealt{Avara2016}, \citealt{Ricarte}) have found that the MAD state is possible in the thin disc, with jet efficiencies given by the same BZ formula as for the thick and slim discs. However, the magnetization, and therefore the jet efficiency, is lower, and heavily dependent on the Eddington ratio. Furthermore, it is unknown if the MAD assumption is correct for thin discs, as they are not advection-dominated. 

Models of galaxy formation, starting with semi-analytical models (\citealt{Bower2006}, \citealt{Croton2006}, \citealt{Lagos2008}) and early hydrodynamical simulations (\citealt{DiMatteo2005}, \citealt{Booth2009}), began incorporating AGN feedback after X-ray observations showed that it is important in offsetting cooling flows in galaxy clusters (e.g.~\citealt{Birzan2004}, \citealt{McNamara2005}, \citealt{Wise2007}). Many cosmological hydrodynamical simulation models have included AGN feedback using a single, typically thermal AGN feedback mode, corresponding to the high-Eddington ratio, `quasar' mode  (e.g.~OWLs: \citealt{Schaye2010}, Magneticum: \citealt{Hirschmann2014}, EAGLE: \citealt{Schaye2015}, MassiveBlack-II: \citealt{Khandai2015}, Romulus: \citealt{Tremmel2017}, Astrid: \citealt{Bird2022}, \citealt{Ni2022}). Other simulation models have often used two-mode AGN feedback where at low-Eddington ratios, the `quasar' mode is replaced by either winds representing the ADIOS outflow (IllustrisTNG; \citealt{Weinberger2017b}, \citealt{Nelson2019}), manual bubble injection representing lobes inflated by jets (e.g.~\citealt{Sijacki2007}, Illustris; \citealt{Vogelsberger2014}), or direct kinetic jet feedback (HorizonAGN; \citealt{Kaviraj2017}, NewHorizon; \citealt{Dubois2021}, SIMBA; \citealt{Dave2019}). Some simulation models have included additional feedback modes (e.g~X-ray heating in SIMBA). Sophisticated subgrid models have been developed to more realistically represent AGN feedback, including accretion disc modeling and BH spin evolution (e.g~\citealt{Volonteri2005}, \citealt{Lagos2009}, \citealt{Fanidakis2011}, \citealt{Dubois2012}, \citealt{Steinborn2015}, \citealt{Fiacconi2018}, \citealt{Bustamante2019}, \citealt{Griffin2019a}, \citealt{Husko2022_spin_driven}, \citealt{Koudmani2024}, \citealt{Sala2024}), as well as inclusion of super-Eddington accretion and its feedback effects (e.g.~\citealt{Rennehan2024}, \citealt{Bennett2024}, \citealt{Husko_2025_SE}).

Motivated by observational evidence of the complexity of AGN feedback, as well as advances in theoretical modeling, we have developed a hybrid AGN feedback model for use in cosmological, hydrodynamical simulations of galaxy formation. The model synthesizes different aspects of AGN activity that have been studied with accretion disc theory, including recent developments from GRMHD simulations. We include three accretion disc states: the thick, thin and slim disc, at low, moderate and super-Eddington ratios, respectively. Two feedback modes are included: 1) thermal isotropic, representing the effects of radiation and winds, and 2) kinetic jets, representing the effects of relativistic jets. We also model BH spin to more realistically implement AGN feedback, especially for jets. In addition to its usefulness as an AGN feedback model, galaxy formation simulations run with the model can be used for detailed studies of BH and accretion physics in a cosmological context.

Aspects of the model have already been introduced in previous studies, where we coupled it with the SWIFT-EAGLE model (\citealt{Schaye2015}, \citealt{Schaller2024}). In \cite{Husko2022_spin_driven} we studied BH spin and jet feedback from the thick disc using idealized galaxy clusters. We found that spin-driven jets can quench BCGs, even in very massive galaxy clusters. In \cite{Husko_winds} we compared this mode with the thermal isotropic one from the thin disc, as well as studied a simple hybrid model that used kinetic jets at low Eddington ratios and thermal isotropic feedback at large Eddington ratios. The hybrid model performed best, yielding moderate star formation rates in the BCGs (but not permanent quenching), and the lowest central entropies of the ICM. In \cite{Husko_2025_SE} we introduced the super-Eddington aspect of the model (the slim disc), with both jet and thermal feedback, and also considered jets in the thin disc. In \cite{Husko2022_self_similar} we performed the first idealized jet simulations using smoothed particle hydrodynamics (SPH) to validate the numerical behaviour of jets and the lobes they inflate. We found that most aspects of their behaviour are captured well with as few particles as a few hundred. 

\begin{figure*}
\includegraphics[width=1\textwidth, trim = 0 10 0 0]{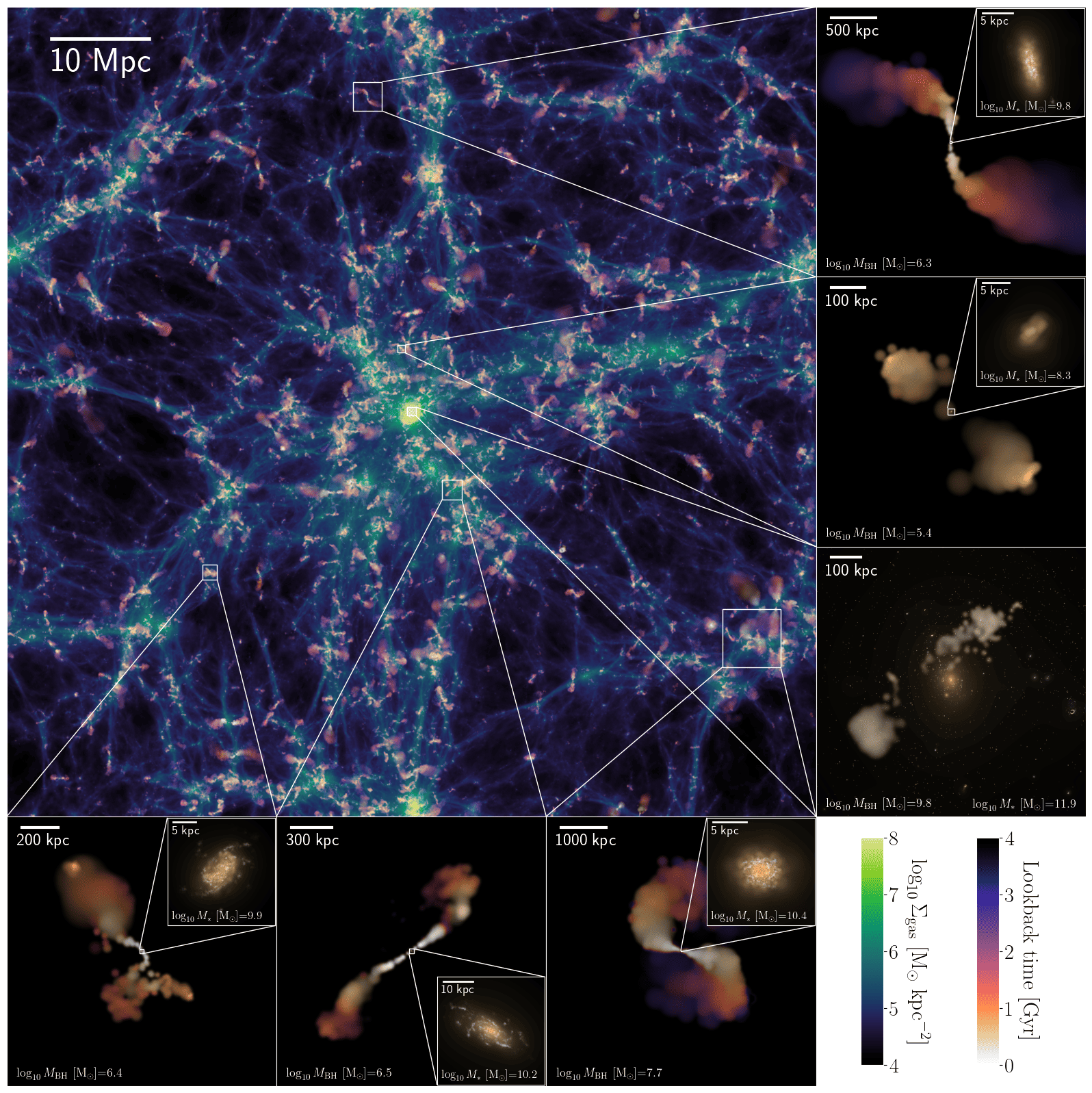}
\caption{Visualization of jet activity in the L100m6h simulation at $z=0.2$ in proper coordinates. The background image shows the cosmic web using the gas surface density, while particles kicked into jets by black holes are displayed using both information on the elapsed time since when they were kicked into jets (colour) and their surface density (opacity). Side panels zoom in on individual jets of interest, with further zoom-ins onto their host galaxies (except in the case of a cluster and its BCG) displayed using luminosities in the SDSS i, r and g bands assigned to RGB colours.}
\label{fig:cosmic_web_jets}
\end{figure*}

Here we combine the different aspects of the hybrid AGN feedback model discussed above into a three-state model. We also add winds to the thick disc at low Eddington ratios, and introduce accretion efficiencies, which take account of the fact that the vast majority of the mass accreting through a thick or slim disc is blown away by these winds. The model builds upon the AGN feedback model used for the OWLs and EAGLE simulations (\citealt{Booth2009}, \citealt{BoothSchaye2010}, \citealt{Schaye2015}). It has been integrated into the \textsc{swift} simulation code and included in the COLIBRE\footnote{\url{https://colibre-simulations.org}} galaxy formation simulation project (\citealt{Schaye2025}), as an alternative AGN feedback model alongside its primary thermal-only model. In Fig.~\ref{fig:cosmic_web_jets} we show the widespread and large-scale impact of jet activity at $z=0.2$ in one of our hybrid AGN feedback simulations run as part of the COLIBRE project. We describe COLIBRE in \S~\ref{sec:COLIBRE}. The main purpose of this study is to provide a comprehensive and self-contained description of the hybrid AGN feedback model (\S~\ref{sec:hybrid_model}), as well as to describe its calibration as part of the COLIBRE project (\S~\ref{sec:calibration_first_results}). We also provide predictions for BH spin (\S~\ref{sec:BHspin}) and a first analysis of the operation of AGN feedback in the hybrid simulations (\S~\ref{sec:feedback}). We compare with the simulations using purely thermal AGN feedback, focusing on the energetics of feedback. In \S~\ref{sec:conclusions} we summarize and conclude.

\section{The COLIBRE model and simulations}
\label{sec:COLIBRE}

Here we will describe the simulation methods and the COLIBRE galaxy formation model. We provide a summary, and refer the reader to \cite{Schaye2025}, as well as individual methods papers, for further details. The simulations are performed using a version of the \textsc{swift}\footnote{\url{https://swiftsim.com}} code (\citealt{Schaller2024}) in which the COLIBRE model has been included. In \S~\ref{sec:simulation_methods} we first describe the broad aspects of the simulation methods. In \S~\ref{sec:runs} we present the different resolutions and simulation runs that are used in this study. In \S~\ref{sec:COLIBRE_subgrid} we summarize the COLIBRE subgrid model. BHs and AGN feedback are also included in the fiducial thermal model, but we describe these separately in \S~\ref{sec:black_holes}, and we do so in more detail than for other aspects of COLIBRE, since BH physics is particularly relevant for this study. 

\subsection{Simulation methods}
\label{sec:simulation_methods}

\subsubsection{Initial conditions}

The simulations include four particle types: DM, gas, stars and BHs. The first two are included in the initial conditions (ICs), which are created at redshift $z=63$, while the second two are created during the simulation. Four times as many DM particles as gas particles are included in the ICs, which results in the particle masses of the two types being nearly equal. The large number of DM particles suppresses spurious energy transfer from DM to baryons in galaxies (\citealt{Ludlow_2019}, \citealt{Ludlow2023}), which can affect galaxy properties negatively. The ICs are created using {\scshape monofonic} (\citealt{Hahn_2020}, \citealt{Michaux2021}).

\subsubsection{Gravity and hydrodynamics}

Gravitational forces are computed using a fourth-order fast multipole method (\citealt{Greengard1987}, \citealt{Dehnen2014}), and softened using a softening length $\epsilon$. The effects of massive neutrinos, which to first order influence the expansion rate of the universe, are included using the linear response method (\citealt{Ali-Hamoud2013}). The equations of hydrodynamics are solved using the SPHENIX SPH scheme (\citealt{Borrow2022}). SPHENIX is a density-energy SPH scheme that includes artificial viscosity and conduction, as well as respective limiters of the two. 



\subsubsection{Structure finding}

As the simulation progresses, a friends-of-friends (FoF) algorithm is run to identify groups, but also to seed BHs (\S~\ref{sec:seeding}). The FoF algorithm is run on the DM particles starting from $a=0.05$, after every $\Delta a = 0.00751a$, using a linking length of 0.2 times the mean distance between them. Other particles are `attached' to FoF groups based on their nearest DM particles within the linking length. The \textsc{hbt-herons} (\citealt{Forouhar2025}) subhalo finder, an updated version of \textsc{hbt+} (\citealt{Han2018}), is then run using the catalogues and the snapshots. FoF groups with fewer than 32 particles of all types are discarded when constructing halo catalogues. \textsc{hbt-herons}, unlike most other halo finders, uses temporal information to assign hierarchical (sub)halo membership, alongside dynamical boundness criteria. Finally, we run the \textsc{soap} post-processing tool (\citealt{McGibbon2025}) to calculate all relevant galaxy and halo properties in spherical and projected apertures of fixed proper size, inclusive or exclusive of unbound particles and self-bound substructures. The fiducial choice for galaxy properties in this study is spherical exclusive apertures of 50 proper kpc (hereafter pkpc) in radius. These are centred on the most bound particle of each subhalo.

\subsection{Runs}
\label{sec:runs}

The COLIBRE simulations assume the ‘3×2pt + All Ext.’, spatially-flat $\Lambda$CDM cosmology from the Dark Energy Survey (DES) year three results (\citealt{Abbott_2022}). The parameter values are $h = 0.681$, $\Omega_\mr{m}=0.306$, $\Omega_\mr{CDM}=0.256011$, $\Omega_\mr{b}=0.0486$, $\Omega_\mr{\Lambda}=0.693922$, $\Omega_\mr{\gamma}=0.000053$. For the definitions of these parameters, as well as other cosmological parameters and their values, see either \cite{Schaye2025} or \cite{Abbott_2022}. 

In this study, we mostly use simulations with a (50 Mpc)$^3$ comoving volume (hereafter L050) at m7 resolution (herafter L050m7). Here, `m7' refers to the initial mean particle mass (i.e.~the mass resolution), which is $m_\mr{g}=1.47\times10^7$ M$_\odot$ for gas particles, while DM particles have $m_\mr{DM}=1.94\times10^7$ M$_\odot$. The gravitational softening length for all particles is $\epsilon=1.4$ kpc in proper coordinates at $z<1.57$, and $\epsilon=3.6$ kpc in comoving coordinates at higher redshifts. We also run simulations at m6 and m5 resolutions, for which the masses of all particles are decreased by successive factors of 8; the gas particle masses are $m_\mr{g}=1.84\times10^6$ M$_\odot$ and $m_\mr{g}=2.3\times10^5$ M$_\odot$, respectively. The gravitational softening lengths are reduced by successive factors of two relative to m7.

In addition to L050, we used L025 and L012.5 simulations at m6 and m5 resolution during the calibration process. In this paper we also present results from L100 (at m6 and m7) and L200 (at m7) volumes run with the hybrid AGN feedback model, which we refer to as L100m6h, L100m7h and L200m7h. Here, `h' refers to the fact that these were run with the hybrid model, unlike the fiducial simulations that use purely thermal AGN feedback.

\subsection{Subgrid physics}
\label{sec:COLIBRE_subgrid}

\subsubsection{Radiative cooling and heating}

Unlike in many previous large-volume cosmological simulations, gas is allowed to cool down to $T\approx10$ K, i.e.~no pressure floor is applied to prevent cooling below warm interstellar medium (ISM) temperatures ($\approx10^4$ K). The radiative cooling and heating rates are calculated with \textsc{hybrid-chimes} (\citealt{Ploeckinger2025}), which uses the \textsc{chimes} reaction network (\citealt{Richings2014a, Richings2014b}) to calculate the species abundances of electrons and nine hydrogen and helium species in non-equilibrium. The contribution to cooling and heating due to an additional 147 metal species is accounted for using tables that assume steady-state chemistry and ionization equilibrium. In \textsc{hybrid-chimes}, the pre-tabulated metal cooling rates are corrected by the non-equilibrium electron density throughout the simulation, therefore also including important non-equilibrium effects in the metal cooling rates. The parametrization of the interstellar radiation field, the cosmic ray rate, and the shielding column density is linked to the local Jeans length, as introduced in \cite{Ploeckinger2020}, with the updates  described in \cite{Ploeckinger2025}.

\subsubsection{Chemistry, dust, and diffusion}

Chemical enrichment of gas by heavy elements (described in detail in Correa et al.~2025, submitted.) is assumed to occur due to mass loss of Asymptotic Giant Branch (AGB) stars, massive stars, core-collapse supernovae (CCSN), SNIa, neutron star mergers, common-envelop SN jets and collapsars. 14 elements are tracked, 11 of which contribute to radiative cooling and heating: H, He, C, N, O, Ne, Mg, S, Si, Ca, Fe (S and Ca are not explicitly tracked, but solar abundances relative to Si are assumed for these two elements). In addition, Ba, Sr and Eu are also tracked. The chemical yields of the processes that generate these elements depend on the metallicity of star particles (inherited from the gas metallicities of particles from which stars are born in COLIBRE). Gas particles gain mass in the simulation as they are enriched. Gas particles that reach a mass of $4m_\mr{g}$, where $m_\mr{g}$ is the mean initial baryonic particle mass, are split into two.

The dust grain model, which is described in detail in \cite{Trayford2025}, includes 3 grain compositions (graphite, forsterite and fayalite) and 2 grain sizes (0.01 and 0.1 $\mu$m), and is coupled to the radiative cooling module. Gas particles track their fractions of mass contained in the relevant dust grains types. Grains are assumed to be seeded by AGB stars and CCSN using metallicity-dependent dust yields. The growth and destruction of dust grains in the ISM are modeled on the fly. The model includes accretion onto grains, as well as sputtering (destruction by collisions with hot gas particles), grain collisions (shattering and coagulation), and astration. Dust is also directly destroyed in gas particles that are affected by thermal SN and AGN feedback.

The turbulent diffusion model (Correa et al.~2025, submitted) accounts for mixing below the resolution scale using the diffusion equation, with a diffusion coefficient that depends on the local turbulent velocity dispersion. It is applied to elemental mass fractions, mass fractions of dust grains of different compositions and sizes, and diagnostic tracers that follow material emerging through different stellar mass loss channels. A time step condition is applied to ensure that diffusion is sampled well.

\subsubsection{Star formation and mass loss}

The star formation model is detailed in \cite{Nobels2024}. Star formation follows the Schmidt law: $\dot{m}_*=\varepsilon_\mr{ff} m_\mr{g}/t_\mr{ff}$, where $t_\mr{ff}$ the gravitational free-fall time-scale and $\varepsilon_\mr{ff}=0.01$ the star formation efficiency per free-fall time. Only gas that is gravitationally unstable on the scale of the SPH kernel is eligible to form stars, with both the local velocity dispersion and thermal pressure contributing to the stability of the gas against collapse. Stellar particles represent simple stellar populations (SSPs) of subgrid stars whose masses are distributed according to the \cite{Chabrier2003} initial mass function (IMF) in the range $0.1-100$ M$_\odot$. Once formed, star particles lose mass due to winds and SN. Most of the mass loss occurs in the first 40 Myr once a stellar particle has formed. To sample this mass loss phase well, star particles are limited to time steps no longer than 1 Myr for the first 40 Myr of their evolution.

\subsubsection{Stellar feedback}
\label{sec:stellar_feedback}

The main channel of stellar feedback in COLIBRE is SN (SNIa and CCSN, the latter being dominant). The energy associated with SNe is released stochastically, mostly in thermal form and using relatively large temperature increments so that numerical overcooling is suppressed (\citealt{DallaVecchia2012}). The gas neighbours targeted for feedback are chosen isotropically (see \citealt{Chaikin2022}). The heating temperatures used are given by
\begin{equation}
\Delta T_\mr{SN} = 10^{6.5}\hspace{1mm} \mathrm{K} \hspace{1mm}\bigg(\frac{n_\mr{H}}{n_\mr{H,pivot}}\bigg)^{2/3},
\label{eq:SN_dT}
\end{equation}
where $n_\mr{H}$ is the SPH-smoothed gas number density at the location of the star particle, and $n_\mr{H,pivot}$ a free parameter (different values are used for the thermal and hybrid AGN feedback models; see \S~\ref{sec:calibration_results}). The index of 2/3 is motivated by the derivation by \cite{DallaVecchia2012} for a minimum heating temperature required to prevent numerical overcooling. Minimum and maximum heating temperatures are also used, and they depend on resolution; see \cite{Chaikin2025} and \cite{Schaye2025}.

In addition to the variable heating temperature, the model includes variable SN energies (for CCSN feedback, but not for SNIa). The energies injected per unit time by CCSN are not equal to those expected from the SSP represented by the particle (given its age and metallicity), but rather $f_\mr{E}$ times that value. $f_\mr{E}$ for CCSN can be both smaller and larger than unity, and varies with the stellar birth pressure, with minimum and maximum values between $0.1$ and $4$ (but dependent on resolution). This variation is required because the stochastic thermal feedback may result in an over- or underestimate of the true radiative losses, which cannot be predicted robustly at our numerical resolution (\citealt{Chaikin2025}). Therefore, calibration of $f_\mr{E}$ is necessary if one wishes to reproduce both galaxy masses and sizes simultaneously. For CCSN feedback, a fraction $f_\mr{kin}=0.1$ of the energy is injected kinetically in the form of low-velocity kicks, using a target velocity of 50 km~s$^{-1}$. The kicks are done in pairs, in random but opposite directions in a way that conserves energy, momentum and angular momentum (\citealt{Chaikin2023}). The role of kinetic feedback is to generate turbulence in the ISM, helping to stabilize the gas against star formation.

COLIBRE also includes early stellar feedback alongside CCSNe (despite calling these `early', these processes continue to operate after the SN phase). This includes: 1) stellar winds, 2) radiation pressure and 3) H~\textsc{ii} regions. Early stellar feedback is detailed in \cite{BenitezLlambay2025}. Stellar winds and radiation pressure are both modeled as momentum injections, which depend on the metallicity and age of the SSP. These momentum injections are modeled as kinetic feedback with velocity increments of 50 km~s$^{-1}$.


\subsection{Black holes}
\label{sec:black_holes}

Here we summarize how supermassive BHs are treated in the fiducial COLIBRE model, which uses thermal AGN feedback. These prescriptions are described in more detail in \cite{Schaye2025}; see also \cite{Nobels2022} and \cite{Bahe2022}. In \S~\ref{sec:hybrid_model}, where we introduce the hybrid model of AGN feedback, we explain the differences of that model relative to the one presented in this section. 

\subsubsection{Seeding}
\label{sec:seeding}

BH seeding follows the approach introduced by \cite{DiMatteo2003}: they are seeded in haloes identified by the FoF algorithm (\S~\ref{sec:simulation_methods}) whose mass exceeds $M_\mr{FoF,seed}$ and that do not yet have a BH. BHs are seeded with a mass of $M_\mr{BH,seed}$ at the location of the densest gas particle in the FoF group, which is effectively converted into the collisionless BH particle. BHs carry a subgrid mass, $M_\mr{BH}$, and a particle mass, $M_\mr{BH,part}$ (initially equal to the mass of the gas particle that created the BH). The subgrid mass is the physical, true mass of the BH. In the early evolution of the BH, its subgrid mass is generally much lower than the mass of the other particles in the simulation. The particle mass is thus used for all N-body gravitational interactions with other particles within the code, since large differences between masses used in the gravity calculations lead to undesirable effects. As the BH grows through accretion, it eventually reaches $M_\mr{BH}\approx M_\mr{BH,part}$.

For m7 resolution, $M_\mr{FoF,seed}=5\times10^{10}$ M$_\odot$, while for m6 and m5, $M_\mr{FoF,seed}=10^{10}$ M$_\odot$. The BH seed mass, $M_\mr{BH,seed}$, is a free parameter with values of order $\approx10^4-10^5$ M$_\odot$, corresponding to the direct collapse seed formation scenario. Its value affects the onset of BH growth and AGN feedback. It is used in the calibration of the model at each resolution (\citealt{Chaikin2025}), and different values are used for the thermal and hybrid AGN feedback simulations (\S~\ref{sec:calibration_results}). 

In the hybrid simulations, we seed BHs with an initial spin value of $\vert a \vert=0.01$, directed randomly (see \S~\ref{sec:hybrid_model}, and more specifically \S~\ref{sec:BH_spin}, for details on BH spin). This effectively assumes that BHs are not spinning at the time they form, but we avoid using $\vert a \vert=0$, since this can lead to numerically unstable expressions in our model. We tested seeding BHs with higher spins, up to $\vert a \vert=1$, but found that this had no appreciable affect on either BH or galaxy evolution.

\subsubsection{Gas accretion}

Black holes are assumed to only accrete from gas in COLIBRE. In the hybrid AGN feedback model, there are two relevant mass accretion rates: $\dot{M}_\mathrm{acc,d}$, the accretion rate onto a subgrid accretion disc (hereafter the disc-scale accretion rate), and $\dot{M}_\mathrm{acc,H}$, the accretion rate onto the BH itself (i.e.~its event horizon, hereafter the horizon-scale accretion rate). The first of these accretion rates is generally agnostic to what occurs on subgrid (accretion-disc) scales, while the second can be heavily affected by assumptions about subgrid accretion discs and feedback. These two accretion rates can be very different in the hybrid model (see \S~\ref{sec:accretion_rates}), since we include subgrid processes occurring on accretion disc scales in more detail in the hybrid model.

In the thermal AGN feedback model, mass is assumed to accrete unimpeded from disc scales all the way into the event horizon. In other words, $\dot{M}_\mathrm{acc,d}=\dot{M}_\mathrm{acc,H}$. The BH accretion rate in the thermal AGN feedback model (and the disc-scale accretion rate in the hybrid model) is given by
\begin{equation}
\dot{M}_{\text {acc,d }}=\dot{M}_{\mathrm{BHL}} \frac{f_{\text {turb }} f_{\text {ang }}}{\left(f_{\text {turb }}^{2}+f_{\text {ang }}^{2}\right)^{1 / 2}},
\label{eq:accr_rate}
\end{equation}
where $\dot{M}_{\mathrm{BHL}}$ is the Bondi-Hoyle-Lyttleton (BHL) accretion rate (\citealt{Bondi}, \citealt{Hoyle1939}), given by
\begin{equation}
\dot{M}_{\mathrm{BHL}}=4 \pi G^{2} \frac{\rho_{\mathrm{g}}}{c_{\mathrm{s}}^{3}} M_{\mathrm{BH}}^{2},
\label{eq:BHL}
\end{equation}
with $\rho_\mr{g}$ and $c_\mr{s}$ the kernel-weighted gas density and sound speed at the location of the BH, and $f_\mathrm{turb}$ and $f_\mathrm{ang}$ are correction terms that account for the suppression of the accretion rate due to turbulence and vorticity, respectively (\citealt{Krumholz2006}). These corrections are due to turbulence and vorticity on resolved scales, well outside a subgrid accretion disc. The modified BHL formula may work better with a realistic multiphase ISM that has non-zero angular momentum with respect to the BH, and perform better than the traditional formula in the context of galaxy formation (\citealt{Hopkins2011}, \citealt{Hobbs2012}, \citealt{Angles2021}, \citealt{Gaspari2013}).

The turbulence limiter is given by 
\begin{equation}
f_\mr{turb}=\left[\frac{\lambda^{2}+\mathscr{M}^{2}}{\left(1+\mathscr{M}^{2}\right)^{4}}\right]^{1 / 2},
\label{eq:f_turb}
\end{equation}
where $\lambda=1.1$ and $\mathscr{M}=\sqrt{(v^2+\sigma_\mr{g}^2)/c_\mr{s}^2}$ is the Mach number, with $v$ the kernel-interpolated bulk velocity of gas relative to the BH, and $\sigma_\mr{g}$ the non-thermal velocity dispersion of the gas in the BH frame. For low Mach numbers, $f_\mr{turb}$ asymptotes to $\lambda$, while for high Mach numbers, it asymptotes to $1/\mathscr{M}^3$. In the latter case, the accretion rate scales as $\dot{M}_\mr{BHL}\propto1/(v^2+\sigma_\mr{g}^2)^{3/2}$, with no dependence on the sound speed. In the case that $v\gg\sigma_\mr{g}$, the accretion rate corresponds to the classical limit $\dot{M}_\mr{BHL}\propto1/v^3$ for supersonic bulk flow (\citealt{Hoyle}).

The vorticity limiter is given by (\citealt{Krumholz2006})
\begin{equation}
f_{\text {ang }}=\frac{0.34}{1+\omega_{\star}^{0.9}},
\label{eq:f_ang}
\end{equation}
where $\omega_{\star}=\omega R_{\mathrm{B}} / c_{\mathrm{s}}$, $R_{\mathrm{B}}=G M_{\mathrm{BH}} / c_{\mathrm{s}}^{2}$ is the Bondi radius and $\omega=|\nabla \times \mathbf{v}|$ is the kernel-weighted vorticity of the ambient gas. For a negligible amount of vorticity, $f_\mr{ang}$ asymptotes to $0.34$, and the accretion rate is suppressed by that factor relative to BHL rate with no vorticity. This suppression is a result of accretion proceeding through a torus (rather than spherically), which is always expected to be present on some scales, even if the angular momentum is very small. For a large vorticity, the accretion rate is suppressed even more.

The accretion rate is limited to 100 times the Eddington rate\footnote{This limit is a compromise between allowing effectively uncapped BH accretion, and not allowing extremely high accretion rates that may lead to numerical issues.}, which is given by:
\begin{equation}
\dot{M}_{\mathrm{Edd}}=\frac{4 \pi G m_{\mathrm{p}} M_{\mathrm{BH}}}{\epsilon_{\mathrm{rad}} \sigma_{\mathrm{T}} c},
\label{eq:Edd}
\end{equation}
where $\sigma_\mr{T}$ is the Thomson cross-section and $\epsilon_\mr{rad}$ the accretion efficiency of a thin accretion disc. We assume $\epsilon_\mr{rad}=0.1$ in the thermal AGN feedback model and use the BH spin-dependent formula in the hybrid AGN feedback model (\citealt{NovikovThorne1973}, Eqn.~\ref{eq:eps_rad_NT}).

The (subgrid) black hole mass evolves according to
\begin{equation}
\dot{M}_\mr{BH}=(1-\epsilon_\mr{rad})\dot{M}_\mr{acc,d},
\label{eq:mass_evolution_thermal}
\end{equation}
in the thermal AGN feedback model, since a fraction $\epsilon_\mr{rad}$ of the accreting mass is radiated away as energy. In the hybrid AGN feedback model, we replace this equation by a similar equation including additional efficiency terms corresponding to the accretion disc winds and jets (see Eqn.~\ref{eq:net_accr_rate}). Furthermore, in that version, the accretion rate onto the disc, $\dot{M}_\mr{acc,d}$, is replaced with the accretion rate through the inner boundary of the accretion disc, into the BH event horizon, which is suppressed relative to $\dot{M}_\mr{acc,d}$ by an accretion efficiency factor, $\epsilon_\mr{acc}\leq1$ (accounting for mass loss due to accretion disc winds, as mass flows through an accretion disc from larger scales).

BHs grow in the simulation by `nibbling' rather than swallowing entire gas particles (\citealt{Bahe2022}). BHs are seeded on a gas particle that is converted into a BH particle, with the BH subgrid mass, $M_\mr{BH}$, much smaller than the initial particle mass, $M_\mr{BH,part}$. The subgrid BHs therefore initially `accrete' from the particle that they are attached to since being seeded, as long as $M_\mr{BH}<M_\mr{BH,part}$. In this phase, the particle mass does not grow, while the subgrid mass does. Once the subgrid mass $M_\mr{BH}$ exceeds the particle mass $M_\mr{BH,part}$, BHs `nibble' on surrounding particles in order to conserve mass and energy. The nibbling is done in a kernel-weighted fashion, but only on gas particles whose mass is larger than half their initial baryonic mass. In the case that the BH grows from its particle mass reservoir rather than through nibbling on gas neighbours, in the thermal AGN feedback model we reduce the particle mass at the rate $\dot{M}_\mr{BH,part}=-\epsilon_\mr{rad}\dot{M}_\mr{acc,d}$, which is consistent with Eqn.~(\ref{eq:mass_evolution_thermal}), and which accounts for energy losses. In the hybrid AGN feedback model, we replace the radiative efficiency in this equation with the sum of all energy release efficiencies, and the disc-scale accretion rate with the accretion rate into the BH horizon.

\subsubsection{Mergers}

In order to merge, BHs must be located within 3 softening lengths of each other, the less massive BH needs to be within the SPH kernel of the more massive BH, and the relative velocity of the BHs must not be too large: $\Delta v < \sqrt{2G(M_\mr{BH,part,1}+M_\mr{BH,part,2})/r}$, where $M_{\mr{BH,part},i}$ is the particle mass and $r$ the separation between the two BHs. This condition ensures that the relative velocity does not exceed the escape velocity of the system. When merging the BHs, we also conserve linear momentum. During a merger, a fraction of the total mass-energy of the two BHs is lost to gravitational waves. For the thermal AGN feedback model, this fraction is equal to
\begin{equation}
\frac{E_\mr{GW}}{(M_\mr{BH,1}+M_\mr{BH,2})c^2}=0.057191\nu+0.543763\nu^2,
\label{eq:E_GW}
\end{equation}
which is accurate to order $\nu^2$, where $\nu=M_\mr{BH,1}M_\mr{BH,2}/(M_\mr{BH,1}+M_\mr{BH,2})^2$ is the symmetric mass ratio. This formula is based on numerical relativity simulations (\citealt{Barausse2012}), is valid for non-spinning BHs, and accounts for the losses during the initial BH inspiral, the final plunge, the merger, and the ringdown. In the hybrid AGN feedback model, we use a BH spin-dependent formula (\S~\ref{sec:BH_mergers}, Eqn.~\ref{eq:E_rad}), taken from the same work, of which the above is a special case. In the non-spinning case, the mass loss can reach $\approx5$ per cent for equal-mass mergers. In the maximally spinning, counter-aligned case, this can reach $\approx12$ per cent. We do not implement gravitational-wave recoils.

\subsubsection{Repositioning}

Instead of modeling dynamical friction acting onto BHs, whose effects we cannot reliably resolve in our simulations, we reposition them manually to the local minimum of the potential well, following the prescription detailed in \cite{Bahe2022}. At every time step, the BH is instantaneously moved to the position of the neighbouring gas particle with the lowest gravitational potential that satisfies the following criteria. The particle must: 1) be within the SPH kernel of the BH, 2) be within 3 gravitational softening lengths of the BH and 3) have a lower potential than that at the BH's location (excluding the BH's own contribution). If no such particle is found, the BH does not reposition. When repositioning, we do not alter the BH's velocity.

\subsubsection{Thermal AGN feedback}
\label{sec:thermal_feedback}

The thermal AGN feedback model follows the prescription of \cite{Booth2009}. Instead of continuous energy injection at every time step, energy from accretion is stored in a reservoir until it reaches a value large enough to heat a single gas neighbour by a large amount, to prevent numerical overcooling and drive powerful outflows (for similar reasons as for SN feedback, \S~\ref{sec:stellar_feedback}). The rate at which the energy reservoir is incremented at every time step is the thermal feedback power
\begin{equation}
    P_\mr{th}=\epsilon_\mr{f}L_\mr{bol}=\epsilon_\mr{f}\epsilon_\mr{rad}\dot{M}_\mr{acc,d}c^2,
\label{eq:P_th}
\end{equation}
where $L_\mr{bol}$ is the (emitted) bolometric luminosity of the AGN, $\epsilon_\mr{rad}=0.1$ is the assumed thin disc radiative efficiency, and $\epsilon_\mr{f}$ the coupling efficiency, a parameter that determines what fraction of the emitted radiation couples to gas (either on accretion disc or larger scales, and either by directly heating it or driving a wind through radiation pressure that then shocks on subgrid scales). The coupling efficiency is calibrated on the BH mass$-$stellar mass relation (\S~\ref{sec:calibration_results}), with different values (between 0.02 and 0.1) adopted in the thermal and hybrid AGN feedback models, and at different resolutions. In the hybrid AGN feedback model, a version of the above equation is applied only at moderate Eddington ratios (corresponding to the thin disc state), while otherwise (in the thick and slim disc), the factor $\epsilon_\mr{f}\epsilon_\mr{rad}$ is replaced by $\epsilon_\mr{wind}$, which corresponds to an accretion disc wind that is shocked on subgrid scales and does energy-driven feedback on larger scales (\S~\ref{sec:thermal_feedback}). In the hybrid AGN feedback model, the accretion rate $\dot{M}_\mr{acc,d}$ is also suppressed by the accretion efficiency factor, $\epsilon_\mr{acc}$.

The energy required for a single thermal AGN feedback event is
\begin{equation}
    \Delta E_\mr{AGN}=\frac{m_{\mathrm{g}} k_{\mathrm{B}} \Delta T_{\mathrm{AGN}}}{(\gamma-1) \mu m_{\mathrm{p}}},
\label{eq:delta_E}
\end{equation}
where $\gamma=5/3$ is the adiabatic index for monatomic gas, and $\mu=0.6$, valid for ionized gas of primordial composition. The heating temperature for a single feedback event scales with BH mass:
\begin{equation}
    \Delta T_\mr{AGN}=10^9 \hspace{1mm} \mathrm{K} \hspace{1mm} \bigg(\frac{M_\mr{BH}}{10^8\hspace{0.5mm}\mathrm{M}_\odot}\bigg),
\label{eq:delta_T_AGN}
\end{equation}
with a minimum of $\Delta T_\mr{AGN,min}=10^{6.5}$ K and a maximum of $\Delta T_\mr{AGN,max}=10^{9.5}$ K at m7 resolution and $\Delta T_\mr{AGN,max}=10^{10}$ K at m6 and m5 resolutions. While the feedback efficiencies associated with thermal feedback change in the hybrid AGN feedback model, we use the same heating temperatures as for the thermal model. 

The variation of the heating temperature with BH mass, as opposed to the use of a large, constant value such as in EAGLE ($\Delta T_\mr{AGN}=10^{9}$ K, \citealt{Schaye2015}), has a few benefits, all of which are connected. First, it allows better sampling of AGN feedback in low-mass galaxies. In other words, a constant value biases the simulations towards better resolving AGN feedback in more massive systems. While it is typically assumed that that regime is the only one where AGN feedback is important, variable feedback temperatures (and jet kick velocities; \S~\ref{sec:AGN_implementation}) is what allows the hybrid AGN feedback model to feature significant AGN effects in galaxies with masses of order $M_*=10^9$ M$_\odot$ or even lower, rather than only at $M_*\gtrsim10^{10.5}$ M$_\odot$ (as we show in \S~\ref{sec:velocity_variations}). Second, using the specific formulation $\Delta T_\mr{AGN}\propto M_\mr{BH}$ ensures that, as BHs grow, they do the same number of heating events per unit of growth on a logarithmic scale. This again means that feedback from higher-mass BHs is not preferred over feedback from lower-mass ones, in terms of how well resolved it is. Finally, using variable heating temperatures allows one to implement feedback more gently in the early stage of BH growth. 


In addition to gravitational time steps, BHs use a time step condition $\Delta t_\mr{AGN} \leq\Delta E_\mr{AGN}/P_\mr{th}$, which ensures that BHs are woken up by the time they need to heat one gas neighbour. In the hybrid AGN feedback model, we apply an equivalent time step limiter for jet feedback, and two additional time step conditions related to BH spin evolution (\S~\ref{sec:timesteps}). A further minimum of 100 yr is applied to BH time steps (both in the thermal-only and hybrid AGN feedback simulations). This is very short, but it is sometimes reached in the m5 simulations, since extremely high densities are reached at that resolution.

\subsection{Parameters and calibration}

In \S~\ref{sec:calibration_first_results}, we discuss the strategy adopted for calibrating the parameter values of the hybrid AGN feedback model (\S~\ref{sec:calibration_first_results}). The strategy is different to the one used for calibrating the thermal AGN feedback simulations, which is discussed in detail in \cite{Chaikin2025}. Briefly, the final calibration involved simulations with a comoving volume of (50 cMpc)$^3$ at m7 resolution, in a Latin hypercube configuration of parameter values. Gaussian process emulation was used to infer the best-fitting parameter values. These parameters (see below for the list) are present also in the hybrid AGN feedback model. The best-fitting parameter values found by \cite{Chaikin2025} for all three resolutions were used as initial guesses for the hybrid AGN feedback model. They are varied by hand, alongside new parameters specific to the hybrid AGN feedback model, and calibrated as discussed in \S~\ref{sec:calibration_first_results}.

The Gaussian process emulation employed by \cite{Chaikin2025} aimed to find the parameter values that simultaneously yield the best agreement with the $z=0$ galaxy stellar mass function (GSMF) data from \cite{Driver2022} (GAMA survey year 4 release) and $z=0$ galaxy size data from \cite{Hardwick2022}. Equal weights were given to both datasets, and the fitting was done only over the range $10^9<M_*<10^{11.3}$ M$_\odot$, due to resolution constraints at the low-mass end and box-size limitations at the high-mass end. 

Most of the calibration process involved varying four subgrid parameters: 1) $M_\mr{BH,seed}$, the BH seed mass, 2) $f_\mr{kin}$, the fraction of CCSN energy injected in kinetic form, 3) $n_\mr{H,pivot}$, the pivot density for SN heating temperatures, and 4) $P_\mr{E,pivot}$, the pivot stellar birth pressure at which the SN energy fraction ($f_\mr{E}$), which depends on stellar birth pressure, is halfway between its minimum and maximum value.

For the m6 and m5 resolutions, instead of employing emulation, slight adjustments to parameters were chosen by hand relative to their m7 values, to achieve similar agreement with the observational data. This process was done using L025 and L012p5 boxes for m6 and m5, respectively. In short, SN heating temperatures and pivot pressures increase, in order to suppress numerical overcooling, since higher densities are reached in higher-resolution simulations. For the same reason, BHs also grow more in higher-resolution simulations. To adjust for this effect, BH seed masses are reduced at higher resolution. For all three resolutions, the thermal AGN feedback coupling efficiency, $\epsilon_\mr{f}$, is adjusted independently of other parameters, to yield BH masses similar to $z=0$ observations.

\section{Hybrid AGN accretion and feedback model}
\label{sec:hybrid_model}

Here we introduce our hybrid AGN model for BH accretion, spin evolution and AGN feedback. We provide a full and self-consistent description. In \S~\ref{sec:overview} we provide a brief description of the model, while other subsections present the model in detail. For readers interested only in the outline of the model, we suggest to read \S~\ref{sec:overview} and skip the rest of this section.

In \S~\ref{sec:model_comparison} we compare our model to similar previously published ones. Different aspects of our model were already introduced and discussed in \cite{Husko2022_spin_driven, Husko_winds, Husko_2025_SE}. In Appendix~\ref{sec:appendix_model} we discuss how these different aspects are combined and what is different about our final model compared to these previously published, simpler versions. 

\subsection{Overview of the model}
\label{sec:overview}

\begin{figure*}
\includegraphics[width=1\textwidth, trim = 40 30 40 30]{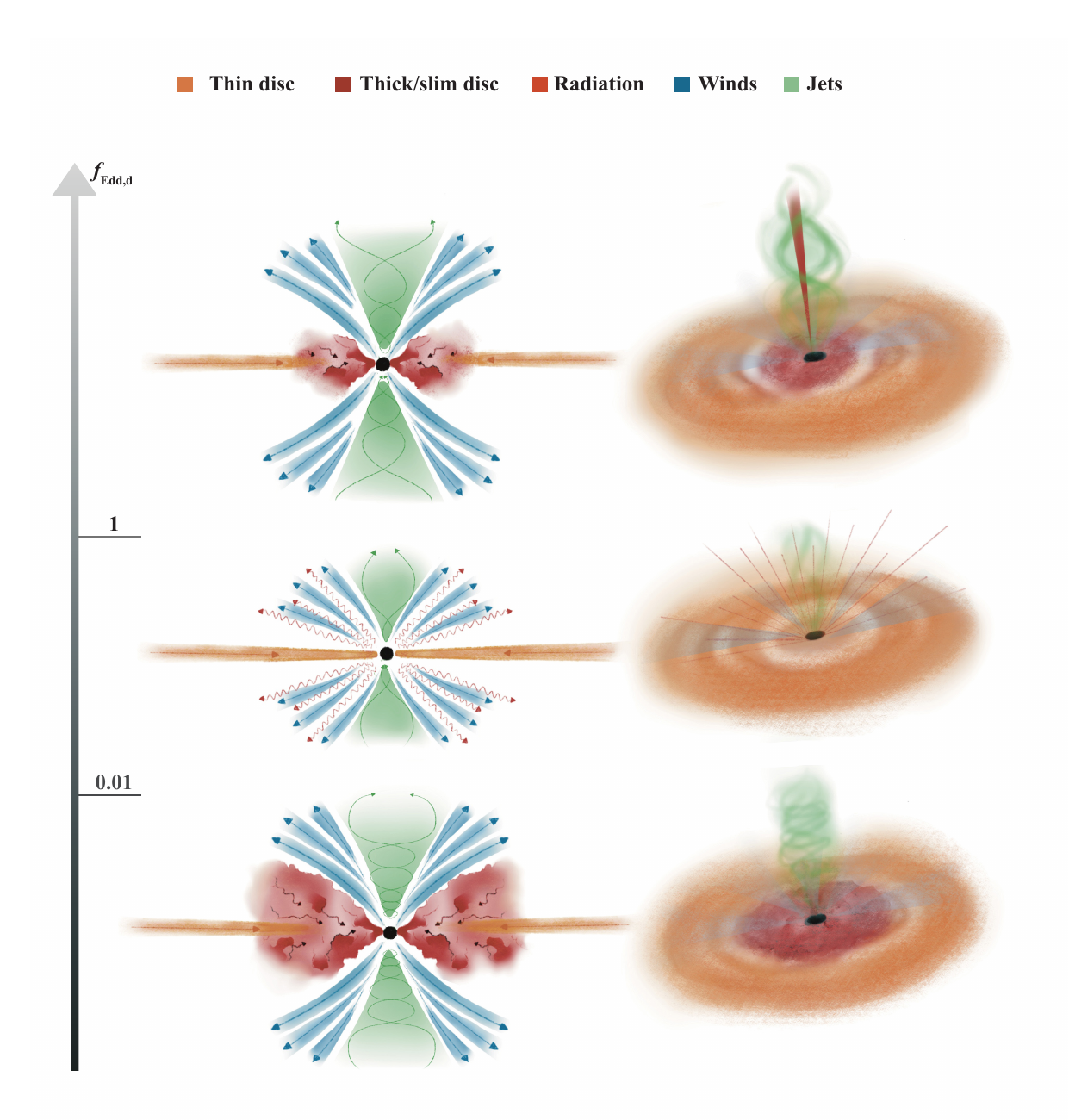}
\caption{The three different accretion disc states assumed in our model, corresponding to different Eddington ratio regimes, from bottom to top displaying the thick, thin and slim discs, with the thin disc component present at large radii. The accretion disc state changes depending on the disc-scale Eddington ratio, $f_\mr{Edd,d}$, defined in \S~\ref{sec:accretion_states}, at two critical values of $0.01$ and $1$. The left and right illustrations show edge-on and inclined views of the accretion disc, respectively. Radiation and winds are assumed to originate on accretion disc scales, and the effects of both are implemented using thermal isotropic feedback in the simulations. Jets, which consist purely of Poynting flux on these accretion disc scales, are implemented as bipolar velocity kicks.}
\label{fig:accretion_schematic}
\end{figure*}%

The BH accretion and feedback subgrid model that we present here has three accretion disc states, representing qualitatively different properties of accretion discs. These different states, namely the thick, thin and slim disc, occur at different Eddington ratios (low, moderate and super-Eddington, respectively). These accretion disc states are illustrated schematically in Fig.~\ref{fig:accretion_schematic}. We discuss them in detail in \S~\ref{sec:accretion_states}. It is important to emphasize that there is no subgrid accretion disc that we evolve (i.e.~no subgrid mass or angular momentum reservoir); we instead assume that accretion occurs from large scales onto the BH instantaneously. However, assumptions about accretion disc properties can significantly affect mass accretion, BH spin evolution and feedback even in this simplified scenario.

The model includes two AGN feedback modes: thermal isotropic and kinetic jets. The thermal isotropic mode represents the effects of winds that are adiabatically shocked on subgrid scales, as well as effects of radiation (which may couple to gas on larger than accretion disc scales). Both feedback modes are active in all three accretion disc states, resulting in a total of six effective feedback channels. In practice, we find that only three of the six channels are actually relevant for the galaxy and BH population as a whole: 1) jets at low Eddington ratios (at low redshifts), 2) jets at moderate Eddington ratios (at high redshifts), and 3) thermal feedback at moderate Eddington ratios (at all redshifts).

In our model, BH growth can be significantly impeded by the direct effects of feedback, for two separate reasons. First, the feedback efficiencies can be very high and result in net mass loss by BHs. Second, our model assumes that subgrid accretion disc winds strip away most of the accreting material at low and super-Eddington accretion rates, which we represent using an accretion efficiency parameter that quantifies this mass loss. This parameter relates the accretion rate onto a subgrid accretion disc to the accretion rate through the BH event horizon. The BH mass evolution and various accretion rates related to the above are discussed in \S~\ref{sec:accretion_rates}.

In order to realistically employ feedback (especially concerning jets), we model the evolution of the angular momentum of a BH, $\mathbf{J}_\mr{BH}$ (see \S~\ref{sec:BH_spin} and \S~\ref{sec:BH_mergers}). The dimensionless BH spin parameter (hereafter the BH spin) has a magnitude defined as 
\begin{equation}
    a = \frac{J_\mr{BH}c}{M_\mr{BH}^2G},
\label{eq:BH_spin}
\end{equation}
where $J_\mr{BH}=\vert\mathbf{J}_\mr{BH} \vert$ is the angular momentum magnitude, and a sign which indicates whether accretion is prograde or retrograde (either an aligned or counter-aligned inner accretion disc). The BH spin direction is used as the direction for jet launching. Since the BH spin vector is effectively a time-integrated quantity, it is typically much more stable in time than the instantaneous direction of the angular momentum of gas near the BH. This facilitates stable jet launching and results in large-scale jet activity, as shown in Fig.~\ref{fig:cosmic_web_jets}. In addition to being used for jet directions, the BH spin also affects the strength of feedback (i.e.~the efficiencies).

Some of the feedback efficiencies in our model (the jet efficiency especially) depend on the magnetisation state of the accretion disc. We assume the largest possible (steady-state) magnetisation. In other words, we adopt the magnetically-arrested disc (MAD) assumption (\citealt{Narayan2003}), as discussed in \S~\ref{sec:magnetisation}. We stress that the feedback efficiencies in our model are largely not free parameters; they instead depend on BH spin and Eddington ratio, as discussed in \S~\ref{sec:feedback_effs}. The exception to this is the coupling efficiency of radiation from the thin accretion disc, but this same parameter exists in the purely thermal AGN feedback simulations. On the other hand, our model includes two new important parameters that are related to feedback, which set the values of accretion efficiencies in the thick disc state and jet velocities.

Due to the inclusion of jet feedback and BH spin, additional time-step conditions need to be adopted (see \S~\ref{sec:timesteps}). Our jet launching scheme is discussed in \S~\ref{sec:jet_scheme}, while the implementation of AGN feedback more generally is discussed in \S~\ref{sec:AGN_implementation}.


\subsection{Accretion disc states}
\label{sec:accretion_states}

We assume different accretion disc states depending on the Eddington ratio defined using the accretion rate onto the subgrid accretion disc (the disc-scale accretion rate), before mass loss due to winds. This Eddington ratio is defined as
\begin{equation}
    f_\mathrm{Edd,d}=\frac{\dot{M}_\mathrm{acc,d}}{\dot{M}_\mathrm{Edd}},
\label{eq:Eddington_ratio}
\end{equation}
where $\dot{M}_\mathrm{acc,d}$ is given by Eqn.~(\ref{eq:accr_rate}). For most calculations within our model, we use the Eddington ratio\footnote{To define the Eddington ratios in the hybrid AGN feedback model, following recent conventions, we use the BH spin-dependent Eddington accretion rate $\dot{M}_\mathrm{Edd}$ (Eqn.~\ref{eq:Edd}), in which the constant radiative efficiency $\epsilon_\mr{rad}=0.1$ is replaced with a BH spin-dependent one that is appropriate for a thin, relativistic disc (Eqn.~\ref{eq:eps_rad_NT}). This can lead to differences in Eddington ratios of up to a factor of 3.} defined as
\begin{equation}
    f_\mathrm{Edd,H}=\frac{\dot{M}_\mathrm{acc,H}}{\dot{M}_\mathrm{Edd}},
\label{eq:Eddington_ratio_net}
\end{equation}
which is the same as Eqn.~(\ref{eq:Eddington_ratio}), but with the disc-scale accretion rate, $\dot{M}_\mathrm{acc,d}$, replaced by the accretion rate into the BH (event horizon), $\dot{M}_\mathrm{acc,H}$ (see \S~\ref{sec:accretion_rates} for subtleties associated with this difference). This second Eddington ratio includes the effects of mass loss from the accretion disc due to winds and corresponds more closely to observed Eddington ratios. When deciding the nature of an accretion disc, one could use either of the two Eddington ratios. We discuss the merits of both options in Appendix~\ref{sec:appendix_transition_large_vs_net}, and based on the discussion therein, we choose to use the disc-scale Eddington ratio (Eqn.~\ref{eq:Eddington_ratio}) to decide the accretion disc state. 

The three accretion disc types we include in the model are the following:
\begin{enumerate}
    \item \textbf{Thick disc} for $f_\mathrm{Edd,d}<0.01$. This accretion flow is also known as the ADAF (advection-dominated accretion flow), ADIOS (advection-dominated inflow-outflow solution if winds are included), hot accretion flow, RIAF (radiatively-inefficient accretion flow), the hard state (in terms of X-ray spectra) and the low state (in terms of accretion rate). The disc is geometrically thick (with an aspect ratio $H/R\approx0.5$, $H$ being the scale height of the disc at a cylindrical radius $R$) and optically thin. The gas in this disc is very hot and relatively diffuse, with low radiative efficiencies and strong advection that traps most of the thermal energy. Gas orbits are not fully circular, instead having a significant radial component. The gas flow is continuous all the way down to the event horizon, with no abrupt change in properties at the radius of the innermost stable circular orbit ($R_\mathrm{ISCO}$). The poloidal magnetic flux at the event horizon of the black hole is very large, leading to strong jets if the BH is spinning. Winds driven by thermal pressure and MHD effects are also present, taking away most of the accreting mass. We take the solution for this type of disc from \cite{Narayan1995}, and we also use results from numerical GRMHD simulations (\citealt{Tchekhovskoy2010}, \citealt{Narayan2021}). BH spin evolution and AGN jet feedback in this accretion disc state was studied in detail in \cite{Husko2022_spin_driven}.
    \item \textbf{Thin disc} for $0.01<f_\mathrm{Edd,d}<1$. This accretion disc state is also known as the radiatively-efficient, standard or \cite{ShakuraSunyaev1973} disc (SD or SSD), as well as the soft or high state. The disc is geometrically thin (usually $H/R\in[0.001,0.01]$) and optically thick, with a high radiative efficiency of $\sim10$ per cent. Gas orbits are almost fully circular and extend down to $R_\mathrm{ISCO}$. Within this inner radius, they quickly become unstable. Jets are likely present, but their strength depends sensitively on the accretion rate and magnetic state of the disc. This state is effectively the one that is assumed at all accretion rates in the thermal AGN feedback model. We take the solution for this disc from \cite{ShakuraSunyaev1973}. We studied feedback in this accretion mode in detail in \cite{Husko_winds}, alongside the thick disc that was used at low Eddington ratios.
    \item \textbf{Slim disc} for $f_\mathrm{Edd,d}>1$. This accretion flow appears in the super-Eddington regime of accretion. The disc is geometrically thick ($H/R$ of order $0.1$ to $0.5$), but less so than the thick disc. It shares some features with both the thick and thin discs: it is both advection-dominated and radiatively efficient. It launches strong jets and is very luminous (the efficiencies for jets and radiation are lower than in the thick and thin disc, respectively, but the accretion rate is higher, resulting in potentially higher jet powers and luminosities). Radiation pressure and MHD effects drive a wind that may take away most of the accreting mass. We take the solution for this type of disc from \cite{Wang1999}, and also use numerical general-relativistic radiation magneto-hydrodynamical (GRRMHD) simulation results from \cite{Ricarte}. We introduced this disc state and the feedback from it in our model in \cite{Husko_2025_SE}.
\end{enumerate}

The choices for the critical Eddington ratios ($f_\mathrm{Edd,d}=0.01$ and $f_\mathrm{Edd,d}=1$) that separate the above states are discussed in Appendix~\ref{sec:appendix_transition_crit}. The modeling of accretion discs is highly dependent on assumptions about the kinematic viscosity $\nu$ in the discs, whose properties are largely unknown. All of the above accretion disc models are of so-called $\alpha-$discs, where $\alpha$ is a numerical parameter that is related to the kinematic viscosity $\nu$ through $\nu=\alpha c_\mr{s}H$. Here, $c_\mr{s}$ and $H$ are the sound speed and height of the disc at a given radius, respectively. In the $\alpha-$disc models, $\alpha$ does not vary with radius, but is instead given by a constant value. We set a value of $\alpha=0.1$, as motivated by the discussion in Appendix~\ref{sec:appendix_transition_crit}.

\begin{figure*}
\includegraphics[width=1\textwidth, trim = 0 20 0 0]{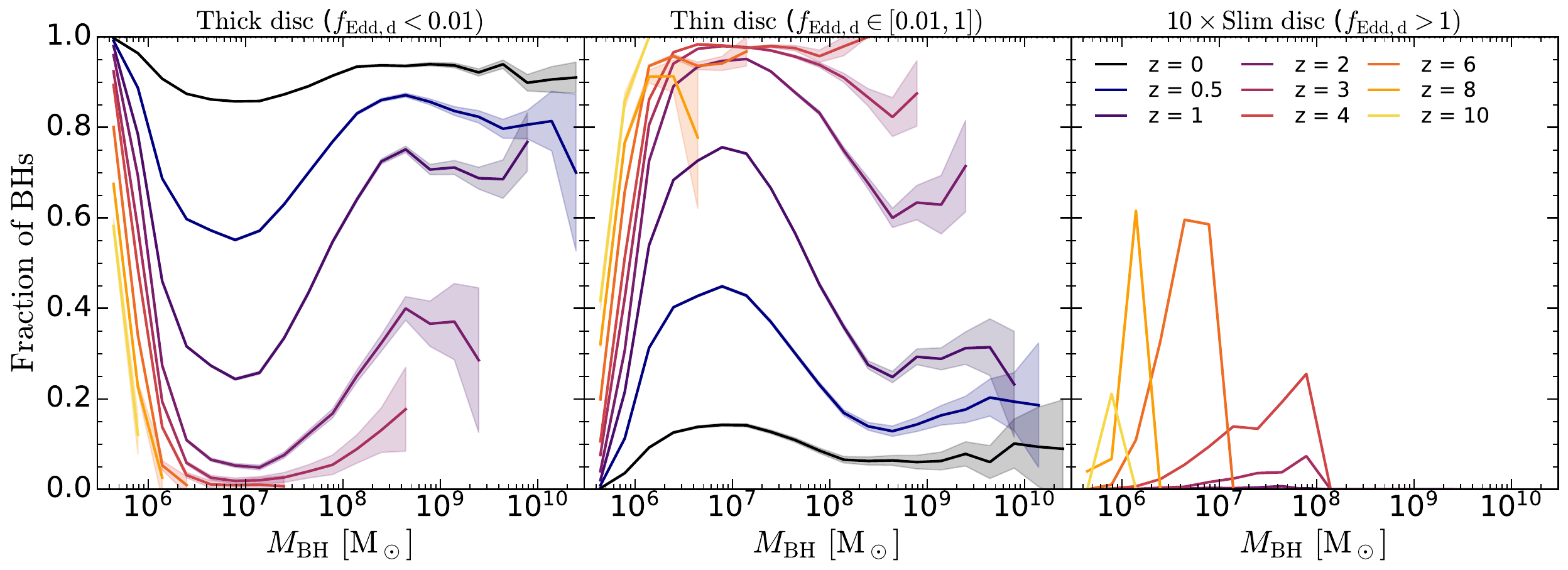}
\caption{The fraction of BHs occupying different accretion disc states in the L200m7h simulation: \textit{left}: thick disc, \textit{middle}: thin disc, \textit{right}: slim disc, multiplied by a factor of 10 for visual comparison purposes. Different colours indicate different redshifts. The shaded regions show uncertainties computed assuming binomial proportion confidence intervals.}
\label{fig:accretion_regimes}
\end{figure*}

In order to illustrate the importance of the three accretion disc states, in Fig.~\ref{fig:accretion_regimes} we show the fraction of all BHs that are in a given accretion disc state, as a function of BH mass and for different redshifts. Here we use results from one of our own simulations (specifically, L200m7h) that use the hybrid AGN feedback model in combination with the COLIBRE galaxy formation model. Most BHs are in the thick disc state at $z=0$, with the remainder ($5-15$ per cent) in the thin disc state. The fraction of low-redshift BHs in the thin disc state peaks at $M_\mr{BH}\approx10^7$ M$_\odot$, corresponding to host galaxies with stellar massses $M_*=10^{10}-10^{10.5}$ M$_\odot$, a mass regime where BHs grow most rapidly. The BHs in the thin disc state are observable as classical optical and X-ray AGN. The $z=0$ thin disc occupation fraction that we find here matches AGN fractions found observationally in the local Universe (e.g.~\citealt{Zhang2021}, \citealt{Steffen2023}), where AGN are identified using the BPT diagram. In these observations, the AGN fraction is non-negligible for galaxies with stellar masses between $10^9$ M$_\odot$ and $10^{11}$ M$_\odot$, with a peak value of $15-20$ per cent at $M_*\approx10^{10.5}$ M$_\odot$. Although Fig.~\ref{fig:accretion_regimes} has the BH mass on the $x-$axis, this overall shape (see black line in the middle panel) and the peak value of 15 per cent found in the simulations, reached at $M_\mathrm{BH}\approx10^{7}$ M$_\odot$, are in good agreement with observations given that galaxies with $M_*\approx10^{10.5}$ M$_\odot$ host BHs with median masses $M_\mathrm{BH}\approx10^{7}$ M$_\odot$ (\S~\ref{sec:calibration_first_results}).

As redshift increases, the fraction of BHs occupying the thin disc state increases, at all BH masses, but most prominently around that same mass scale, $M_\mr{BH}\approx10^7$ M$_\odot$. By $z=1, 2, 4$, only $\approx25$, $\approx5$ and $\approx2$ per cent of such BHs are in the thick disc state, with most of the remainder growing efficiently in the thin disc state. Above this mass scale, the fraction of BHs in the thin disc state drops, since more massive BHs are more likely to have quenched their host galaxy and stifled their own growth. The fractions of BHs occupying different accretion states reflects the evolution of the median Eddington ratio at fixed BH mass, which increases with redshift (Huško et al.~in prep.).

The slim disc state is very rare at all redshifts, and exceedingly so at lower redshifts and/or high BH masses. Low-mass BHs ($M_\mr{BH}\in[10^6,10^8]$ M$_\odot$) are the most likely to be in the slim disc state. The BH mass range where this is likely at a given redshift is even narrower, and moves from higher to lower-mass BHs with increasing redshift. At $z=3$, $z=4$, $z=6$ and $z=8$, the peak in the fraction of BHs occupying the slim disc state is at $M_\mr{BH}\approx10^8$, $\approx10^8$, $\approx10^7$ and $\approx10^6$ M$_\odot$, respectively, with the corresponding peak occupation fractions equal to $\approx1$, $\approx2.5$, $\approx6$ and $\approx6$ per cent. 

\subsection{Black hole mass evolution}
\label{sec:accretion_rates}

The black hole mass evolves according to
\begin{equation}
    \dot{M}_\mr{BH}=(1-\epsilon_\mr{tot})\dot{M}_\mr{acc,H}=(1-\epsilon_\mr{rad}-\epsilon_\mr{jet}-\epsilon_\mr{wind})\epsilon_\mr{acc}\dot{M}_\mr{acc,d},
\label{eq:net_accr_rate}
\end{equation}
where $\epsilon_\mr{tot}$ is the sum of all energy release efficiencies (radiative efficiency $\epsilon_\mr{rad}$, jet efficiency $\epsilon_\mr{jet}$ and wind efficiency $\epsilon_\mr{wind}$), and the individual efficiency terms are defined using the relation $P_i=\epsilon_i\dot{M}_\mr{acc,H}c^2$, where $P_i$ is the feedback power into a given channel $i$. Eqn.~(\ref{eq:net_accr_rate}) is analogous to that used for the thermal AGN feedback model (Eqn.~\ref{eq:mass_evolution_thermal}), but it has additional feedback terms alongside radiation. The radiation and wind terms are both associated with the thermal isotropic feedback channel used in our model, with the radiation term used in the thin disc state (and $\epsilon_\mr{wind}=0$), and the wind term in the thick and slim disc states (with $\epsilon_\mr{rad}=0$ for the purpose of Eqn.~\ref{eq:net_accr_rate}). The sum of the three efficiencies may exceed unity, in which case the BH can reduce its mass as it accretes (due to losing more mass-energy associated with its rotational energy, than it gains through accretion).


In addition to these feedback terms, Eqn.~\ref{eq:net_accr_rate}) also features the accretion efficiency, $\epsilon_\mr{acc}<1$, which relates the disc-scale accretion rate, $\dot{M}_\mr{acc,d}$, and the accretion rate into the event horizon, $\dot{M}_\mr{acc,H}$, through $\dot{M}_\mr{acc,H}=\epsilon_\mr{acc}\dot{M}_\mr{acc,d}$. This relation encapsulates mass loss due to accretion disc winds that we assume to be present in the thick and slim discs. In the next subsection, we discuss our assumptions for the accretion efficiency in different accretion disc states.

\subsubsection{Accretion efficiency}
\label{sec:accretion_eff}

\begin{figure}
\includegraphics[width=1\columnwidth, trim = 0 10 0 0]{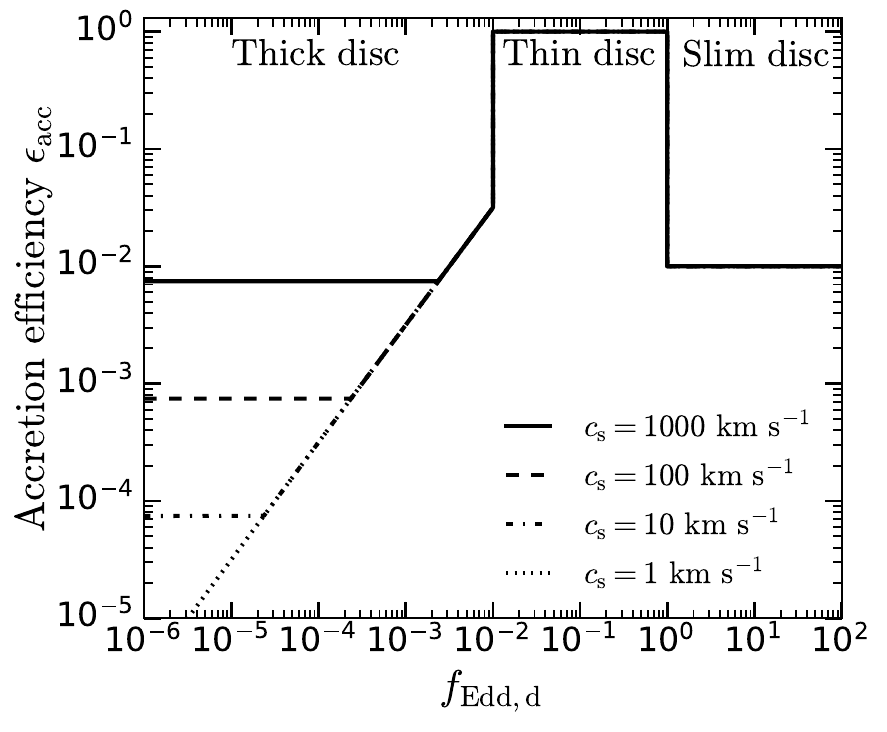}
\caption{The assumed accretion efficiencies as a function of the disc-scale Eddington ratio. The accretion efficiency connects the accretion rate onto a subgrid accretion disc (Eqn.~\ref{eq:net_accr_rate}) with the accretion rate into the BH itself, and it represents the effects of disc winds that take away most of the mass. In the thin disc we assume a $100$ per cent accretion efficiency, while in the slim disc we assume 1 per cent. In the thick disc, we combine results of GRMHD simulations and analytical calculations (Eqns.~\ref{eq:acc_eff} and \ref{eq:R_th}). For the free parameter $r_\mr{tr,0}$, we take the value $10^4$, corresponding to the value we found through calibration on the bolometric AGN luminosity function (see \S~\ref{sec:calibration_first_results}). The accretion efficiency in the thick disc is limited by the ratio of the Bondi radius and the gravitational radius of the BH, which depends only on the gas sound speed. Using different line styles, we show the accretion efficiency for several values of the sound speed, ranging from values appropriate for the hot ICM in galaxy clusters to those appropriate for cold, dense gas in the ISM.}
\label{fig:accretion_efficiencies}
\end{figure}

The accretion efficiency, $\epsilon_\mathrm{acc}$, represents the fraction of mass accreting onto a subgrid accretion disc that actually reaches the event horizon of a BH. We assume that it is less than unity only in the thick and slim disc, where we assume strong winds to be operating. We show our assumed accretion efficiencies in Fig.~\ref{fig:accretion_efficiencies} and in the first row of Table \ref{tab:tab1}.

\captionsetup[table]{skip=0pt} 
\begin{table*}
\begin{center}
\caption{The accretion efficiency, dimensionless magnetic flux, and feedback efficiencies assumed for the three accretion disc states. The values (which depend on the BH spin, $a$) of the (dimensionless) event horizon angular velocities $\Omega_\mr{H}$ and radius of the innermost stable circular orbit, ISCO, are given shortly after Eqn.~(\ref{eq:eps_thermal}) in the text and in Appendix~\ref{sec:appendix_ISCO}, respectively. The thick disc size, $r_\mr{th}$, is given by Eqn.~(\ref{eq:R_th}). $\epsilon_\mr{f}$ is the coupling efficiency of thin disc radiative feedback, while $r_\mr{tr,0}$ is the dimensionless size of the thick accretion disc at a critical Eddington ratio $f_\mr{Edd,d}=0.01$. Both $\epsilon_\mr{f}$ and $r_\mr{tr,0}$ are free parameters.}
\label{tab:tab1}
\end{center}

\begin{tabular*}{1.\textwidth}{@{\extracolsep{\fill}}lccc}
  
    &  \textbf{Thick disc} ($f_\mr{Edd,d}<0.01$) &  \textbf{Thin disc} ($0.01<f_\mr{Edd,d}<1$) & \textbf{Slim disc}  ($f_\mr{Edd,d}>1$) \\
  \hline 
  $\epsilon_\mr{acc}$ (accretion efficiency) & $\big(10/r_\mr{th}\big)^{0.5}$, $r_\mr{th}=\min\big[r_\mr{tr,0}\big(\frac{0.01}{f_\mathrm{Edd,d}}\big)^2, r_\mr{B}\big]$  & $1$ & $0.01$  \\
  \hline 
  $\phi$ (magnetic flux)  & $\phi_\mr{0}(a)=-20.2a^3-14.9a^2+34a+52.6$  & \multicolumn{2}{c}{\multirow{1}{*}{$\phi_\mr{0}(a)(f_\mathrm{Edd,H}/1.88)^{1.29}/[1+(f_\mathrm{Edd,H}/1.88)^{1.29}]$ }} \\
  \hline
  $\epsilon_\mr{th}$ (thermal efficiency)  & $0.005\big[1+0.3\big(\frac{\phi}{50}\big)\big(\frac{\Omega_\mr{H}}{0.2}\big) \big]$ & $\epsilon_\mr{f}\big(1-\sqrt{1-2/3r_\mr{ISCO}}\big)$ & $0.065\big[1+\big(\frac{\phi}{50}\big)^2\big] \max(0, 1+\Omega_\mathrm{H}-8\Omega_\mathrm{H}^2)$ \\
  \hline 
  $\epsilon_\mr{jet}$ (jet efficiency)  & \multicolumn{3}{c}{\multirow{1}{*}{$(\kappa/4\pi)\phi^2\Omega_\mr{H}^2\big(1+1.38\Omega_\mr{H}^2-9.2\Omega_\mr{H}^4\big)$}}  \\
  
\end{tabular*}

\end{table*}

We assume the accretion efficiency to be given by
\begin{equation}
\epsilon_\mr{acc}= \begin{cases} 
      \Big(\frac{10}{r_\mr{th}}\Big)^s, &   f_\mr{Edd,d}<0.01 \hspace{1mm}(\mathrm{thick}\hspace{1mm}\mathrm{disc}) \\
      1, &  0.01<f_\mr{Edd,d}<1 \hspace{1mm}(\mathrm{thin}\hspace{1mm}\mathrm{disc})\\
      0.01, & f_\mr{Edd,d}>1 \hspace{1mm}(\mathrm{slim}\hspace{1mm}\mathrm{disc})
   \end{cases}
\label{eq:acc_eff}
\end{equation}
For the thin disc we assume a 100 per cent accretion efficiency, while we assume 1 per cent for the slim disc, as motivated below. For the thick accretion disc state, the accretion efficiency depends on the 
(dimensionless) radius of the thick component of the accretion disc $r_\mr{th}=R_\mr{th}/R_\mr{G}$, where $R_\mr{th}$ is the physical radius of the thick disc, and $R_\mr{G}=M_\mr{BH}G/c^2$ is the gravitational radius of the BH. At radii larger than $R_\mr{th}$, the disc may be present, but it is in the thin state and is thus assumed in our model to not be contributing to the launching of a wind. The scaling of the accretion efficiency with the size of the thick disc, $\epsilon_\mr{acc}= (10/r_\mr{th})^{s}$, is well motivated by theory (\citealt{YuanNarayan2014}), where larger discs have smaller accretion efficiencies, implying that the exponent $s>0$. Its value is also limited to $s<1$ on theoretical grounds (\citealt{Blandford1999}), but it cannot be predicted analytically. Observations suggest $0.3<s<1$ (\citealt{Nemmen2015}), while recent MHD and GRMHD simulations have converged on $s\approx0.5$ (\citealt{Guo2023}, \citealt{Cho2024}), so we assume that value. The normalization constant equal to 10 (dimensionless radii) is present in the scaling since accretion disc winds are not thought to be launched from within $\approx10$ gravitational radii (\citealt{YuanNarayan2014}).

For the size of the thick disc, we take the scaling from \cite{Narayan1995}, which assumes that the transition from the thin disc at large radii to the thick disc at small radii occurs at a radius where half of the energy is radiated away, and half advected inwards\footnote{Other calculations, such as from the disc evaporation model (e.g.~\protect\citealt{Taam2012} and references therein), give quantitatively different expressions for the transition radius. However, the qualitative behaviour of a larger thick disc component at smaller Eddington ratios also appears in such models.}. We also assume that the disc can be no larger than the Bondi radius. For the dimensionless size of a thick disc we thus assume
\begin{equation}
    r_\mr{th}=\min\bigg[r_\mr{tr,0}\bigg(\frac{0.01}{f_\mathrm{Edd,d}}\bigg)^2, r_\mr{B}\bigg].
\label{eq:R_th}
\end{equation}
where $r_\mr{tr,0}=R_\mr{tr,0}/R_\mr{G}$ is the dimensionless size of the thick accretion disc that is accreting at the critical Eddington ratio of $0.01$, and $r_\mr{B}=R_\mr{B}/R_\mr{G}=(c/c_\mr{s})^2$ is the dimensionless Bondi radius. The value of $r_\mr{tr,0}$, which serves as the normalization of the dependence of $r_\mr{th}$ on $f_\mr{Edd,d}$, is in principle predicted by theory (\citealt{Narayan1995}). However, it is highly dependent on the accretion disc viscosity $\alpha$, more steeply than $r_\mr{tr,0}\propto \alpha^4$. The value of $\alpha$ could be in the range $0.1-0.4$, so the value of $r_\mathrm{tr,0}$ itself is highly uncertain. We hence leave it as a free parameter, and calibrate its value (see \S~\ref{sec:calibration_procedure}). 

Given the assumptions discussed above, the accretion efficiency for the thick disc, i.e.~at $f_\mr{Edd,d}<0.01$, can be written as
\begin{equation}
\epsilon_{\mathrm{acc,th}} \approx
\begin{cases}
 
       3.16\times10^{-2} \,\left( \frac{r_{\mathrm{tr,0}}}{10^4} \right)^{-0.5} \left( \frac{f_{\mathrm{Edd,d}}}{0.01}\right),\;
 & r_{\mathrm{B}} \ge r_{\mathrm{tr,0}} \left(
 \frac{f_{\mathrm{Edd,d}}}{0.01}\right)^{-2}\\
       
 10^{-3}\left(\frac{c_{\mathrm{s}}}{100\,\mathrm{km\,s}^{-1}}\right), \;
 & r_{\mathrm{B}} < r_{\mathrm{tr,0}} \left (
 \frac{f_{\mathrm{Edd,d}}}{0.01}\right)^{-2}, \;
 \end{cases}
 \label{eq:eps_thick}
\end{equation}
for our fiducial values. The first of these cases is valid if the transition radius (between an outer thin and an inner thick disc) is smaller than the Bondi radius, i.e.~if an outer thin disc is present. In this case, the accretion efficiency is linear with $f_\mr{Edd,d}$ and reaches values no larger than $3.16\times10^{-2}$. In the second case, the entire disc (out to the Bondi radius) is thick. The dimensionless Bondi radius depends only on the sound speed of the gas, so the accretion efficiency scales as $\epsilon_{\mathrm{acc,th}}\propto c_\mr{s}$. For accretion from hot gas (e.g.~$c_\mr{s}=1000$ km~s$^{-1}$), the accretion efficiency reaches values of $10^{-2}$, but it will generally be lower.

For the slim disc, a similar picture as above likely holds. However, slim discs become relatively more advective and more efficient at launching winds at higher accretion rates (\citealt{Ricarte}), rather than at lower ones like for the thick disc. This implies that the accretion efficiency for the slim disc should have the form $\epsilon_\mr{acc}\propto 1/f_\mr{Edd,d}^a$, with $a>0$. In other words, it should decline with increasing Eddington ratio, which is opposite behaviour to that in the thick disc state. Since no simulations of slim discs have, to our knowledge, yet studied wind mass loading as a function of the Eddington ratio, we instead assume a constant slim disc accretion efficiency parameter, $\epsilon_\mr{acc,slim}$. We do not calibrate it on any particular quantity or relation, but instead assume a value of $\epsilon_\mr{acc,slim}=0.01$, which is similar to the values of the accretion efficiency in the thick disc (although lower than in some GRRMHD simulations, see e.g. \citealt{Jiang2019b}).

\subsection{Accretion disc magnetisation}
\label{sec:magnetisation}

The magnetic flux threading the BH event horizon affects the efficiency of thermal feedback in the thick and slim disc states, as well as jet efficiencies in all accretion states. In general, accretion discs may occupy two qualitatively distinct states in terms of how strong their magnetic fields are: the `standard and normal evolution' (SANE) state (\citealt{Narayan2012}), or the magnetically-arrested disc (MAD) state (\citealt{Narayan2003}). In the SANE state, magnetic fields are unconstrained and dynamically unimportant, while in the MAD state, they are dynamically important and have reached a saturated value, that depends exclusively on accretion rate and BH spin. This state is reached as a result of a strong magnetic dynamo effect, in combination with advection of the field from larger radii.

We assume that all of our accretion discs are in the MAD state, and summarize our choice for the dimensionless magnetic flux, $\phi=\Phi_\mathrm{H}/\sqrt{\dot{M}_\mathrm{acc,H}R_\mathrm{G}^2c}$, $\Phi_\mathrm{H}$ being the magnetic flux threading the event horizon, in the second row of Table~\ref{tab:tab1}. The dimensionless magnetic flux in the MAD state is given by
\begin{equation}
\phi(f_\mr{Edd,H},a)= \phi_\mr{0}(a)\times\begin{cases} 
      1, &  f_\mr{Edd,d}<0.01 \\
      \frac{(f_\mathrm{Edd,H}/1.88)^{1.29}}{1+(f_\mathrm{Edd,H}/1.88)^{1.29}}, &  f_\mr{Edd,d}>0.01
   \end{cases}
\label{eq:phi_a}
\end{equation}
Here, $\phi_\mr{0}(a)$ is the dimensionless magnetic flux of a thick or sufficiently super-Eddington slim disc, which depends only on BH spin and is given by (\citealt{Narayan2021})
\begin{equation}
    \phi_\mr{0}(a)=-20.2a^3-14.9a^2+34a+52.6.
\label{eq:phi_0}
\end{equation}
For thin discs and mildly super-Eddington slim discs, the value of $\phi$ is lower than $\phi_\mr{0}(a)$. The functional form given by Eqn.~(\ref{eq:phi_a}) for those discs gives a strong dependence on the Eddington ratio, $\phi\propto f_\mr{Edd,H}^{1.29}$ if $f_\mr{Edd,H}\ll1$, and saturation to $\phi_\mr{0}(a)$ if $f_\mr{Edd,H}\gg1$.

The MAD assumption is supported by observations for thick discs (\citealt{EHT2021}, \citealt{EHTSagA5}). It is also strongly supported on theoretical grounds for advection-dominated discs (\citealt{YuanNarayan2014}). We thus expect it to be valid at least in the thick disc state and for very super-Eddington slim discs. However, for thin discs and accretion near the Eddington rate, our assumed magnetic flux may be too high.

\subsection{Feedback efficiencies}
\label{sec:feedback_effs}

\begin{figure*}
\includegraphics[width=1\textwidth, trim = 0 10 0 0]{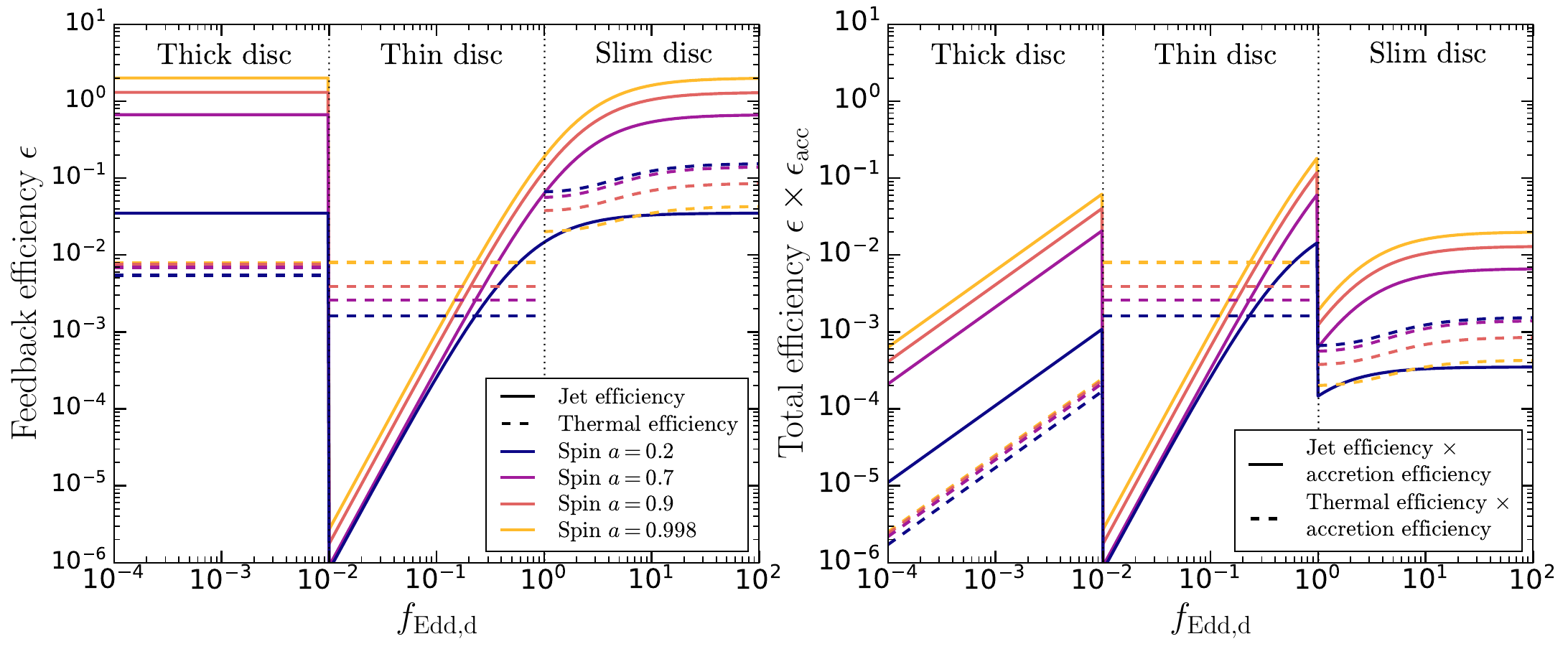}
\caption{The dependence of feedback efficiencies used in the hybrid AGN feedback model on the accretion disc-scale Eddington ratio $f_\mr{Edd,d}$ for the three accretion and feedback states. Solid lines show the jet efficiencies (see \S~\ref{sec:jet_efficiencies}), and dashed the thermal efficiencies (see \S~\ref{sec:thermal_eff}), assuming a coupling efficiency $\epsilon_\mr{f}=0.025$. We show these for several values of BH spin using different colours (see legend); note that the jet efficiency is $\epsilon_\mr{jet}=0$ for $a=0$. In the left panel we show the pure feedback efficiencies, while in the right panel, we further multiply these by accretion efficiencies (see \S~\ref{sec:accretion_eff}) to emphasize the net feedback efficiencies and to facilitate comparison between the different modes. For this figure, we used $r_\mr{tr,0}=10^4$ as found through calibration (\S~\ref{sec:calibration_results}).}
\label{fig:feedback_eff}
\end{figure*}

We include two modes of AGN feedback: thermal isotropic and kinetic jets, both of which are used at the same time, and in each accretion disc state. The first of these modes encapsulates the effects of radiation and accretion disc winds, while the second one represents the effects of jets driven by the \cite{Blandford1977} mechanism. The feedback powers are given by 
\begin{equation}
P_\mr{th/jet}=\epsilon_\mr{th/jet}\dot{M}_\mr{acc,H}c^2=\epsilon_\mr{th/jet}\epsilon_\mr{acc}\dot{M}_\mr{acc,d}c^2,
\label{eq:P}
\end{equation}
where $\epsilon_\mr{th/jet}$ is either the thermal or jet efficiency. The feedback efficiencies are shown in Fig.~\ref{fig:feedback_eff}, and are introduced in the rest of this section. In the left panel we show only the feedback efficiencies, while in the right panel, we also multiply them by the accretion efficiencies, to make the net energy release efficiencies more comparable between different accretion disc states. The feedback efficiencies that we assume are also summarized in the 3rd and 4th rows of Table~\ref{tab:tab1}, for thermal and jet feedback, respectively.

\subsubsection{Thermal feedback}
\label{sec:thermal_eff}


We assume that AGN emit radiation at the rate
\begin{equation}
L_\mr{AGN}=\epsilon_\mr{rad}\dot{M}_\mr{acc,H}c^2,
\label{eq:L_AGN}
\end{equation}
where $\epsilon_\mr{rad}$ is the radiative efficiency and $\dot{M}_\mr{acc,H}$ is the accretion rate into the event horizon. For the thin disc, we assume that a fraction $\epsilon_\mr{f}$, the coupling efficiency, of that radiation couples to gas (either on accretion disc or larger scales), causing feedback\footnote{Our model is agnostic as to the exact mechanism by which radiation causes feedback. Radiation may launch a wind on accretion disc scales, or on larger scales. It may also directly heat gas on large scales. In either case, we assume that the thermal isotropic feedback employed in the simulations is representative the effects of whichever of these effects that is dominant.}, and the observable AGN luminosity is thus equal to $(1-\epsilon_\mr{f})L_\mr{AGN}$. The coupling efficiency, $\epsilon_\mr{f}$, is a free parameter. Its value is calibrated by ensuring our simulated BH masses match observations at $z=0$ (\S~\ref{sec:calibration_first_results}).

For the radiative efficiency, for the thin disc, we assume the expression for a relativistic thin accretion disc (\citealt{NovikovThorne1973}):
\begin{equation}
\epsilon_\mr{rad,thin}(a)=1-e_\mr{ISCO}(a)=1-\sqrt{1-\frac{2}{3r_\mr{ISCO}(a)}}.
\label{eq:eps_rad_NT}
\end{equation}
Here, $e_\mr{ISCO}$ is the specific binding energy (in units of $c^2$) at the innermost stable circular orbit (ISCO). $r_\mr{ISCO}=R_\mr{ISCO}/R_\mr{G}$, is the dimensionless radius of the ISCO. We give the radius of the ISCO as a function of BH spin in Appendix~\ref{sec:appendix_ISCO}. The above formulation gives a radiative efficiency for the thin disc that varies between $\epsilon_\mathrm{rad}\approx4\%$ at $a=-0.998$ and $\epsilon_\mathrm{rad}\approx30\%$ at $a=0.998$, which is to be compared with the constant value of $10$ per cent in the thermal AGN feedback model. 

In the thick and slim discs, we assume different, but very small radiative efficiencies (see Appendix~\ref{sec:appendix_rad_eff}), since both of those disc types are radiatively-inefficient. We also assume that $\epsilon_\mr{f}=0$ in those discs, since the slim disc winds may be partly driven by radiation on the accretion disc scales. The radiative efficiencies for those discs are thus used exclusively for calculating AGN luminosities.

For the thick and slim discs, we instead assume the presence of accretion disc winds that are shocked on subgrid scales (\citealt{Costa2014}), so we implement the effects of these winds using the thermal feedback channel (in other words, we assume $\epsilon_\mr{th}=\epsilon_\mr{wind}$ for these disc states). The thermal feedback efficiencies are thus given by
\begin{equation}
\epsilon_\mr{th}(f_\mr{Edd,H},a)= \begin{cases} 
      0.005\bigg[1+0.3\left(\frac{\phi}{50}\right)\left(\frac{\Omega_\mr{H}}{0.2}\right) \bigg], &  f_\mr{Edd,d}<0.01 \\
      \epsilon_\mr{f}\bigg(1-\sqrt{1-\frac{2}{3r_\mr{ISCO}}}\bigg), &  0.01<f_\mr{Edd,d}<1 \\
      0.065\bigg[1+\left(\frac{\phi}{50}\right)^2\bigg] M(\Omega_\mathrm{H}), & f_\mr{Edd,d}>1
   \end{cases}
\label{eq:eps_thermal}
\end{equation}
where $M(\Omega_\mathrm{H}) \equiv \max\big(0, 1+\Omega_\mathrm{H}-8\Omega_\mathrm{H}^2\big)$. The second expression corresponds to $\epsilon_\mr{f}\epsilon_\mr{rad,thin}$ for the thin disc. The first and third expressions correspond to wind efficiencies for the thick and slim discs obtained from \cite{Sadowski2013} and \cite{Ricarte}\footnote{The thick disc wind efficiency is taken directly from \protect\cite{Sadowski2013}, while the slim disc wind efficiency is a fitting function to the results of \protect\cite{Ricarte} that we derived and presented in \protect\cite{Husko_2025_SE}; see Appendix A therein.}, respectively. Here, $\Omega_\mr{H}(a)$ is the dimensionless angular velocity of the BH event horizon, given by $\Omega_\mr{H}(a)=a/2r_\mr{H}(a)$, where $r_\mr{H}(a)=1+\sqrt{1-a^2}$ is the dimensionless radius of the event horizon. 


The thermal feedback efficiency for the thin disc, due to radiation (assuming $\epsilon_\mr{f}=0.025$, in the mid-range of values we find by calibration for the hybrid AGN feedback model, see \S~\ref{sec:calibration_first_results}), is shown in Fig.~\ref{fig:feedback_eff} using dashed horizontal lines at moderate Eddington ratios. It varies from $\approx0.2$ per cent at low BH spin ($a=0.2$) to $\approx1$ per cent at maximal BH spin ($a=0.998$). The wind efficiency for the thick disc is shown with horizontal lines at low Eddington ratios in Fig.~\ref{fig:feedback_eff}. It is generally similar to or larger than the thermal efficiency in the thin disc. However, one needs to keep in mind that these winds are applied at lower accretion rates than in the thin disc, and that we apply an accretion efficiency in the thick disc (resulting in much lower net accretion rates than in the thin disc, see the right panel of Fig.~\ref{fig:feedback_eff}). As a result, thick disc winds have a very small effect in practice, at least at the level of calibration of galaxy formation simulations. This is also related to the fact that jets are the dominant feedback mechanism in this regime, which is also visible in Fig.~\ref{fig:feedback_eff} (we introduce the jet efficiencies in the next subsection). The wind efficiency in the slim disc is shown using horizontal dashed lines at super-Eddington accretion rates in Fig.~\ref{fig:feedback_eff}. The wind efficiencies have values of order $5-20$ per cent, and are thus generally the largest among the thermal efficiencies used in our model. However, as in the thick disc, the winds from the slim disc are generally subdominant to its jets (unless the BH spin is quite low, $\vert a\vert<0.4$).

\subsubsection{Jet efficiencies}
\label{sec:jet_efficiencies}

We assume that all accretion discs launch jets on account of energy (and angular momentum) extraction through the \cite{Blandford1977} (BZ) process. There have been many studies on jet launching from MAD thick discs, and recently it has been shown that the results from these studies may also be applied to slim and thin discs, once differences in the dimensionless magnetic flux, $\phi$, between these disc types are accounted for (\citealt{Ricarte}). We assume the jet efficiency formula from \cite{Tchekhovskoy2010}:
\begin{equation}
    \epsilon_\mr{jet}=\frac{\kappa}{4\pi}\phi^2\Omega_\mr{H}^2\big(1+1.38\Omega_\mr{H}^2-9.2\Omega_\mr{H}^4\big),
\label{eq:epsilon_jet}
\end{equation}
where $\kappa$ is a numerical factor that depends on the initial geometry of the magnetic field (e.g.~0.054 for split-monopole vs.~0.044 for parabolic, we assume $\kappa=0.05$). 
For a non-spinning BH ($a=0$), the jet efficiency given by the above formula is $\epsilon_\mr{jet}=0$. All three of our accretion disc states use Eqn.~(\ref{eq:epsilon_jet}). They differ only in their values of $\phi$, which introduces large differences in the jet efficiencies. For strongly super-Eddington rates, the jet efficiency of the slim disc is the same as that of the thick disc, since they have the same values of $\phi$. For the thin disc, on the other hand, $\phi$ is generally much lower, except at near-Eddington rates (see Eqn.~\ref{eq:phi_a}).

The jet efficiency given by Eqn.~(\ref{eq:epsilon_jet}) is shown in Fig.~\ref{fig:feedback_eff} using solid lines. The jet efficiencies can be of order unity (or exceed it) for highly-spinning BHs. For moderate and large values of BH spin, jets dominate over thermal feedback at all Eddington ratios except in the thin disc for $f_\mr{Edd,H}\in[0.01,0.3]$ (jets dominate even in the thin disc for $f_\mr{Edd,H}\gtrsim0.3$, since the value of $\phi$ is large enough at those near-Eddington rates to yield jet efficiencies larger than $\approx1$ per cent).

\subsection{Black hole spin evolution}
\label{sec:BH_spin}

The spin of every BH, given by Eqn.~(\ref{eq:BH_spin}), evolves due to the following set of physical mechanisms: 1) gas accretion, 2) BH-BH mergers) 3) jet-induced spindown, 4) radiation torques, and 5) Lense-Thirring torques. These mechanisms affect both the magnitude and direction of the BH spin. However, in practice, the inner region of the accretion disc is usually either perfectly aligned (spin positive) or counter-aligned (spin negative) with respect to the BH spin vector (\citealt{BardeenPetterson}), or precessing around that same vector (\citealt{Stella1998}). Thus, almost all quantities that depend on BH spin and that appear in our model can be assumed to depend exclusively on its magnitude and sign.

Generally, the BH spin could occupy the full range $a\in[-1,1]$, with magnitudes larger than $1$ not allowed in general relativity (\citealt{Kerr}). In reality, however, the BH spin is thought to have an effective upper limit of $a_\mathrm{max}\approx0.998$ due to torques from photon emission during gas accretion (\citealt{Thorne1974}). We thus impose that upper limit (which does have a non-negligible impact on the maximum values of some of the efficiencies that we use).

In \S~\ref{sec:spin_magnitude} we discuss how accretion and jet-induced spindown impact the evolution of the BH spin magnitude. The choice of the sign of BH spin and the evolution of its direction as a vector are discussed in \S~\ref{sec:spin_sign} and \S~\ref{sec:spin_direction}, respectively. Lense-Thirring torques have a great impact on both. 

\subsubsection{Evolving the magnitude of black hole spin}
\label{sec:spin_magnitude}

The BH spin magnitude evolves according to  to the spinup/spindown function $s$, given by (\citealt{Bardeen1970}, \citealt{Moderski1996}, \citealt{Fanidakis2011})
\begin{equation}
    s \equiv \frac{\mr{d}a}{\mr{d}M_\mr{acc,H}/M_\mr{BH}}=\ell_\mr{in}-2a (1-\epsilon_\mr{tot}) + s_\mr{EM},
\label{eq:da_dlnMSMBH}
\end{equation}
where $\ell_\mr{in}=L_\mr{in}/(R_\mr{G}c)$ is the dimensionless specific angular momentum at the inner boundary of the accretion disc ($L_\mathrm{in}$ being the specific angular momentum in physical units). $\mr{d}M_\mr{acc,H}$ is the rest mass of the material accreted through the BH event horizon, and $\epsilon_\mr{tot}$ is the total efficiency of all energy loss processes as defined in Eqn.~(\ref{eq:net_accr_rate}). The first term in Eqn.~(\ref{eq:da_dlnMSMBH}) is due to gas accretion onto the BH, the second originates from the definition of the BH spin $a$ through the presence of the BH mass, while the last term encapsulates angular momentum loss due to electromagnetic effects (the launching of winds and mostly jets).

\begin{figure}
\includegraphics[width=1\columnwidth, trim = 0 10 0 0]{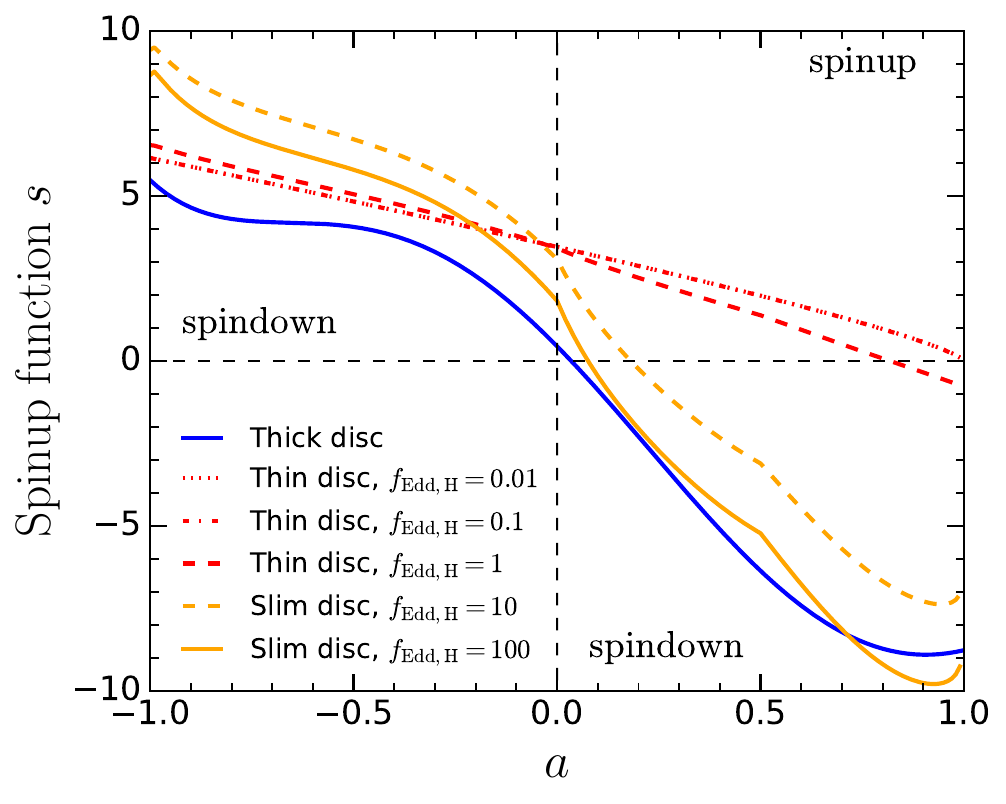}
\caption{The BH spinup/spindown function. This is given by Eqn.~(\ref{eq:da_dlnMSMBH_th}) for the thick disc at $f_\mr{Edd,d}<0.01$ (blue line) and Eqn.~(\ref{eq:da_dlnMSMBH_Ricarte}) for the thin and slim discs ($f_\mr{Edd,d}>0$, red and orange lines, respectively). For the latter, we show the spinup/spindown function for several values of $f_\mathrm{Edd,H}$. Positive or negative values of the spin $a$ correspond to the inner region of the accretion disc co-rotating or counter rotating relative to the BH spin. The bottom left and top right quadrants correspond to the magnitude of the spin increasing, and the top left and bottom right quadrants to the magnitude of the spin decreasing.}
\label{fig:spinup_func}
\end{figure}

For the thick disc, instead of using Eqn.~(\ref{eq:da_dlnMSMBH}) directly, we use results from GRMHD simulations, to be consistent with our assumptions about feedback efficiencies. We assume the fitting function of \cite{Narayan2021} for discs in the MAD state:
\begin{align}
    s_\mathrm{thick}&=\bigg(\frac{\mr{d}a}{\mr{d}M_\mr{acc,H}/ M_\mr{BH}}\bigg)_\mathrm{thick} \nonumber \\
    &=0.45 - 12.53a - 7.8a^2 +9.44a^3 + 5.71a^4 -4.03a^5.
\label{eq:da_dlnMSMBH_th}
\end{align}
We show this spinup/spindown rate in Fig.~\ref{fig:spinup_func} using the blue line. In this figure, spinup (an increase in BH spin magnitude) corresponds to the bottom left and top right quadrants, while spindown (a decrease in BH spin magnitude) corresponds to the top left and bottom right quadrants. BHs in the thick accretion disc state are almost always spinning down, except in a very narrow region $a\in[0,0.05]$. This is a signature of strong jet spindown, and it means that jets will effectively fully spin down a BH given enough time. In practice, however, as the BH spin becomes low enough, feedback efficiencies also drop to low values, making it more likely that the BH will increase its accretion rate due to self-regulation (e.g.~\citealt{BoothSchaye2010}) and transition to the thin disc state, where spinup is more efficient (see below). Furthermore, BH-BH mergers (see \S~\ref{sec:BH_mergers}) will typically leave remnant BHs with non-zero spin, so we generally do not expect to find many BHs with spin values that are very low (e.g.~$\vert a\vert<0.3$).

For the slim and thin discs, we use results from \cite{Ricarte}, who provide a fitting formula that smoothly interpolates between the thin disc state without significant jet feedback (for $f_\mathrm{Edd,H}\ll1$), and a highly-super-Eddington slim disc where jet feedback essentially matches the thick disc (and so jet spindown should also be similar). Their formula, which is analogous to Eqn.~(\ref{eq:da_dlnMSMBH}), is
\begin{equation}
    s_\mathrm{thin/slim}=\bigg(\frac{\mr{d}a}{\mr{d}M_\mr{acc,H}/M_\mr{BH}}\bigg)_\mathrm{thin/slim}=s_\mathrm{HD} + s_\mr{EM},
\label{eq:da_dlnMSMBH_Ricarte}
\end{equation}
where the first term is a pure hydrodynamical term and the second is an electromagnetic term. $s_\mathrm{HD}$ is given in their approach by
\begin{equation}
    s_\mathrm{HD}=\frac{s_\mathrm{thin}+s_\mathrm{min}\xi}{1+\xi},
\label{eq:s_HD}
\end{equation}
where $\xi=0.017f_\mathrm{Edd,H}$, $s_\mathrm{min}=0.86-1.94a$ and $s_\mathrm{thin}=\ell_\mathrm{ISCO}-2a e_\mathrm{ISCO}$ is the spinup/spindown function of the thin disc (without outflows and outside the MAD regime), in which $\ell_\mathrm{ISCO}$ and $e_\mathrm{ISCO}$ are the (dimensionless) specific angular momentum and binding energy, respectively, at the ISCO (see Appendix \ref{sec:appendix_ISCO}). The EM term, $s_\mathrm{EM}$, is given by\footnote{We removed a superfluous minus sign in the original equation provided by \protect\cite{Ricarte}.}
\begin{equation}
    s_\mathrm{EM}=-\mathrm{sgn}(a)\epsilon_\mathrm{EM}\bigg(\frac{1}{k\Omega_\mathrm{H}}-2a\bigg),
\label{eq:s_EM}
\end{equation}
where $\mathrm{sgn}$ is the sign function, $\epsilon_\mathrm{EM}=\epsilon_\mathrm{jet}+\epsilon_\mathrm{wind}$ is the total (jet+wind) EM efficiency of the slim disc, and $k$ is given by 
\begin{equation}
        k=
\begin{dcases}
    0.23 ,& a \leq 0 \\
    \min(0.35,0.1+0.5a),              & a>0
\end{dcases}
\label{eq:k}
\end{equation}
The slim disc spinup/spindown rate is shown in Fig.~\ref{fig:spinup_func} for several values of $f_\mathrm{Edd,H}$. For small values, $0.01<f_\mathrm{Edd,H}\ll1$, the spinup/spindown function shows monotonic spinup to $a\approx1$. As $f_\mathrm{Edd,H}$ approaches and exceeds unity (super-Eddington accretion), the impact of jets and winds on spindown becomes stronger, and the spinup/spindown function becomes similar to that of the thick disc (albeit there is a small residual difference at $a<0$, even for strongly super-Eddington slim discs).

\subsubsection{Choosing the sign of BH spin}
\label{sec:spin_sign}

We assume that the effects of LT torques (\citealt{LenseThirring}) are effectively instantaneous in the inner regions of the accretion disc (compared to the time-scale of accretion). In this picture, the inner accretion disc is defined as the region where the BH's angular momentum dominates over that of the angular momentum of the accretion disc. As a result, the accretion disc becomes either aligned or counter-aligned with the BH's spin through LT torques, or it precesses around the BH spin. In the case of counteralignment, we consider accretion to be retrograde and the BH spin negative; otherwise BH spin is positive. The thin disc thus develops a warp due to LT torques and is assumed to be perfectly aligned or counteraligned with the BH within a warp radius $R_\mr{warp}$ (see Appendix \ref{sec:appendix_warp} for the expressions that we use), which is the radius out to which the ‘communication' of the BH and the disc is effective in terms of torques (\citealt{BardeenPetterson}). Outside this radius, the accretion disc is undisturbed and aligned with the larger-scale accretion flow. For the thick and slim discs, the assumption of exact (counter-) alignment is invalid. Instead, the disc precesses about the BH spin vector. This precession occurs on very short time-scales, much shorter than those we are simulating. For this reason we also assume (counter-) alignment of the thick disc, in a time-averaged sense. 

The sign of the BH spin (i.e.~whether the disc aligns or counteraligns) is decided based on the criterion developed by \cite{King2005}. In this prescription, the BH and the inner accretion disc are assumed to come into (counter-) alignment in a way that the magnitude of the BH spin does not change, and the total angular momentum (of the BH + inner accretion disc out to $R_\mathrm{warp}$) is conserved. In this process, the BH spin aligns with the total angular momentum. The condition for counteralignment (and for BH spin to be negative) in this approach can be stated as follows:
\begin{equation}
\cos \theta<-\frac{J_{\mathrm{warp}}}{2 J_{\mathrm{BH}}},
\label{eq:counteralignment}
\end{equation}
where $\cos \theta=\bf{\hat{J}_\mr{BH}}\cdot\bf{\hat{J}_\mr{d}}$ is the misalignment between the BH spin and the larger-scale angular momentum of the disc (i.e.~outside $R_\mathrm{warp}$), whose direction is $\bf{\hat{J}_\mr{d}}$. $J_{\mathrm{warp}}$ is the total angular momentum of the inner accretion disc out to $R_\mathrm{warp}$ (before alignment occurs). In our cosmological application of the model, $\bf{\hat{J}_\mr{d}}$ is the kernel-weighted direction of the gas angular momentum, computed from the same gas neighbours that are used to compute the disc-scale accretion rate. This assumption of alignment of the gas angular momentum between our simulated scales and the outer regions of a subgrid accretion disc is quite strong, and discussed in detail in \cite{Husko2022_spin_driven}. 

\subsubsection{Evolving the direction of the black hole spin}
\label{sec:spin_direction}

From a numerical standpoint, the direction of the BH spin is evolved in the following way. For each increment of mass $M_\mr{warp}$ (the mass within $R_\mr{warp}$) consumed by the BH, the BH-inner accretion disc system is assumed to come into equilibrium (with the inner accretion disc aligned or counter-aligned with the BH), so that the direction of the angular momentum of both the BH and the inner accretion disc is parallel or anti-parallel to the direction of the total angular momentum, $\bf{J_\mr{tot}}=\bf{J_\mr{BH}}+\bf{J_\mr{warp}}$. Here, $\mathbf{J_\mr{warp}}=J_\mr{warp}\mathbf{\hat{J}_\mr{d}}$, is the angular momentum of a single warp increment, which is assumed to be aligned with the angular momentum of the outer accretion disc (i.e.~the large-scale accretion flow, which we calculate directly from the simulation). In practice we do not evolve the BH one warp increment at a time, but instead a number of increments, $N_\mr{warp}=\Delta M/M_\mr{warp}$, are used over each time-step, where $\Delta M = \dot{M}_\mr{acc,H}\Delta t$ is the total mass to be consumed over a single time-step of length $\Delta t$. This approach is valid as long as none of the relevant quantities (BH mass, BH spin magnitude and direction) change appreciably during a single time-step. We ensure this by implementing additional time-step limiters related to BH spin (\S~\ref{sec:timesteps}).

To calculate $M_\mr{warp}$ and $J_\mr{warp}$, we require knowledge of 1) the warp radius, $R_\mr{warp}$, 2) the surface density, $\Sigma(R)$, and 3) the specific angular momentum, $L(R)$, of the accretion disc. The warp mass, $M_\mr{warp}$, is calculated as the integral over the surface density out to $R_\mr{warp}$, while the warp angular momentum, $J_\mr{warp}$, includes the specific angular momentum, $L(R)$, in the same integral. For the thin disc we assume Keplerian rotation, $L(R)=\sqrt{M_\mr{BH}GR}$, which is appropriate if pressure (both radiative and thermal) is negligible compared to gravity. For the thick and slim discs, the specific angular momentum is smaller by a numerical factor of order $0.5$, and this is due to thermal and radiation pressure, respectively, being important in those two accretion disc states. We discuss our choices for $R_\mr{warp}$ and $\Sigma$ in Appendix~\ref{sec:appendix_warp}. These choices then specify $M_\mr{warp}$ and $J_\mr{warp}$. We also take into account the possible truncation of the accretion discs at their self-gravity radius, as discussed in Appendix~\ref{sec:appendix_warp}.

\subsection{Final BH spin and mass loss to gravitational waves after BH mergers}
\label{sec:BH_mergers}

The BH spin of a remnant after a BH-BH merger depends on the mass ratio of the BHs, $q=M_\mr{BH,2}/M_\mr{BH,1}<1$ and three vectors: the two BH spin vectors of the BHs, $\mathbf{a}_1$ and $\mathbf{a}_2$, as well as the orbital angular momentum of the system around their centre of mass, $\mathbf{L}$. We use a fitting formula from \cite{Barausse2009}, derived from numerical relativity simulations:
\begin{equation}
    \mathbf{a}_\mr{fin} = \frac{1}{(1+q)^2}(\mathbf{a}_1+\mathbf{a}_2q^2+\mathbf{l}q),
\label{eq:a_fin}
\end{equation}
where $\mathbf{l}$ is a vector whose direction is the same as that of the orbital angular momentum $\mathbf{L}$ (in the centre-of-mass frame), while its magnitude is given by
\begin{equation}
\begin{array}{lll}
|\mathbf{l}|=&\frac{s_{4}}{\left(1+q^{2}\right)^{2}}\left(\left|\mathbf{a}_{1}\right|^{2}+|\mathbf{a}_{1}|^{2} q^{4}+2\left|\mathbf{a}_{1} \| \mathbf{a}_{2}\right| q^{2} \cos \phi\right)+ \\
&+\left(\frac{s_{5} \mu+t_{0}+2}{1+q^{2}}\right)\left(\left|\mathbf{a}_{1}\right| \cos \theta+\left|\mathbf{a}_{2}\right| q^{2} \cos \xi\right)+ 
2 \sqrt{3}+t_{2} \mu+t_{3} \mu^{2}.
\label{eq:l}
\end{array}
\end{equation}
Here, $\mu=q/(1+q)^2$ is the symmetric mass ratio, while $s_4 = -0.1229$, $s_5 = 0.4537$, $t_0 = -2.8904$, $t_2 = -3.5171$, $t_3 = 2.5763$ are fitting parameters. The cosines that appear in Eqn.~(\ref{eq:l}) are given by $\cos \phi=\hat{\mathbf{a}_{1}} \cdot \hat{\mathbf{a}_{\mathbf{2}}}$, $\cos \theta=\hat{\mathbf{a}_{1}} \cdot \hat{\mathbf{l}}$ and $\cos \xi=\hat{\mathbf{a}_{2}} \cdot \hat{\mathbf{l}}$.

Given the information available within the model, we could in principle calculate the recoil velocity of the remnant, as well as the total mass fraction lost to gravitational waves. We do not implement the former at this stage, since we cannot reliably track the movement of black holes in their host galaxies due to our use of BH repositioning. However, we do implement the latter. We use results from numerical relativity simulations (\citealt{Barausse2012}) and calculate the final mass of the remnant BH as:
\begin{equation}
\begin{array}{ll}
    M_\mr{BH,fin} = &(M_\mr{BH,1}+M_\mr{BH,2})\Big\{1 - [1 - e_\mr{ISCO}(\Tilde{a})]\mu  \\
    &- 4\mu^2[4p_0+16p_1\Tilde{a}(\Tilde{a}+1)+e_\mr{ISCO}(\Tilde{a})-1]\Big\},
\label{eq:E_rad}
\end{array}
\end{equation}
where $p_0=0.04827$, $p_1=0.01707$ and $e_\mr{ISCO}(\Tilde{a})$ is the dimensionless specific binding energy (the expression can be read off from Eqn.~\ref{eq:eps_rad_NT}) at the innermost stable circular orbit calculated using an effective BH spin variable defined as 
\begin{equation}
    \Tilde{a} = \frac{|\mathbf{a}_1|\cos\theta+|\mathbf{a}_2|\cos\xi}{(1+q)^2}.
\label{eq:a_tilde}
\end{equation}

\subsection{Additional time-step conditions}
\label{sec:timesteps}

Given the changes to the BH model in the form of AGN jets and BH spin evolution, a few additional time-step conditions need to be implemented. The minimum of these time-steps is taken to actually evolve the BH, alongside other time-step conditions already used for the BH in the code. We introduce a jet-related time-step condition that is given by:
\begin{equation}
\Delta t_\mr{jet}=\frac{\Delta E_\mr{jet}}{P_\mr{jet}},
\label{eq:jet_time_step}
\end{equation}
where $P_\mathrm{jet}$ is the current, instantenous jet power, while $\Delta E_\mathrm{jet}$ is the energy to be given to a pair of particles (see \S~\ref{sec:AGN_implementation}). This time-step limiter ensures that the BH is woken up by the time it needs to `hand out' a pair of kicks. 

We also introduce two time-step conditions related to the angular momentum of the BH. The first of these ensures that the magnitude of the BH spin does not change much over a single time-step, and it is given by
\begin{equation}
\Delta t_\mr{a}=0.1\frac{\vert a\vert M_\mr{BH}}{s \dot{M}_\mr{acc,H}},
\label{eq:spin_time_step}
\end{equation}
where $s=\mathrm{d}a/(\mathrm{d}M_\mathrm{acc,H}/M_\mathrm{BH})$ is the spinup/spindown function discussed in \S~\ref{sec:spin_magnitude}, given by Eqns.~(\ref{eq:da_dlnMSMBH_th}) and (\ref{eq:da_dlnMSMBH_Ricarte}). The numerical factor $0.1$ quantifies how finely we want to evolve BH spin; it ensures that the value of BH spin changes by no more than $10$ per cent (relative to the current value) over the next time-step.

The second angular momentum time step condition is related to the redirection of the BH spin vector. Since the BH spin vector may be redirected very quickly relative to its magnitude (due to LT torques), this condition is separate to that mentioned above. This time-step condition is given by
\begin{equation}
\Delta t_\mr{a}=0.1\frac{M_\mr{warp}J_\mr{BH}}{\dot{M}_\mr{acc,H}J_\mr{warp}\sin\theta},
\label{eq:redirection_time_step}
\end{equation}
where $\theta$ is the angle between the current BH spin vector and the angular momentum of gas in the accretion disc on large scales. The numerical prefactor is again present to ensure a sufficiently fine evolution of the BH spin direction. In particular, in the case that the BH spin vector and the gas angular momentum are perpendicular ($\sin\theta=1$), this criterion will lead to a change of no more than $\approx5\degree$ in the BH spin vector direction per time-step.

\subsection{Jet launching scheme}
\label{sec:jet_scheme}

For each BH, we parametrize the choice of when and how to kick gas particles in the form of a target jet velocity, $v_\mathrm{jet}$. We express the energy received by the particles through the target jet velocity as $\Delta E=2\times \bar{m}_\mathrm{ngb}v_\mathrm{jet}^2/2$, where $\bar{m}_\mathrm{ngb}$ is the mean gas particle mass of neighbours in the SPH kernel of the BH, and the multiplication by two is present since we always kick in pairs. For every BH, jet energy is funneled to a reservoir until it surpasses the value $\Delta E$, at which point a pair of particles are kicked.

We do not kick particles perfectly along the jet direction (which is parallel or anti-parallel with the BH spin vector for each BH), but instead implement a finite half-opening angle $\theta_\mr{jet}=7.5\degree$ (we found little difference when varying this from $0\degree$ to $30\degree$). This is accomplished by assigning a new kick direction every time a kick event occurs; this direction is given by a unit vector $\mathbf{n}_\mr{jet}$ that is drawn randomly and uniformly in solid angle within a cone with half-opening angle $\theta_\mr{jet}$ directed along the BH spin axis. Since we always kick in pairs, the above procedure is done for one particle in the ‘positive' direction (along one side of the jet) and for another particle in the ‘negative' direction (the other side of the jet). We draw only one direction for every pair of kicks, so that the imparted momentum is perfectly opposite in direction for the two particles. 

We kick particles by increasing their velocity (in the frame of each BH) by $\Delta \mathbf{v} = \Delta v \mathbf{n}_\mr{jet}$, where $\mathbf{n}_\mr{jet}$ is the kick direction for a given particle. The magnitude of the velocity increase, $\Delta v$, is chosen in such a way that the kinetic energy of each particle increases exactly by $\Delta E_\mathrm{jet}/2$. $\Delta v$ thus depends on the initial particle velocity (in the frame of the BH) and varies for every particle. Conservation of kinetic energy gives
\begin{equation}
\frac{1}{2}m_\mr{g}(\mathbf{v}_i+\Delta\mathbf{v})^2 - \frac{1}{2}m_\mr{g}\mathbf{v}_i^2 = \frac{\Delta E_\mr{jet}}{2},
\label{eq:conservation}
\end{equation}
where $\mathbf{v}_i$ is the initial velocity in the BH frame. This equation can be solved for the magnitude of the velocity increase $\Delta v$, yielding
\begin{equation}
\Delta v = \sqrt{v_{i,\mr{jet}}^2 + v_\mr{jet}^2} - v_{i,\mr{jet}},
\label{eq:conservation2}
\end{equation}
where $v_{i,\mr{jet}}=\mathbf{v}_i \cdot \mathbf{n}_\mr{jet}$ is the initial velocity projected onto the kick direction. 

Ideally, one would want to apply radial kicks to particles with respect to the BH. In our scheme, the relative position vector pointing from the BH to each particle may be significantly misaligned with respect to the (new) relative velocity vector. For example, a particle that is near the equatorial plane (defined by the BH spin vector) would be kicked with a velocity that is near-perpendicular to the relative position vector that connects the BH and the particle. While this situation is not desirable, it is our favoured approach for the following reasons. In order to ensure that jets are being launched rather than winds, small opening angles have to be used. For a spherical kernel with $\approx60$ neighbours, a cone with an opening half-angle of $7.5\degree$ occupies only $\approx13$ per cent of the kernel volume. Thus, if we were kicking particles radially, on average, we would expect to find only a few neighbours in this region to pick from, assuming an isotropic density distribution. Jet feedback would then become more prone to noise, as we would not find any neighbours to kick in many cases. With non-isotropic particle distributions, e.g.~one planar with respect to the BH spin vector, this issue would be exacerbated.


\subsection{The numerical implementation of AGN feedback}
\label{sec:AGN_implementation}

The thermal mode of AGN feedback is implemented in the same way as in the purely thermal AGN feedback model (see \S~\ref{sec:thermal_feedback}), with the only difference being that different feedback efficiencies are used. We implement jets as described in the previous subsection. Both feedback modes use their own energy reservoirs that are independent of each other.

The reader may wonder whether it is reasonable to describe AGN jet feedback using a prescription as simple as pure kinetic kicks, especially given the complexity of jet and lobe physics (e.g.~\citealt{Bourne2023}), which includes instabilities, relativistic effects, magnetic fields and cosmic rays. These processes are either not resolved or not included in our model. Nonetheless, we argue that the effects of kinetic jets indeed represent the first-order effects of jet feedback as it is thought to be occurring in the real Universe: the driving of anisotropic bow shocks that heat the ambient medium. This issue is discussed further in Appendix \ref{sec:kinetic_jets}.

The velocity in our model represents both a physical and a numerical parameter. On the physical side, increasing it leads to stronger shocks, and hotter and less dense lobes inflated by jets. Feedback thus becomes more effective. However, increasing it should not affect the dynamical evolution of jets and lobes for very supersonic jets that inflate self-similar lobes (\citealt{Falle1991}, \citealt{Kaiser1997}). In idealized simulations, we found the opposite to be the case: varying jet velocities can have a major impact (\citealt{Husko2022_self_similar}, \citealt{Husko2022_bubbles}, \citealt{Husko2022_spin_driven}). This is largely caused by numerical sampling effects. Consider the relation between the total kinetic energy of a jet event and the velocity: $E_\mr{kin,}=0.5N_\mr{jet}m_\mr{g}v_\mr{jet}^2$, where $N_\mr{jet}$ is the number of particles kicked into the jet. For a jet event of a fixed energy, increasing the jet velocity by some factor $X$ causes the number of particles by which the jet is sampled to drop as $N_\mr{jet}\propto 1/X^2$. Therefore, the choice of jet velocities heavily impacts the sampling of AGN jet feedback, and it needs to be made carefully.

We assume the following relation between the target jet velocity and BH mass:
\begin{equation}
v_\mr{jet}(M_\mr{BH}) = v_\mr{jet,0}\times\min \bigg(1, \sqrt{\frac{M_\mr{BH}}{10^9\hspace{0.5mm}\mr{M}_\odot}}\bigg),
\label{eq:jet_velocity_scaling}
\end{equation}
where $v_\mr{jet,0}$ is the normalization of the scaling (and also the maximum achieved value, for BHs with $M_\mr{BH}\geqslant10^9$ M$_\odot$). We also apply a minimum to the above relation, which is $v_\mr{jet,min}=10^{2.5}\approx316$ km~s$^{-1}$. The choice of the normalization, $v_\mr{jet,0}$, is discussed in \S~\ref{sec:calibration_first_results}. The form of Eqn.~(\ref{eq:jet_velocity_scaling}) is chosen to satisfy the same argument as for thermal feedback (see \S~\ref{sec:thermal_feedback}). In particular, given the above form, the energy per feedback event (individual particle kick) scales linearly with BH mass, which means that for every unit of BH growth on a logarithmic scale, BHs are expected to experience a similar number of feedback events\footnote{This is not strictly true in the hybrid AGN feedback model, for neither the thermal nor jet feedback modes. The reason is that feedback efficiencies depend on BH spin and Eddington ratio, which introduces an indirect dependence on BH mass. This implies that the amount of energy injected by BHs does not scale linearly with BH mass, as it does in the purely thermal AGN feedback model. However, these effects are relatively small.}. This in turn means that there is no BH mass scale below which AGN feedback is not sampled sufficiently well. With variable velocities, we sample AGN feedback better than we would be able to if a constant velocity were used. 

\subsection{Comparison with other models}
\label{sec:model_comparison}

Accretion disc and BH spin modeling, in the context of galaxy formation and evolution, has previously been included in numerical calculations that use Monte Carlo-generated hierarchical merger trees (e.g.~\citealt{Volonteri2005}, \citealt{Shapiro2005}), semi-analytical models (e.g.~\citealt{Lagos2009}, \citealt{Fanidakis2011}, \citealt{Barausse2012b}, \citealt{Griffin2019a}) and hydrodynamical simulations (e.g.~\citealt{Dubois2014_spin}, \citealt{Fiacconi2018}, \citealt{Bustamante2019}, \citealt{Talbot2020}, \citealt{Dubois2021}, \citealt{Koudmani2024}, \citealt{Sala2024}). Most such models have included only the thin accretion disc, while the thick accretion disc, with its strong jet spindown effects, is typically neglected. Recent models have begun including it, however (\citealt{Griffin2019a}, \citealt{Talbot2020}, \citealt{Dubois2021}, \citealt{Koudmani2024}). The slim disc, which is relevant for super-Eddington accretion, is also usually neglected.

Our subgrid accretion disc and BH spin evolution model inherits many aspects of the \cite{Griffin2019a} semi-analytical model, adapted for use in hydrodynamical simulations by assuming that the angular momentum of a subgrid accretion disc is aligned with that of the gas in the BH kernel. We improve upon that model by explicitly including the thick disc and its associated jet-induced spindown. We also model super-Eddington accretion using the slim disc, where jets are also very important. Our model is similar to the one employed in the NewHorizon simulations (\citealt{Dubois2021}), with two main differences: 1) it includes super-Eddington accretion and the slim disc, and 2) it includes an accretion efficiency for the thick disc (also for the slim disc), which is much smaller than unity. In practice, this vastly reduces accretion, feedback and jet-induced spindown from the thick disc in our model compared to NewHorizon. The strong jets from the thick disc in NewHorizon are probably the reason why the BH masses and spins are generally too low in that model.

Our model is, in some aspects, quite similar to the unified accretion disc model presented by \cite{Koudmani2024}, although they do not model the slim disc. Their model includes an accretion efficiency for the thick disc, which varies with the unsuppressed Eddington ratio in the same way as in our model ($\epsilon_\mathrm{acc}\propto f_\mathrm{Edd}$), although with a much different normalization (so that accretion in the thick disc is less efficient in our model). Their model assumes that the inflow rates of mass and angular momentum onto the accretion disc are well-resolved, which they achieve using super-Lagrangian refinement (e.g.~\citealt{Bourne2024}, \citealt{Shin2025}). This means they can explicitly track the total mass and angular momentum in the subgrid accretion disc$-$BH system. Paired with assumptions on the disc structure, their model yields an accretion rate through the disc and onto the BH that is generally much different from the accretion rate onto the disc itself. Due to viscous time delays the accretion rate onto the BH can be very different in magnitude and much less variable. 

Including a subgrid accretion disc (as a separate reservoir of mass and angular momentum that modulates the accretion rate) as described above is hard to do in cosmological galaxy formation simulations such as the ones presented in this paper, since we are generally very far from resolving the inflow rates onto the subgrid accretion disc-BH system. For this reason, we assume that mass flows through the disc unimpeded (modulo accretion disc winds that strip away some of the mass) and with no time delay with respect to the inflow onto the disc. In practice, some assumptions on the size of a subgrid accretion disc (see e.g.~\citealt{Cenci2021}) could be sufficient, when paired with the assumed mass inflow rate (e.g.~Eqn.~\ref{eq:accr_rate}), to model the mass and angular momentum of the disc. This would then yield an accretion rate through the disc and onto the BH, as in the \cite{Koudmani2024} model, which might represent a much more realistic accretion model than the one where the BH immediately accretes the material that is falling onto a subgrid disc.

Many previous models have used an accretion efficiency or a concept similar to it (e.g.~\citealt{AnglesAlcazar2017}). Mass loading factors for jets and winds (e.g.~\citealt{Choi2012}, \citealt{Rennehan2024}) may produce a similar effect in that only a small fraction of the mass accreting from large scales makes it down to the BH event horizon, with the rest being expelled into feedback by a wind or jet. In these models, it is assumed that the mass being affected by feedback on resolved scales originates entirely from a subgrid wind or jet. This means there is an explicit link between the wind or jet velocity and the mass loading (accretion efficiency). We do not make such an assumption: the mass being heated by thermal feedback or kicked by jets in our model on resolved scales could be the mass in the ambient medium that is swept up by a subgrid wind or jet. Therefore, our accretion efficiencies are independent of the feedback prescription.

Our model is also significantly different compared to others in relation to AGN feedback. Many galaxy formation models use multiple feedback modes, with jets often being used at low Eddington ratios and thermal feedback at high Eddington ratios (e.g.~\citealt{Kaviraj2017}, \citealt{Dubois2021}). Compared to such models, ours differs most in also including jets at high Eddington ratios alongside thermal feedback (in the thin disc), which bears similarity to SIMBA's low-velocity directed winds at high Eddington ratios (\citealt{Dave2019}). It also differs in including thermal feedback at low Eddington ratios alongside jets (in the thick disc, albeit here the jets are completely dominant). Furthermore, our model includes the slim disc, with both thermal and jet feedback being important (see the similar model by \citealt{Massonneau2023}). Unlike most other hydrodynamical galaxy formation models (except that of NewHorizon; \citealt{Dubois2021}), all of our AGN feedback efficiencies depend on BH spin. Our AGN feedback is in some respects similar to that of \cite{Rennehan2024}, who also use three accretion states, both types of spin-dependent feedback across the different states, and feedback mass loadings that effectively behave like our accretion efficiencies (although they are explicitly linked with the feedback prescription).

\section{Calibration and initial cosmological results}
\label{sec:calibration_first_results}

Here we give a brief overview of our calibration efforts, before describing the procedure and its relevant ingredients in more detail. Readers not interested in details of the calibration procedure can skip to \S~\ref{sec:calibration_results}, where we discuss the results of our calibration.

In principle, the hybrid AGN feedback model could be calibrated in a similar way as the thermal AGN feedback model used for the flagship COLIBRE simulations (\citealt{Chaikin2025}). This would involve Gaussian-process emulation and a hypercube spanning many parameters, including 1) stellar feedback parameters in the COLIBRE model, 2) thermal AGN feedback parameters present in both the thermal and hybrid AGN feedback models, and 3) new hybrid AGN feedback-related parameters. This process was beyond the scope of our current efforts, but will be done in the future. We opted to instead perform a manual calibration varying a set of parameters as small as possible.

For calibrating the hybrid AGN feedback model, we chose to start from the fiducial set of parameter values found by \cite{Chaikin2025} for the thermal AGN feedback model at each resolution. Out of all parameters that vary with resolution, we chose a minimal set that are present in the thermal AGN model and varied these independently at every resolution, alongside a minimal set of parameters present only in the hybrid AGN model. The parameter variation and calibration process, which is described in more detail in \S~\ref{sec:calibration_procedure}, was used to first calibrate the hybrid AGN feedback model at the m7 resolution. That set of parameter values was then used as the best guess for the calibration at the m6 resolution. We found that less than half of our calibration parameters needed changing at higher resolution, none of which were the new parameters introduced in the hybrid AGN feedback model. Once a satisfactory model (when compared with calibration observables) was found for the m6 resolution by tuning these few parameters, this model was used as a best guess for m5, and they were then tuned again until an acceptable model was found.

In \S~\ref{sec:param_choices} we provide an overview of the parameters that we varied, and a brief motivation for why we chose to vary them. In \S~\ref{sec:calibration_data} we discuss which galaxy and BH relations and properties were chosen for calibration, as well as details of observational data used for comparison during calibration. \S~\ref{sec:param_variations} includes the variations of several of the important calibration parameters, and the impact of these variations on the calibration relations. Informed by these variations, in \S~\ref{sec:calibration_procedure}, we discuss how we carried out the actual calibration. Finally, in \S~\ref{sec:calibration_results} we discuss the calibrated parameter values that result from our calibration procedure. We also discuss the results of the largest hybrid AGN simulations in terms of the calibration observables, and the convergence between resolutions.

\subsection{The choice of calibration parameters}
\label{sec:param_choices}

The five parameters that were varied are:
\begin{itemize}
    \item $v_\mr{jet,0}$: the normalization (which is also the maximum) of the jet velocity - BH mass scaling (Eqn.~\ref{eq:jet_velocity_scaling}),
    \item $r_\mr{tr,0}$: the dimensionless size of the thick accretion disc at a critical Eddington ratio $f_\mr{Edd,d}=0.01$, which determines the accretion efficiency at $f_\mr{Edd,d}<0.01$ (see \S~\ref{sec:accretion_rates}), such that $\epsilon_\mr{acc,th}\propto f_\mr{Edd,d}/\sqrt{r_\mr{tr,0}}$,
    \item $\epsilon_\mr{f}$, the coupling efficiency for thermal AGN feedback in the thin accretion disc state at $0.01<f_\mr{Edd,d}<1$ (\S~\ref{sec:thermal_eff}),
    \item $M_\mr{BH,seed}$, the BH seed mass (\S~\ref{sec:seeding}),
    \item $n_\mr{H,pivot}$, the pivot density for SNIa and CCSN feedback heating temperatures (see Eqn.~\ref{eq:SN_dT}).
\end{itemize}
The first two parameters are present only in the hybrid AGN feedback model, and relate to jet velocities and accretion efficiencies, while the remaining three parameters are present both in the hybrid and in the fiducial, thermal-only AGN feedback model. In principle, we could have varied other hybrid AGN parameters. However, these parameters (whose values were set in \S~\ref{sec:hybrid_model}) were not varied for two reasons: 1) the effects of varying them are generally smaller than for the above two hybrid AGN parameters (as confirmed by many variations not shown here), especially for our target calibration observables, and 2) they are better constrained by either theory or observations. In most cases, both of these reasons were true at the same time. The two new AGN parameters that we vary have both unconstrained values and have a strong effect on our target calibration observables across the ranges we considered (see \S~\ref{sec:velocity_variations} and \S~\ref{sec:efficiency_variations}). 

We found that the modifications introduced in the hybrid AGN feedback model have a large effect on how feedback operates and on how BHs self-regulate their growth. For this reason, varying the coupling efficiency, $\epsilon_\mr{f}$, is necessary to reproduce the observed $M_\mr{BH}-M_*$ relation at the massive end ($M_*>10^{10}$ $\mr{M}_\odot$). Allowing the variation of the BH seed mass, on the other hand, is necessary since that parameter affects how quickly BHs grow and quench their host galaxies, a process that is inherently different in the hybrid AGN feedback model compared to the thermal one. We opted not to vary the AGN heating temperature, $\Delta T_\mr{AGN}$, since it has a widespread impact on nearly all galaxy and BH properties, and there is currently no single relation or property that can be used to constrain its value, especially not when there is additionally a second mode of AGN feedback operating (whose strength is modulated by jet velocities).

The final parameter, the SNIa and CCSN pivot density, $n_\mr{H,pivot}$, is varied since we find AGN feedback to have stronger effects in the low-mass regime ($M_*<10^{10}$ M$_\odot$) in the hybrid AGN feedback model than in the thermal one (see \S~\ref{sec:param_variations}). It is thus also necessary to change at least one stellar feedback-related parameter, since the interplay between AGN feedback and stellar feedback is different. We chose to vary $n_\mr{H,pivot}$ since the effects of varying it are strongest for intermediate-mass galaxies ($M_*\approx10^{10}$ M$_\odot$), where we also found the differences between the thermal and hybrid AGN feedback models to be maximal in terms of the observables used for calibration. 

\subsection{Calibration relations and observational data}
\label{sec:calibration_data}

The observables considered for calibration are largely the same as for the thermal AGN feedback model (\citealt{Chaikin2025}). We calibrate on: 1) the $z=0$ galaxy stellar mass function (GSMF), 2) the $z=0$ projected galaxy stellar half-mass radius$-$stellar mass relation and 3) the $z=0$ black hole mass$-$stellar mass relation. However, to this list we add a fourth observable: 4) the bolometric AGN luminosity function (LF) (also at $z=0$). This is added since the accretion efficiency that we have introduced has a strong effect on the growth history of BHs and how many of them are in the thick disc, radiatively-inefficient regime ($f_\mr{Edd,d}<0.01$) at $z=0$, and how many in the thin disc, radiatively-efficient regime ($f_\mr{Edd,d}>0.01$). This parameter thus strongly affects the balance of accretion disc states and feedback modes, and this is inherently tied to the AGN LF (see \S~\ref{sec:efficiency_variations}).

The observational data, which we introduce in the following subsections, are all converted to the cosmology we adopted for COLIBRE, and to the Chabrier initial mass function (IMF), also used in COLIBRE. For calibration against observational data, we only consider galaxies with stellar masses above 100 resolution elements, which corresponds to $M_*\approx10^9$, $10^8$ and $10^7$ M$_\odot$ for m7, m6 and m5, respectively. Galaxies with fewer stellar particles are poorly resolved, and we do not consider the simulation predictions to be reliable in that stellar mass regime\footnote{The actual particle limit down to which a simulation prediction is reliable depends on what is being considered. For example, we find that the galaxy stellar mass function can be correctly predicted with as few particles as $10$ or even less. On the other hand, galaxy sizes may only start being reliable once resolved with a few hundred or even $1000$ particles. For simplicity, we use a single threshold of $100$ particles for all quantities considered here, but discuss the role of the threshold where relevant.}. We do not consider observational comparisons at $M_*\gtrsim10^{11.3}$ M$_\odot$, since the effects of aperture sizes, projection, and observational sensitivity become more important for central galaxies of galaxy groups and clusters. When plotting any of our calibration relations that involve the stellar mass, we use fixed bin widths of 0.25 dex. Next, we detail individual observational datasets used for calibration.

\subsubsection{Galaxy stellar mass function}

The GSMF data that we compare with is from \cite{Driver2022} (their table 6), which corresponds to the fourth data release of the GAMA survey. We increase the normalization of their GSMF by 0.0807 dex (as recommended by the authors) to account for two effects: 1) the data is at a median redshift of $z=0.03$, whereas we are interested in comparing at $z=0$, and 2) the small GAMA volume is biased and not representative of the cosmic mean. 

Galaxy masses determined observationally exhibit both random (statistical), as well as systematic uncertainties. While we cannot correct for systematic uncertainties, it is important for our comparison to account for the effects of Eddington bias that appear in the these observations, which can shift the GSMF towards different values (\citealt{Schaye2025}). The stellar masses in GAMA have a 0.1 dex random uncertainty. We thus apply a 0.1 dex random log-normal scatter to our simulated stellar masses, which mimics Eddington bias, i.e.~the effects of random observational uncertainties in determining stellar masses. We refer the reader to fig.~D1 in \cite{Schaye2025} for the effects of applying different amounts of this scatter on the main observables, using the thermal AGN feedback model. The effects are very similar for the hybrid model.

\subsubsection{Galaxy size$-$mass relation}

We calibrate galaxy sizes by considering the galaxy size$-$stellar mass relation, which we compare with \cite{Hardwick2022} (their table C1), who reported projected half-mass galaxy sizes from a mass-selected sample of galaxies in the xGASS spectroscopic survey (\citealt{Catinella2018}). We define galaxy sizes as the radii that contain, in projection (along a random direction), half of the 50 kpc 3D mass of a given galaxy. We apply 0.1 dex log-normal scatter to the stellar masses, as for the GSMF. 


\subsubsection{Black hole mass$-$stellar mass relation}

To ensure the simulated BH masses are realistic, we calibrate using the BH mass$-$ stellar mass relation at $z=0$. We primarily compare with data from \cite{Graham2023} (their table 1, with stellar masses corrected as per the erratum \citealt{Graham2024erratum}), who performed morphological decompositions and defined galaxies as elliptical, lenticular or discy. They found that the BH mass$-$stellar mass relation was largely similar for ellipticals and lenticulars, but differs from that of discs, which have significantly lower BH masses at fixed stellar mass. We construct an overall median BH mass$-$stellar mass relation, representative of the entire galaxy population, by binning the data from \cite{Graham2023}, using fractions of different morphological types of galaxies as weights. This procedure is described in more detail in Appendix \ref{app:BH_data}, where we also provide, for convenience, the binned median BH masses with the associated uncertainties on both the median stellar and BH masses in the bins. While only $\approx150$ (in total) BHs were included in these bins, this is sufficient to constrain the median BH masses to within an uncertainty of $\approx0.1$ dex per stellar mass bin. For reference, we also compare with the power law fit from \cite{McConnell2013}, which is based on a smaller sample (with fewer spirals), and which was obtained without weighting the contributions of different morphological types. Therefore, this fit is closer to describing the $M_\mathrm{BH}-M_*$ relation for spheroid-dominated galaxies than for the overall population, and is not expected to agree with our binned weighted median points based on \cite{Graham2023}, nor with our simulations.

The simulated BH mass used to compare with the above data is the subgrid mass of the most massive BH found within a 50 kpc 3D aperture around a given galaxy. We apply 0.15 dex log-normal scatter to the simulated stellar masses of the galaxies hosting these BHs, which is slightly larger than for the GSMF and size comparisons, but more consistent with the stellar mass uncertainties found by \cite{Graham2023}. We only compare at $M_*>10^{10}$ M$_\odot$, since the observational data used here only extends down to such masses.

\subsubsection{Bolometric AGN luminosity function}

For the AGN bolometric LF we use the compilation from \cite{Shen2020}\footnote{Taken directly from the data repository at \url{https://github.com/gkulkarni/QLF/blob/master/Data/allqlfs.dat}.}, who estimated bolometric luminosities from soft X-ray, hard X-ray and mid-infrared AGN observations. We compare to their $z=0.2$ data, using the simulation predictions at the same redshift. We also compare with their `global' fit (i.e.~using redshifts $0<z<6$ to constrain their fitting function). The AGN luminosities, used to construct our predictions for the bolometric AGN LF, are computed using Eqn.~(\ref{eq:L_AGN}). We add 0.3 dex random log-normal scatter to these luminosities, mimicking typical observational uncertainties in inferring bolometric luminosities. When comparing with the observed (inferred) AGN bolometric LF, we only consider luminosities\footnote{Even this limit could be too low. Observed X-ray LFs are typically not reliable below $L_\mr{X}=10^{42}$ erg s$^{-1}$, due to contamination from the host galaxy. This limit corresponds to approximately $L_\mr{bol}=10^{43}$ erg s$^{-1}$, and it may be redshift-dependent.} $L_\mr{AGN,bol}>10^{42}$ erg s$^{-1}$. 

This lower limit is imposed for three reasons. First, observational inferences of the AGN LF are likely very uncertain at these low luminosities, and affected by systematics (\citealt{Shen2020}). Second, we find that nearly all of the BHs that contribute to the bolometric AGN LF at $L_\mr{AGN,bol}>10^{42}$ erg s$^{-1}$ are in the thin disc state. On the other hand, nearly all BHs in the thick disc state ($f_\mr{Edd,d}<0.01$) have luminosities below this threshold value. We are confident about our predictions for the luminosities from the thin disc, since they use radiative efficiencies of order $10$ per cent (but dependent on BH spin) implied by standard accretion disc theory (\citealt{NovikovThorne1973}). On the other hand, thick disc radiative efficiencies (we use those from \citealt{Mahadevan}, Appendix \ref{sec:appendix_rad_eff}) are very uncertain. Thirdly, from a practical standpoint, we find that our predicted AGN LFs converge with resolution for luminosities larger than the above-mentioned critical value, while they increase with higher resolution at values below it (see \S~\ref{sec:convergence}).

\subsection{Single parameter variations}
\label{sec:param_variations}


\captionsetup[table]{skip=0pt} 
\begin{table}
\begin{center}
\caption{The simulations used for our parameter variation study (all in 50$^3$ Mpc$^3$ volume boxes and at m7 resolution). The first column is the simulation name, while the remaining columns are the parameters being varied: 1) the jet velocity normalization (note that $v_\mr{jet}\propto v_\mr{jet,0}\sqrt{M_\mr{BH}}$, see Eqn.~\ref{eq:jet_velocity_scaling}), 2) the accretion efficiency parameter $r_\mr{tr,0}$ (note that $\epsilon_\mr{acc,thick}\propto f_\mr{Edd,d}/\sqrt{r_\mr{tr,0}}$ for $f_\mr{Edd,d}<0.01$, see Eqn.~\ref{eq:acc_eff}), 3) the thermal feedback coupling efficiency in the thin disc ($0.01<f_\mr{Edd,d}<1$). Other parameters are kept fixed at their fiducial values found through calibration (Table \ref{tab:tab2}). The first row is the fiducial thermal AGN feedback simulation (for reference), the second is the fiducial hybrid AGN feedback simulation, while other groups of rows indicate different simulations where a single parameter is varied in the hybrid model (with the varied values highlighted in bold). }
\label{tab:tab3}

\centering
\begin{tabular*}{\columnwidth}{@{\extracolsep{\fill}}lccr}
    Simulation &  $v_\mr{jet,0}$ [km~s$^{-1}$] &  $r_\mr{tr,0}$ & $\epsilon_\mr{f}$  \\
  \hline 
  L050m7 (thermal) & N/A & N/A & $0.1$ \\
  
  L050m7h (fiducial) & $10^{4.5}$ & $10^4$ & $0.03$ \\
  \hline
  L050m7h\_low\_$V\mr{_j}$ & $\mathbf{10^{4}}$ & $10^4$ & $0.03$ \\
  L050m7h\_high\_$V\mr{_j}$ & $\mathbf{10^{5}}$ & $10^4$ & $0.03$ \\
  L050m7h\_const\_$V\mr{_j}$ & \bf{N/A}$^*$ & $10^4$ & $0.03$ \\
  L050m7h\_const\_$V\mr{_j}$\_recal$^\ddagger$ & \bf{N/A}$^*$ & $10^4$ & $0.03$ \\
  \hline
  L050m7h\_low\_$\epsilon_\mr{acc}$ & $10^{4.5}$ & $\mathbf{10^6}$ & $0.03$ \\
  L050m7h\_high\_$\epsilon_\mr{acc}$ & $10^{4.5}$ & $\mathbf{10^2}$ & $0.03$ \\
  L050m7h\_const\_$\epsilon_\mr{acc}$ & $10^{4.5}$ & \bf{N/A}$^\dagger$ & $0.03$ \\
  L050m7h\_const\_$\epsilon_\mr{acc}$\_recal$^\ddagger$ & $10^{4.5}$ & \bf{N/A}$^\dagger$ & $0.03$ \\
  \hline
  L050m7h\_low\_$\epsilon\mr{_f}$ & $10^{4.5}$ & $10^4$ & $\mathbf{0.01}$ \\
  L050m7h\_high\_$\epsilon\mr{_f}$ & $10^{4.5}$ & $10^4$ & $\mathbf{0.1}$ \\
  \hline 
\end{tabular*}
\end{center}
$^*$These simulations did not use the $v_\mr{jet,0}$ parameter; they instead used constant jet velocities equal to $10^4$ km~s$^{-1}$. \\
$^\dagger$These simulations did not use the $r_\mr{tr,0}$ parameter; they instead had constant accretion efficiencies $\epsilon_\mr{acc}=1$ in the thick disc. \\
$^\ddagger$These simulations had further parameters (the BH seed mass and the stellar feedback pivot density) varied, beyond the ones shown here.

\end{table}

Here we vary most of the parameters that are used in our calibration procedure (see \citealt{Chaikin2025} for the effects of variations of the other parameters). These variations are summarized in Table~\ref{tab:tab3}. We perform all variations at m7 resolution, using (50 Mpc)$^3$ simulation volumes (i.e.~L050). The effects of these variations are generally similar at higher resolutions, so we do not show them here. The fiducial hybrid model used as the baseline in these comparisons is that found by calibration as described in \S~\ref{sec:calibration_procedure} and \S~\ref{sec:calibration_results}. Here we do not discuss how well the fiducial model fits the data; this is discussed in \S~\ref{sec:calibration_results}.

We vary: the jet velocity normalization, $v_\mr{jet,0}$ (\S~\ref{sec:velocity_variations}), the accretion efficiency parameter, $r_\mr{tr,0}$ (\S~\ref{sec:efficiency_variations}), and the thermal feedback coupling efficiency, $\epsilon_\mr{f}$ (\S~\ref{sec:coupling_efficiency_variation}). We vary the first two parameters since they are new parameters introduced in the hybrid AGN feedback model, and since they impact nearly all galaxy and BH properties. While the coupling efficiency is not a new parameter, its variation has more significant consequences than it did in the OWLs and EAGLE models (\citealt{BoothSchaye2010}, \citealt{Schaye2015}). We briefly discuss the impact of varying the BH seed mass and the stellar feedback pivot density in \S~\ref{sec:other_variations}. They control the strength of AGN and stellar feedback, respectively.

\subsubsection{The impact of varying jet velocities}
\label{sec:velocity_variations}

\begin{figure*}
\includegraphics[width=1\textwidth, trim = 0 5 0 0]{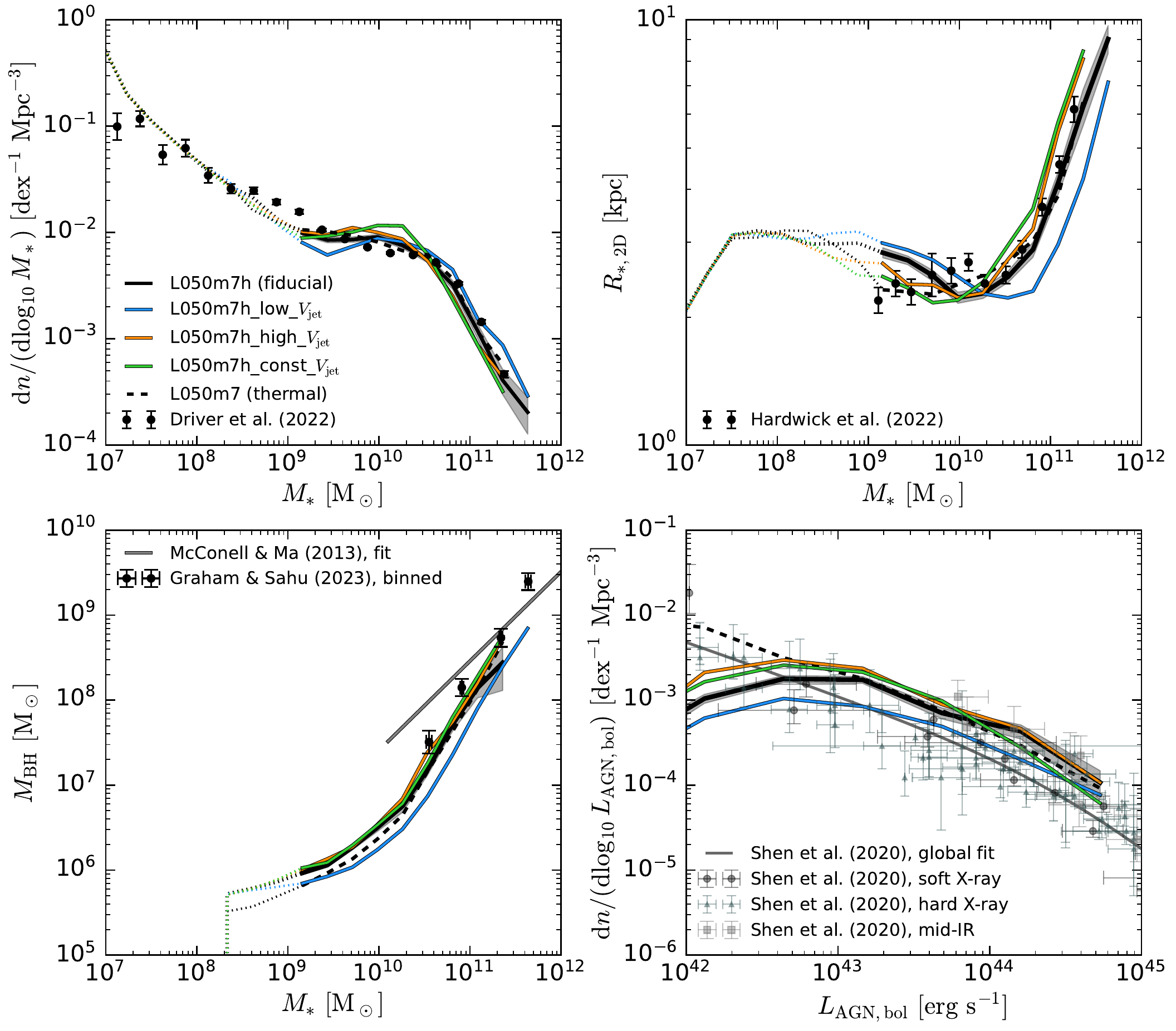}
\caption{The impact of varying the jet velocities (see \S~\ref{sec:AGN_implementation}) on galaxy and BH observables considered for calibration, in L050m7 simulations. From left to right and top to bottom we show, in order: the 1) galaxy stellar mass function, 2) median galaxy projected size$-$stellar mass relation, 3) median BH mass$-$stellar mass relation and 4) AGN bolometric LF. The first three are shown at $z=0$, while the AGN LF is shown at $z=0.2$. The grey dashed line shows the thermal AGN model, the black solid line shows the fiducial hybrid AGN model (with the shaded region showing uncertainties around that curve, which are Poisson errors in the case of the galaxy stellar mass and AGN LF functions, and standard errors of the median in the case of galaxy sizes and BH masses), and coloured curves show variations relative to the fiducial hybrid AGN simulation (see legend). Lines are dotted at $M_*<10^9$ M$_\odot$, which corresponds to roughly 100 star particles, indicating limitations due to resolution. The velocity normalization was varied up and down by factors of $3.16$ (0.5 dex; orange and blue, respectively), corresponding to increasing and decreasing the energies of individual feedback events by a factor of 10. The variations were done relative to the fiducial value of $v_\mr{jet,0}=10^{4.5}\approx31600$ km~s$^{-1}$. We also performed a simulation with a constant velocity of $v_\mr{jet,0}=10^4$ km~s$^{-1}$ (green). Observational data are taken from: 1) \protect\cite{Driver2022}, 2) \protect\cite{Hardwick2022}, 3) \protect\cite{Graham2023} and \protect\cite{McConnell2013}, 4) \protect\cite{Shen2020}. Further details are discussed in \S~\ref{sec:calibration_data}.}
\label{fig:velocity_variations}
\end{figure*}%

In Fig.~\ref{fig:velocity_variations} we show the impact of varying the jet velocities, specifically the velocity normalization parameter $v_\mr{jet,0}$, on the calibration observables that we described in the previous subsection. The velocity normalization, $v_\mr{jet,0}$, is varied up and down by factors of $\approx3.16$ ($0.5$ dex) relative to the fiducial value ($10^{4.5}\approx31600$ km~s$^{-1}$). This corresponds to ranges of energy per feedback event differing by an order of magnitude. These variations change how explosive\footnote{By `explosive' we refer to how much energy a single gas particle receives as part of a feedback event. More explosive feedback events are less prone to numerical overcooling and drive stronger outflows.}, intermittent, and well-sampled the feedback is. For reference, we performed a run with a constant velocity $v_\mr{jet}=10^{4}$ km~s$^{-1}$, as opposed to the fiducial $v_\mr{jet}\propto M_\mr{BH}^{0.5}$ scaling. Note that comparisons of our fiducial model with these higher- and lower-velocity runs, as well as with the one using a constant velocity, cannot be used to rule out these models. A recalibration should first be attempted on such models. The low-velocity variation shown in Fig.~\ref{fig:velocity_variations} was our initial attempt at finding a best-fitting model. We have attempted to recalibrate the constant-velocity run; this is discussed in Appendix \ref{app:recal_models}. As we show there, a constant-velocity model that matches all the observed calibration data can indeed be found. We opted for variable velocities, however, to better sample AGN feedback in low-mass galaxies, and to avoid an unphysical BH mass$-$stellar mass relation, as explained in \S~\ref{sec:thermal_feedback}.

Increasing jet velocities results in a slight reduction of the GSMF (top left) at $M_*>10^{10.5}$ M$_\odot$. This is due to the more explosive nature of feedback, which tends to reduce radiative energy losses and make feedback more efficient. If the jet velocities are reduced, the number of galaxies in this mass range increases. Somewhat counterintuitively, larger velocities lead to an increase in the GSMF at $10^9<M_*<10^{10}$ M$_\odot$, probably because AGN feedback is sampled more poorly and is thus less effective, despite being more explosive. Lower velocities similarly reduce the number of low-mass galaxies. Small differences in galaxy masses between the different runs persist down to $M_*\approx10^8$, indicating that AGN feedback affects dwarf galaxies. The run that has a constant $v_\mr{jet}=10^{4}$ km~s$^{-1}$ features a similar number of high-mass galaxies as the other runs, with the exception of the low-normalization one, but the number of galaxies around the knee ($M_*\approx 10^{10}$ M$_\odot$) is significantly increased. This is probably caused by AGN feedback being poorly sampled, and therefore less effective, in low-mass galaxies. Differences in the GSMF at lower masses could merely indicate a horizontal shift; if there is a greater number of more massive galaxies, there must also be a smaller number of less massive galaxies. In Appendix \ref{app:SHMR} we show the effect of the same jet velocity variations on the stellar mass$-$halo mass relation. There, it is clear that higher jet velocities cause more effective feedback in haloes more massive than $10^{12}$ M$_\odot$, leading to decreased stellar masses at fixed halo mass. The opposite is true for lower-mass haloes.

Galaxy sizes (top right) show a picture consistent with stellar masses in terms of jet velocity variations: increasing (decreasing) jet velocities results in massive galaxies, $M_*>10^{10}$ M$_\odot$, being larger (smaller) due to more effective AGN feedback, and the opposite is true for less massive galaxies. Galaxy sizes show differences between the different simulations down to a mass of $M_*\approx10^8$ M$_\odot$, as for the GSMF. The constant-velocity case has the largest galaxies, despite not having the largest velocities at high masses. This is likely due to the assembly of galaxies and their cores occurring at high redshifts, when the BH mass, and therefore the jet velocity (in the fiducial, variable-velocity case) is not large. 

BH masses (bottom left) are higher (lower) at fixed stellar mass, if higher (lower) jet velocities are used. However, this effect is driven by the fact that galaxy masses are changed when jet velocities are changed, shifting the same BHs horizontally in the $M_\mr{BH}-M_*$ diagram. We checked the BH mass$-$halo mass relation and found it to be unaffected by the velocity variation. This suggests that BH self-regulation is tight and independent of jet velocity, at least to the degree that we varied it here. The case of constant $v_\mr{jet}=10^{4}$ km~s$^{-1}$ has similar BH masses as the other runs (except the run with low velocities). 

Increasing jet velocities causes a slight increase (decrease) in the bolometric AGN LF (bottom right) if jet velocities are increased (decreased), but only at $L_\mr{AGN,bol}<10^{44}$ erg s$^{-1}$. The AGN LF is also slightly higher in the case of constant $v_\mr{jet}=10^{4}$ km~s$^{-1}$. These changes are probably related to how well AGN feedback is sampled.

The impact of these variations demonstrates that jet velocities cannot be calibrated by considering only a single one of the calibration observables. This partly motivates the calibration approach that we outline in \S~\ref{sec:calibration_procedure}.

\subsubsection{The impact of varying the accretion efficiency in the thick disc state}
\label{sec:efficiency_variations}

\begin{figure*}
\includegraphics[width=1\textwidth, trim = 0 5 0 0]{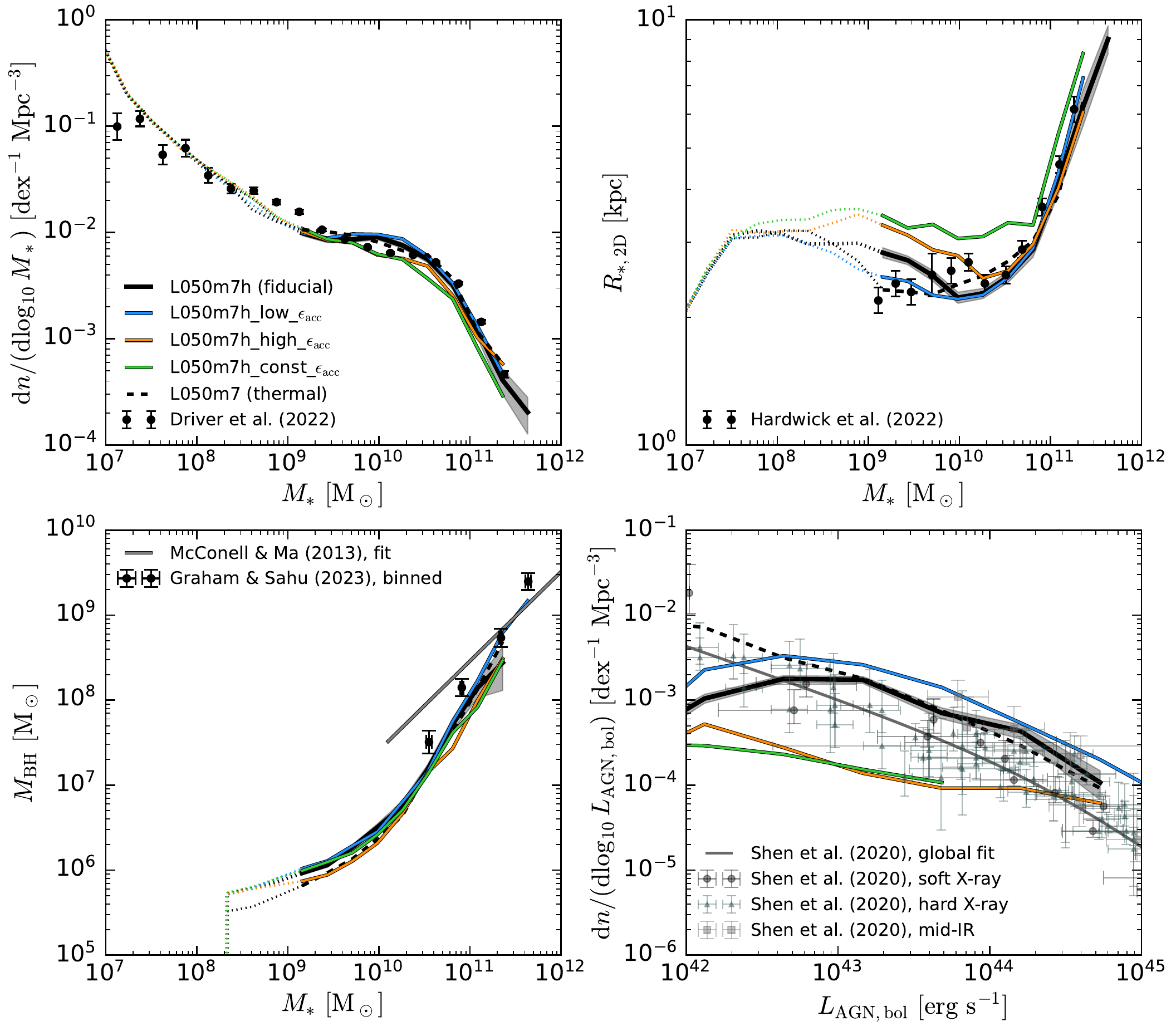}
\caption{As Fig.~\ref{fig:velocity_variations}, but showing the impact of varying the thick disc ($f_\mr{Edd,d}<0.01$) accretion efficiencies through variations of the parameter $r_\mr{tr,0}$ (note that $\epsilon_\mr{acc,th}\propto f_\mr{Edd,d}/\sqrt{r_\mr{tr,0}}$, see \S~\ref{sec:accretion_eff}, Eqn.~\ref{eq:eps_thick}). The fiducial value of ($r_\mr{tr,0}=10^4$) yields the scaling $\epsilon_\mr{acc}\approx0.0316(f_\mr{Edd,d}/0.01)$ for $f_\mr{Edd,d}<0.01$. Here we show the effect of varying $r_\mr{tr,0}$ up and down by factors of 100 (orange and blue, respectively), which changes the normalization of the $\epsilon_\mr{acc}-f_\mr{Edd,d}$ relation by factors of 10. We also compare with a case of fully efficient accretion in the thick disc, i.e.~where $\epsilon_\mr{acc}=1$ (green).}
\label{fig:efficiency_variations}
\end{figure*}%

Fig.~\ref{fig:efficiency_variations} has the same format as Fig.~\ref{fig:velocity_variations}, but shows the effects of varying the thick disc accretion efficiency $\epsilon_\mr{acc}$ through variations of the dimensionless size of the disc ($r_\mr{tr,0}$) at $f_\mr{Edd,d}=0.01$. We vary $r_\mr{tr,0}$ by factors of 100, which changes the accretion efficiency normalization by factors of 10, relative to the fiducial scaling $\epsilon_\mr{acc}=0.0316(f_\mr{Edd,d}/0.01)$. Similar to what we did for the velocities, we also show for reference a simpler case that uses no accretion efficiency (i.e.~$\epsilon_\mr{acc}=1$ in the thick disc). In this figure we do not show a 100 per cent efficiency model that has been recalibrated, but instead show the first-order effects of setting the accretion efficiency to unity. An attempt at recalibration is shown in Appendix \ref{app:recal_models}, where we show that a recalibrated model with a 100 per cent accretion efficiency cannot reproduce the AGN LF.

The GSMF (top left) is significantly affected by changes to the accretion efficiency only at $M_*>10^{9.5}$ M$_\odot$, although small differences exist at lower masses. The number of intermediate-mass galaxies at the knee of the GSMF ($M_*=10^{10}$ M$_\odot$) drops monotonically with increasing $\epsilon_\mr{acc}$. The number of massive galaxies ($M_*\approx 10^{11}$ M$_\odot$) is similar in all cases, except for the $\epsilon_\mr{acc}=1$ case, where the number of such galaxies is slightly lower. As shown in Appendix \ref{app:SHMR}, at fixed halo mass, an increase in the accretion efficiency causes a decrease in stellar masses.

The galaxy sizes (top right) show a picture consistent with stellar masses, with massive galaxies unaffected by variations of the accretion efficiency. For lower-mass galaxies ($M_*<10^{11}$ M$_\odot$) there is a monotonic trend with accretion efficiency, where higher values lead to larger sizes (all the way to $M_*=10^{8}$ M$_\odot$). This shows that thick disc feedback is important in low-mass galaxies, and not just in massive ones. 

BH masses (bottom left) are largely unaffected, although higher accretion efficiencies lead to a small drop in the normalization of the relation. 

The bolometric AGN LF (bottom right) is strongly affected by variations of the accretion efficiency. This is caused by the accretion efficiency also changing the effectiveness of jet feedback in the thick disc, which in turn changes the likelihood that a BH is in the thick disc state ($f_\mr{Edd,d}<0.01$) rather than in the radiatively-efficient thin disc state ($f_\mr{Edd,d}>0.01$). This appears counterintuitive, but can be explained by self-regulation. In the self-regulated growth and feedback picture, BHs will, on average, tend to have the same feedback powers in different simulations (\citealt{Booth2009}, \citealt{BoothSchaye2010}, \citealt{Li2025TNG}). As a result, if the accretion efficiency is decreased, BHs adjust by increasing their accretion rates in order to have the same feedback powers. However, this makes them more likely to enter the thin disc state. BHs in the thin disc state are visible as AGN, and are virtually the only ones contributing to the observed AGN LF at $z=0$, at least at $L_\mr{AGN,bol}>10^{42}$ erg s$^{-1}$. The sensitivity of the bolometric AGN LF to the accretion efficiency was used to calibrate the accretion efficiency parameter, as illustrated in the figure. The strong disagreement of the fully efficient accretion in the thick disc ($\epsilon_\mr{acc}=1$ for $f_\mr{Edd,d}<0.01$) with the observations is one of the main motivations for our introduction of the accretion efficiency.

\subsubsection{The impact of varying the coupling efficiency}
\label{sec:coupling_efficiency_variation}

\begin{figure*}
\includegraphics[width=1\textwidth, trim = 0 5 0 0]{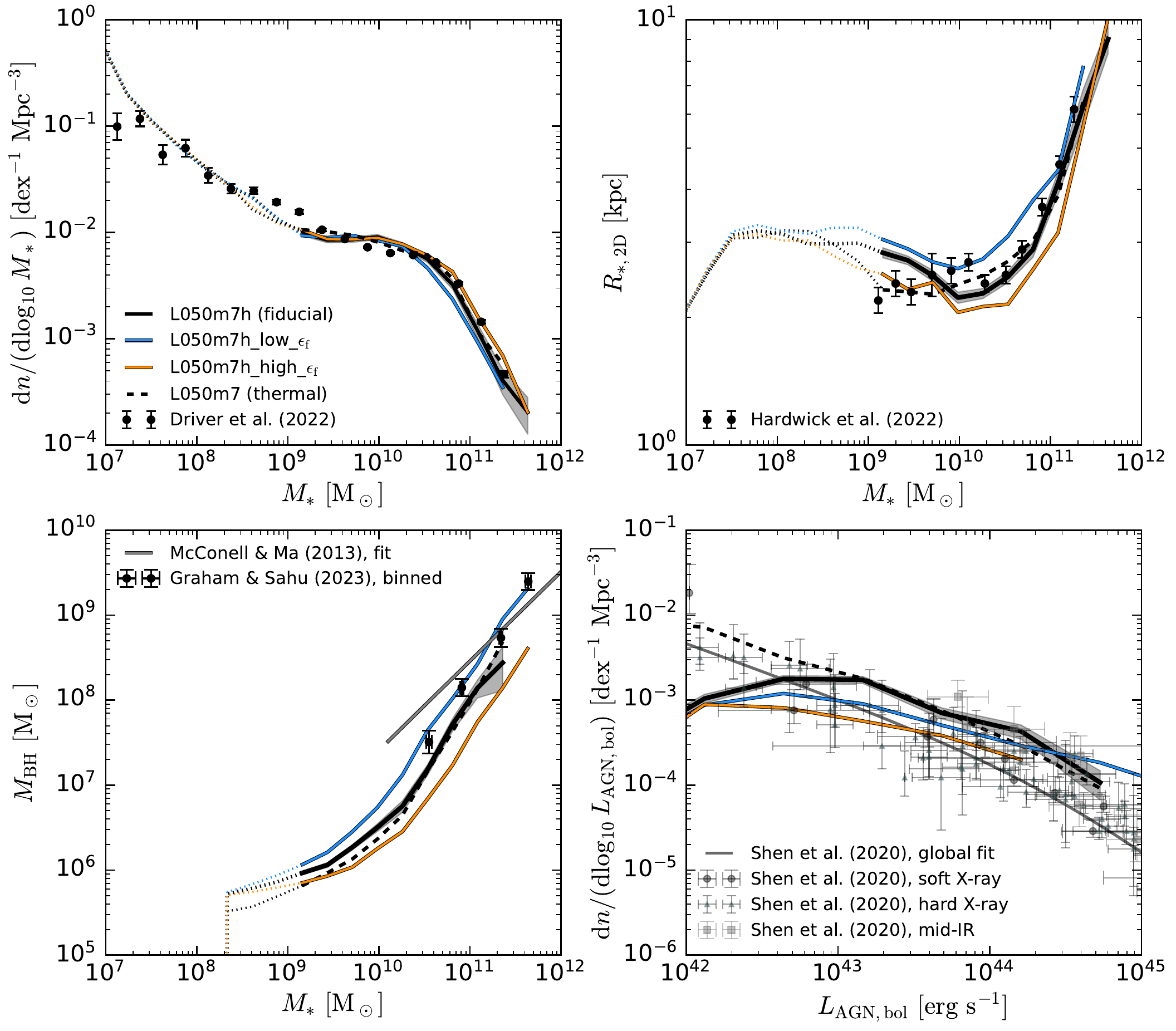}
\caption{As Fig.~\ref{fig:velocity_variations}, but showing the impact of varying the thin disc ($0.01<f_\mr{Edd,d}<1$) coupling efficiency, $\epsilon_\mr{f}$, used for thermal feedback (see \S~\ref{sec:thermal_feedback}), on the galaxy and BH properties considered for calibration. The fiducial value of the coupling efficiency at m7 resolution is $\epsilon_\mr{f}=0.03$; we vary this value by $\approx$ 0.5 dex to $\epsilon_\mr{f}=0.01$ and $\epsilon_\mr{f}=0.1$.}
\label{fig:coupling_efficiency_variations}
\end{figure*}%

In Fig.~\ref{fig:coupling_efficiency_variations} we show the effects on the calibration relations of varying the coupling efficiency of radiation from the thin disc, at $0.01<f_\mr{Edd,d}<1$, which is the main channel of thermal AGN feedback in our simulations (\S~\ref{sec:feedback_energetics}). We vary the value by approximately $0.5$ dex up and down relative to the fiducial value, which is $\epsilon_\mr{f}=0.03$ at m7 resolution. We find that the effects of varying it on galaxy properties are widespread, persisting down to $M_*\approx10^8$ M$_\odot$, and they also have a mild impact on the AGN LF. A higher thin disc coupling efficiency results in higher stellar masses (see also the effects of this variation on the stellar mass$-$halo mass relation, shown in Appendix \ref{app:SHMR}).

The effects on galaxy properties may appear to be counterintuitive; increasing the coupling efficiency (i.e.~making feedback stronger) results in higher stellar masses and smaller galaxies, which one would intuitively expect to happen when making feedback weaker. However, these effects are an indirect consequence of the main impact of varying the coupling efficiency, which is a change in BH masses (visible in the bottom left panel of Fig.~\ref{fig:coupling_efficiency_variations}). Increasing the coupling efficiency reduces BH masses, and vice-versa. This is again a result of self-regulation (e.g.~\citealt{Booth2009}, \citealt{BoothSchaye2010}, \citealt{Li2025TNG}). In short, feedback from BHs adjusts to yield approximately the same outflow powers (on average) in different simulations\footnote{In previous simulations, which used constant heating temperatures for AGN feedback, feedback was not sampled well in low-mass galaxies. Self-regulation was thus only important for massive galaxies ($M_*>10^{10}$ M$_\odot$), where AGN feedback was dominant. In COLIBRE, especially in the hybrid AGN feedback model, AGN feedback is important for a wider range of galaxy masses. The $M_\mr{BH}-M_*$ relation is not as steep around $M_*\approx10^{10}$ M$_\odot$as it was in EAGLE (\citealt{Bower2017}), and self-regulation applies more generally in COLIBRE, not just for massive galaxies dominated by AGN feedback.}. If one uses lower coupling efficiencies, BHs have to grow more in order to have the same feedback powers. The change in BH masses then indirectly impacts galaxy properties, since we tie AGN heating temperatures and jet kick velocities explicitly to BH mass (\S~\ref{sec:AGN_implementation}). Increasing the coupling efficiency decreases the BH masses, which in turn makes AGN feedback weaker on galaxy scales. This demonstrates that in our model, in terms of parameters, changing feedback efficiencies is not dominant, as BHs tend to adjust to inject as much energy as they need to regulate their own growth. Instead, the efficiency of AGN feedback (in terms of how efficiently it quenches galaxies) is mostly governed by how explosive it is, which is instead controlled by heating temperatures and jet velocities.

\subsubsection{The impact of varying other calibration parameters}
\label{sec:other_variations}

We vary neither the BH seed mass nor stellar feedback pivot densities here, as the effects of those variations are similar in the hybrid  and thermal AGN feedback models. We refer the reader to \cite{Chaikin2025} for a detailed discussion of such variations. In short, increasing the BH seed mass causes earlier AGN growth and feedback, and therefore earlier quenching of galaxies (making them less massive and larger). It is most relevant for intermediate-mass and massive galaxies ($M_*>10^{10}$ M$_\odot$). Increasing the pivot density decreases the heating temperatures of SN feedback, making it less effective at low-to-intermediate galaxy masses ($10^9<M_*<10^{10.5}$ M$_\odot$), thus increasing their masses and decreasing their sizes. Variations of both of these parameters have very little effect on the BH mass$-$stellar mass relation and on the AGN LF.

\subsection{Calibration procedure}
\label{sec:calibration_procedure}

In \S~\ref{sec:param_variations} we showed how varying the jet velocities, the accretion efficiencies and the coupling efficiencies affects the main galaxy and BH properties that we calibrate on (\S~\ref{sec:calibration_data}). We also use information on how the variations of the other two parameters (\S~\ref{sec:param_choices}), the BH seed mass and the stellar feedback pivot density, affect the galaxy and BH properties (\S~\ref{sec:other_variations}). Informed by this knowledge, our calibration procedure consists of the following steps:

{\bf Step 1:} Choose a value of the jet velocity normalization, $v_\mr{jet,0}$.

{\bf Step 2:} Given a choice of $v_\mr{jet,0}$, calibrate the coupling efficiency $\epsilon_\mr{f}$ and the accretion efficiency parameter $r_\mr{tr,0}$. These are calibrated on the $M_\mr{BH}-M_*$ relation and the AGN LF at $z=0$, respectively. More details of this are given below.

{\bf Step 3:} Given a $v_\mr{jet,0}$ and the newly-calibrated values of $\epsilon_\mr{f}$ and $r_\mr{tr,0}$, calibrate the seed mass $M_\mr{BH,seed}$ and the SNIa and CCSN pivot density $n_\mr{H,pivot}$, using the $z=0$ GSMF and stellar size$-$mass relation.

{\bf Step 4:} Repeat Steps 1-3 for different values of $v_\mr{jet,0}$. The fully-calibrated models with different choices of $v_\mr{jet,0}$ are then compared, and the one that best fits all of the data is chosen as the final model.

This procedure was performed first at m7 resolution, using (25 Mpc)$^3$ volumes (L025). The final models (in Step 4) were compared using larger (L050) simulations, since they provide better statistics and include more massive galaxies. The best-fitting set of parameter values found at m7 was used as the initial guess for m6, which was then recalibrated. The same was then repeated for m5, using the set of parameter values found for m6 as the initial guess. For m6 and m5, the full procedure was not repeated. We did not recalibrate $r_\mr{tr,0}$ (in Step 2) since we found it to be unnecessary. $\epsilon_\mr{f}$ was only slightly adjusted (in Step 2) at m6 compared to m7, but remained the same at m5 compared to m6. Only the BH seed mass and SN pivot density were significantly varied for m6 and m5 (in Step 3). The above procedure was done with $\approx30$ simulations at m7, and $\approx10$ each at m6 and m5.

The calibration was split into the above steps for the following reasons. The jet velocity affects nearly all observables that we consider in the calibration (see \S~\ref{sec:velocity_variations}). Models with different values of $v_\mr{jet,0}$ thus cannot be calibrated by comparing to a observed relation alone; we must instead fully calibrate such models. There is also no relation or quantity that would uniquely constrain jet velocities. 

Variations of the coupling efficiency, $\epsilon_\mr{f}$ (\S~\ref{sec:coupling_efficiency_variation}), and the accretion efficiency parameter, $r_\mr{tr,0}$ (\S~\ref{sec:efficiency_variations}), affect both galaxy and BH properties, while the BH seed mass, $M_\mr{BH,seed}$, and the SN pivot density, $n_\mr{H,pivot}$, affect mostly galaxy properties, with almost no effect on the high-mass end of the $M_\mr{BH}-M_*$ relation or on the AGN LF (at $z\approx0$). We thus first use the $M_\mr{BH}-M_*$ relation and AGN LF to constrain the values of $\epsilon_\mr{f}$ and $r_\mr{tr,0}$, respectively, and as a subsequent step use the remaining calibration observables (GSMF and galaxy half-mass size$-$stellar mass relation at $z=0$) to calibrate $M_\mr{BH,seed}$ and $n_\mr{H,pivot}$.

The calibration of the coupling efficiency $\epsilon_\mr{f}$ was done in the same way as for the thermal AGN feedback model in COLIBRE. Increasing the value of $\epsilon_\mr{f}$ decreases the normalization of the $M_\mr{BH}-M_*$ relation at the massive end ($M_*>10^{10}$ M$_\odot$), and vice-versa, due to self-regulation (\citealt{BoothSchaye2010}). While changing $\epsilon_\mr{f}$ affects galaxy properties in COLIBRE alongside BH masses, it is important to calibrate the coupling efficiency on the $M_\mr{BH}-M_*$ relation first, and only then calibrate other parameters that directly affect galaxy properties (specifically the BH seed mass and stellar feedback pivot density).

We found that changing accretion efficiencies strongly affects the likelihood that BHs will be in a given accretion disc state (radiatively-inefficient thick disc with $f_\mr{Edd,d}<0.01$ or radiatively-efficient thin disc with $f_\mr{Edd,d}>0.01$), as well as changing galaxy properties, but it does not strongly affect BH masses (see \S~\ref{sec:efficiency_variations}). The accretion efficiency parameter $r_\mr{tr,0}$ (we remind the reader that $\epsilon_\mr{acc,th}\propto f_\mr{Edd,d}/\sqrt{r_\mr{tr,0}}$ at $f_\mr{Edd,d}<0.01$, see Eqn.~\ref{eq:eps_thick}) can thus be calibrated largely independently of the coupling efficiency $\epsilon_\mr{f}$. Increasing the accretion efficiency decreases the normalization of the AGN LF, since more growth and feedback in the thick disc state (mostly jet feedback) makes it more likely that BHs will have $f_\mr{Edd,d}<0.01$ and thus not contribute to the AGN LF. The opposite is true if the accretion efficiency is decreased. 

Having chosen $\epsilon_\mr{f}$ and $r_\mr{tr,0}$, the remaining step of choosing the seed mass $M_\mr{BH,seed}$ and pivot density $n_\mr{H,pivot}$ was relatively simple. We found that the seed mass needed to increase by roughly a factor of two (at all resolutions) in comparison with the thermal AGN feedback model in order for galaxy sizes at the massive end to match observations. The stellar feedback pivot density had to be reduced, decreasing the heating temperature used for SN feedback, thereby weakening it. This was done to counter the effects of a larger seed mass in the hybrid AGN model compared to the thermal one, which results in a relatively more important role of AGN feedback in low-mass and intermediate-mass galaxies ($M_*\lesssim 10^{10}$ M$_\odot$).

\subsection{Calibration results}
\label{sec:calibration_results}


\captionsetup[table]{skip=0pt} 
\begin{table*}
\begin{center}
\caption{The final calibrated values of the parameters in the hybrid AGN feedback model (in parentheses we give the values of these parameters in the thermal AGN feedback model, if they exist). In order, these are 1) the jet velocity normalization (note that $v_\mr{jet}\propto v_\mr{jet,0}\sqrt{M_\mr{BH}}$, see Eqn.~\ref{eq:jet_velocity_scaling}), 2) the accretion efficiency parameter $r_\mr{tr,0}$ (note that $\epsilon_\mr{acc,th}\propto f_\mr{Edd,d}/\sqrt{r_\mr{tr,0}}$ for $f_\mr{Edd,d}<0.01$, see Eqn.~\ref{eq:eps_thick}), 3) the thermal feedback coupling efficiency, 4) the BH seed mass and 5) the SN feedback pivot density.}
\label{tab:tab2}

\centering
\begin{tabular*}{1.\textwidth}{@{\extracolsep{\fill}}lccccr}
    &  $v_\mr{jet,0}$ [km~s$^{-1}$] &  $r_\mr{tr,0}$ & $\epsilon_\mr{f}$ &  $M_\mr{BH,seed}$ [M$_\odot$] & $n_\mr{H,pivot}$ [cm$^{-3}$] \\
  \hline 
  m7 & $10^{4.5}$ & $10^4$ & $0.03$ ($0.1$) & $5\times10^5$ ($3.16\times10^{5}$) & $1.5$ ($0.6$) \\
  \hline
  m6 & $10^{4.5}$ & $10^4$ & $0.02$ ($0.05$) & $5.62\times10^{4}$ ($3\times10^{4}$) & $0.75$ ($0.5$) \\
  \hline
  m5 & $10^{4.5}$ & $10^4$ & $0.02$ ($0.05$) & $4\times10^4$ ($2\times10^{4}$) & $1.2$ ($1.0$) \\
\end{tabular*}
\end{center}
\end{table*}

Here we discuss the final parameter choices for our hybrid AGN feedback simulations at m7, m6 and m5 resolution, as given in Table \ref{tab:tab2}. The jet velocity scaling is the same at all resolutions, and is given by the parameter $v_\mr{jet,0}=10^{4.5}$ km~s$^{-1}$, yielding $v_\mr{jet}\approx31600\sqrt{M_\mr{BH}/10^9\hspace{0.3mm}\mr{M}_\odot}$ km~s$^{-1}$. Jet feedback is thus very explosive, with massive BHs launching particle pairs at a velocity slightly exceeding $0.1c$. This is not far from the observed values of kpc-scale jets, which are typically in the transrelativistic regime ($0.3-0.5c$, \citealt{Jetha2006}, \citealt{Mullin2009}), or even slower (\citealt{Shulevski2019}). The energy associated with each of these kicks corresponds to an equivalent temperature increase of $\approx2.4\times10^{10}$ K (assuming a mean molecular weight of 1.2, appropriate for ionized hydrogen). This is $\approx7.5$ times higher than the maximum heating temperatures used for the thermal feedback channel (in both the thermal and hybrid AGN feedback models) at m7 ($\Delta T_\mr{AGN,max}=10^{9.5}\approx3.1\times10^9$ K) and $2.4$ times higher than the maximum used at m6 and m5 ($\Delta T_\mr{AGN,max}=10^{10}$ K). Note also that this difference is even more stark if one considers an individual jet event to correspond to two particles, since we always kick in pairs. In this case, the total feedback AGN energy for each jet event is a factor $\approx15$ and $\approx4.5$ higher than in a thermal event for m7 and m6/m5, respectively.

Our initial choice for the parameter was $v_\mr{jet,0}=10^{4}$ km~s$^{-1}$, which made jet feedback approximately a factor of two less energetic per feedback event than thermal AGN feedback, at fixed BH mass. This was motivated by a desire to resolve jets as well as possible, while ensuring they are still highly supersonic relative to the circumgalactic/intracluster medium in their host haloes (jets need to be sufficiently supersonic, $v_\mr{jet}\gtrsim10c_\mr{s}$, to inflate lobes, see \citealt{Husko_winds}). However, we found it impossible to calibrate models to a satisfactory degree with such velocities. In particular, massive haloes ($M_\mr{halo}>10^{13}$ M$_\odot$) hosted galaxies that were too massive, star-forming and small compared to observations at $z=0$\footnote{See the blue curve in the top right panel in Fig.~\ref{fig:velocity_variations}; this disagreement with observational data could not be countered by e.g.~increasing the BH seed mass.}. We then attempted a value of $v_\mr{jet,0}=10^{4.25}$ km~s$^{-1}$, which mitigated but did not fully resolve the problem. With our current value, $v_\mr{jet,0}=10^{4.5}$ km~s$^{-1}$, galaxy sizes are in agreement with the data at m7 resolution, but possibly too small at m6 resolution at the high-mass end (see \S~\ref{sec:convergence}).

The fact that the hybrid AGN feedback model has to have such explosive AGN feedback can be explained by it featuring delayed BH growth in massive galaxies at high redshift ($z>3$) (Huško et al.~in prep.). Since heating temperatures and jet velocities scale with BH mass in COLIBRE, delayed BH growth implies less effective feedback for the same host galaxies at the same cosmic time. By making feedback more explosive (at fixed BH mass), we likely compensated for this effect. However, in hindsight, increasing the heating temperatures for the thermal AGN channel would probably have been at least as effective, if not more so, since quenching at high redshifts may be driven more by the thermal channel of feedback. 

The highest values of jet velocities we use, $^{4.5}$ km s$^{-1}$, are significantly higher compared to other cosmological galaxy formation simulations that use jets (\citealt{Kaviraj2017}, \citealt{Dave2019}, \citealt{Dubois2021}), which use values up to $10^4$ km s$^{-1}$. However, our maximum jet velocities are only reached at high BH masses, $M_\mathrm{BH}\geqslant10^9$ M$_\odot$. Galaxies with such high BH masses have already gone through their strongest gas outflow phases and been quenched, which occurs when their progenitor galaxies have a mass of $M\approx10^{10.5}$ M$_\odot$ and their BHs a mass of $M_\mathrm{BH}\approx10^{8}$ M$_\odot$. These galaxies have jet velocities more similar to $10^4$ km s$^{-1}$, used by other galaxy formation simulations. It should be noted, however, that at high redshifts, AGN jets in our simulations are predominantly from the thin disc state at high Eddington ratios (see \S~\ref{sec:feedback}), unlike other simulations.

Our final choice for the accretion efficiency parameter was $r_\mr{tr,0}=10^4$, which yields the scaling (see \S~\ref{sec:accretion_eff}, Eqn.~\ref{eq:eps_thick}) $\epsilon_\mr{acc,th}=0.0316(f_\mr{Edd,d}/0.01)$ for $f_\mr{Edd,d}<0.01$. We find this same scaling to work at all three resolutions: m7, m6 and m5. As is visible, the accretion efficiencies are of order one per cent or lower; this is necessary to significantly dampen the strength of jets from the thick disc state at $f_\mr{Edd,d}<0.01$. Our calibrated coupling efficiency is equal to $\epsilon_\mr{f}=0.03$ at m7 and slightly lower, $\epsilon_\mr{f}=0.02$, at m6 and m5. This is a factor of 3 (2.5) times lower than the coupling efficiencies used in the thermal AGN feedback model (0.1 and 0.05).

Our final BH seed masses are approximately two times larger in the hybrid AGN feedback model than in the thermal AGN feedback model at all resolutions. These values are necessary in order to reproduce the stellar masses and sizes of galaxies in massive haloes ($M_\mr{halo}>10^{13}$ M$_\odot$). This implies that the hybrid AGN feedback model is slightly less effective at quenching these massive galaxies, given the same BH seed mass. Our final pivot densities are higher than the values used in the thermal AGN feedback model at all resolutions. This implies that SN feedback is weaker, which is necessary to reproduce stellar masses and sizes of galaxies around the peak of the stellar mass$-$halo mass relation ($M_\mr{halo}\approx10^{12}$ M$_\odot$). We had to weaken SN feedback (relative to the thermal AGN feedback model) since AGN feedback is stronger in that mass regime due to the higher seed mass.

\subsection{Convergence}
\label{sec:convergence}

\begin{figure*}
\includegraphics[width=1\textwidth, trim = 0 5 0 0]{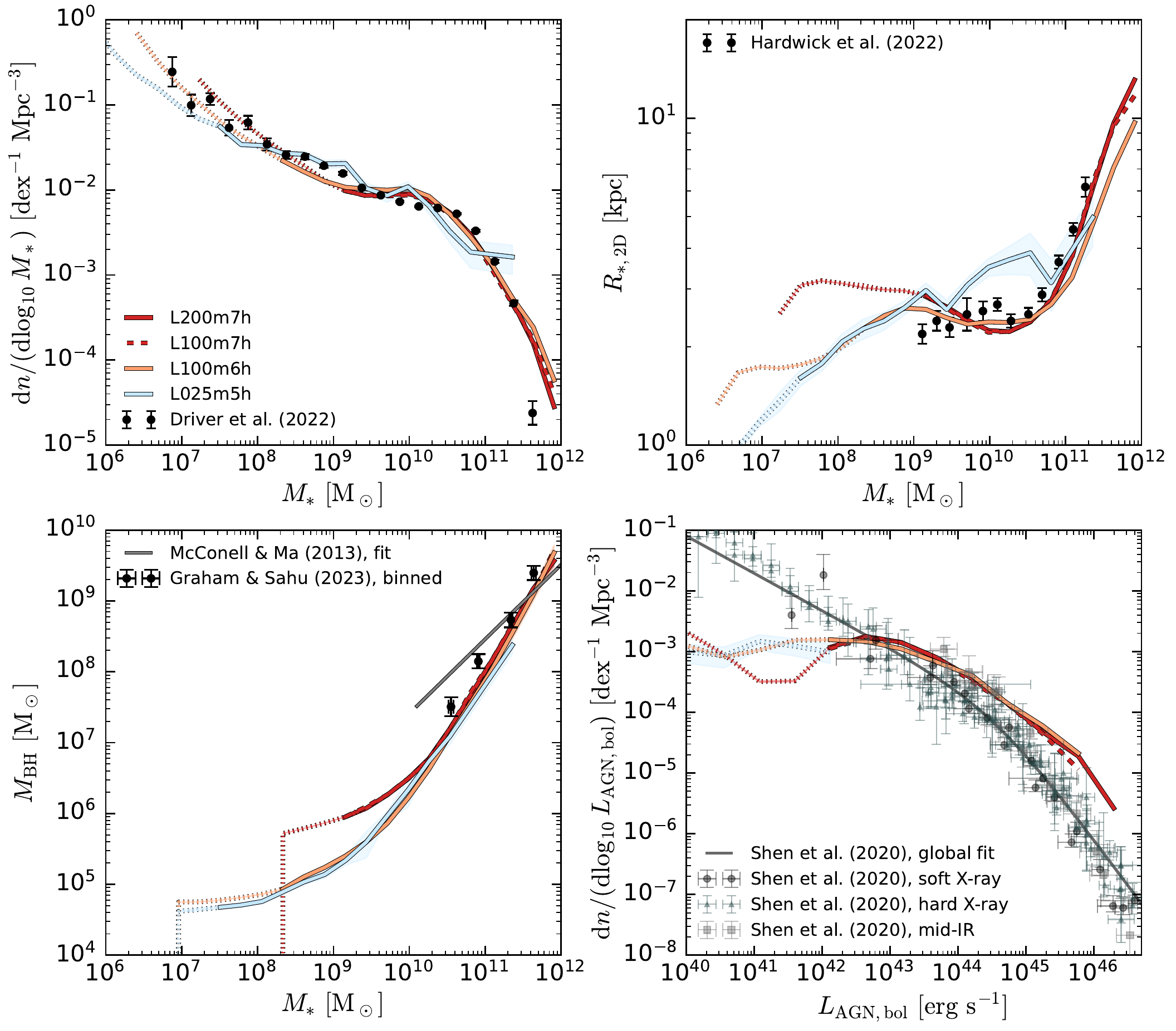}
\caption{The calibration observables for our final calibrated models at m7, m6 and m5 resolutions in the largest simulated volumes that have reached $z=0$ (L200, L100, and L025, respectively). The simulations were calibrated independently at each resolution. The panels that have stellar mass on the $x-$axis extend down to $M_*=10^6$ M$_\odot$, but we stop the curves where we resolve a galaxy with a single stellar particle ($\approx10^7$, $\approx10^6$ M$_\odot$ and $\approx10^5$ M$_\odot$, respectively), and make them dotted where the simulation predictions are unreliable (at a stellar mass corresponding to $\approx100$ particles for each resolution in the first 3 panels, and at an AGN luminosity of $L_\mathrm{bol}=10^{42}$ erg s$^{-1}$ for the last panel). At the high-mass (or luminosity) end we stop the curves at the first bin that has fewer than 10 objects in it. For other details of the comparison, see the caption of Fig.~\ref{fig:velocity_variations}.}
\label{fig:convergence_hybrid}
\end{figure*}%

In Fig.~\ref{fig:convergence_hybrid} we show the calibration relations for our biggest simulations at each resolution that were run with the hybrid AGN feedback model and that have reached $z=0$ (L200m7h, L100m6h, and L025m5h, with L100m7h also included for a box-size comparison), each having been calibrated independently. These are to be compared with observations and with each other. The figure has the same format as Figs.~\ref{fig:velocity_variations}-\ref{fig:coupling_efficiency_variations}, but we extend the axes ranges somewhat, since we are also interested in testing the convergence with resolution. Since the simulations are (re)calibrated at higher resolution, this constitutes a weak convergence test in the terminology of \cite{Schaye2015}. For clarity we do not show the comparison with the thermal AGN feedback model (see fig.~25 and fig.~26 in \citealt{Schaye2025} for that comparison). We note here, however, that the main observables shown here are reproduced to a similar degree of accuracy as with the thermal AGN feedback model. This means that the abundance of galaxies and their sizes, as well as the BH masses, are similar between the two models. This agreement between the two AGN feedback implementations also holds for many other predicted galaxy scaling relations that were not considered during the calibration. Both the thermal and hybrid AGN models at the m5 resolution perform slightly worse than their m6 and m7 counterparts, which is an outcome of more crude calibration due to wall-clock time constraints of running m5 simulations.

\subsubsection{Galaxy stellar mass function}

In the top left of Fig.~\ref{fig:convergence_hybrid}, we compare the simulated and observed GSMFs. We find that the model predicts similar GSMFs for m7 and m6 at all masses where the simulation predictions are reliable for both resolutions ($M_*\gtrsim10^9$ M$_\odot$), with the m6 model having a slightly higher GSMF at $10^9<M_*/M_\odot<10^{10}$. All three GSMFs are in fairly good agreement with the observational data at all stellar masses, within $0.2$ dex (this is comparable to the systematic uncertainty in the observationally inferred stellar masses). The simulated GSMF is $\approx0.15$ dex higher at the knee of the GSMF ($M_*\approx10^{10}$ M$_\odot$) than the data, and $\approx0.2$ dex lower than the data at $M_*\approx10^{9}$ M$_\odot$ for m7 and m6, but in good agreement with the data at m5. At lower masses, the simulated GSMF is in agreement with the increasingly uncertain observational data, even down to $M_*=10^{7}$ M$_\odot$, where the m7 resolution is not reliable. The simulated GSMF is higher than the last data point for the most massive galaxies from \cite{Driver2022} ($M_*\approx10^{11.6}$~M$_\odot$), but the comparison is not reliable here (see \S~\ref{sec:calibration_data}).

\subsubsection{Galaxy size$-$mass relation}

The top right panel of Fig.~\ref{fig:convergence_hybrid} compares simulated galaxy sizes to observations. Both m7 and m6 are generally in good agreement with the data at most stellar masses (within $1\sigma$, where $\sigma$ is the standard error of the median, given by the error bars), while m5 shows slightly too large sizes (at $M_*\approx10^{10}$ M$_\odot$), with some of the disagreement stemming from the L025m5h box not being volume-converged. The m7 model shows signs of disagreement with the data at $M_*=10^{9}$ M$_\odot$, but the sizes are affected by resolution effects at these masses. The m6 model slightly underestimates the observed sizes at the high-mass $M_*>10^{11}$ M$_\odot$ end (at a level of more than $2\sigma$), by $\approx0.1$ dex. However, the size-mass relation is very steep at these masses. A shift in stellar masses by $0.2$ dex could alleviate this tension (by moving the entire relation horizontally), and such a shift would be similar in magnitude to the systematic uncertainties in observational determinations of stellar masses. 

\subsubsection{Black hole mass$-$stellar mass relation}

The bottom left panel of Fig.~\ref{fig:convergence_hybrid} compares BH masses with observations. The BH masses are similar at all three resolutions for $M_*>10^{10}$ M$_\odot$, and in close agreement (within the $1\sigma$ uncertainty) with the data from \cite{Graham2023}. The simulated $M_\mathrm{BH}-M_*$ relation has the same slope as the relation from \cite{Graham2023}, which is much steeper than the power law fit from \cite{McConnell2013}. This is because that fit is more appropriate for spheroid-dominated galaxies, which have higher BH masses at fixed stellar mass. Below $M_*\approx10^{10}$ M$_\odot$, the simulation predictions begin to diverge. At low stellar masses, the BH mass tends smoothly to the value of the seed mass, $M_\mr{BH,seed}$. These values correspond to $5\times10^5$ M$_\odot$, $\approx5.63\times10^4$ M$_\odot$ and $4\times10^4$ M$_\odot$ at m7, m6 and m5, respectively.

\subsubsection{Bolometric AGN luminosity function}

In the bottom right panel of Fig.~\ref{fig:convergence_hybrid}, we compare the AGN bolometric LF to the observational data. We find that the results are well converged above $L_\mr{AGN,bol}=10^{42}$ erg s$^{-1}$, the value below which the observational comparison is not reliable (\S~\ref{sec:calibration_data}). At values below this, the AGN LF for m6 and m5 is significantly higher than m7, and closer to the observed data. At $L_\mr{AGN,bol}>10^{43}$ erg s$^{-1}$, the predicted AGN LF is $0.3-0.7$ dex too high compared with the data. This disagreement implies that our adopted accretion efficiencies in the thick disc state ($f_\mr{Edd,d}<0.01$) are too low. The disagreement is caused by two effects. First, for practical reasons, during calibration we almost always plotted our predictions at $z=0$ (we often did not output snapshots at redshifts other than $z=0$ and $z=2$ during calibration), as for most observational comparisons, the data is at $z\approx0$. However, the data for the AGN bolometric LF is at a median redshift $z\approx0.2$. Mean AGN luminosities grow by nearly $0.3$ dex between these redshifts. Secondly, we mostly used L025 or L050 boxes during calibration, and the AGN LF is not fully converged in terms of volume for these boxes. Despite these mild disagreements with the data, our AGN LF is still in better agreement with it than had we not used sub-unity accretion efficiencies (Fig.~\ref{fig:efficiency_variations}), and this was our main motivation to calibrate on the AGN LF. 

\subsection{The role of AGN feedback in low-mass galaxies}

As shown in \S~\ref{sec:param_variations} (see also Appendix \ref{app:SHMR}), the hybrid AGN feedback simulations feature significant AGN feedback effects in low-mass galaxies. There is no strong observational evidence of AGN feedback operation in low-mass galaxies (e.g.~\citealt{Fabian2012}), but some evidence in selected low-mass galaxies has begun to emerge in recent years (e.g.~\citealt{Penny2018}, \citealt{Carr2025}, \citealt{Salehirad2025}). Around half of the injected AGN energy in the hybrid AGN feedback model is in the form of jets (see \S~\ref{sec:feedback}), which could explain the increased role of AGN feedback in the hybrid model in low-mass galaxies compared to the purely thermal one. This interpretation could be consistent with observations that find evidence of jets in low-mass galaxies (\citealt{Mezcua2019}). Alternatively, the stronger effects of AGN feedback in the hybrid simulations could be due to the use of a higher BH seed mass, which results in overall higher BH masses at fixed halo mass, and therefore stronger feedback. This is consistent with previous simulation studies that find the mass of the BH to be the primary driver of quenching at lower BH masses (e.g.~\citealt{Sharma2020}, \citealt{Koudmani2021}, \citealt{Wellons2023}, \citealt{ArjonaGalvez2024}). \cite{BoothSchaye2013} and \cite{Koudmani2022} have shown that a reduced efficiency of SN feedback can also make AGN feedback stronger by allowing more BH growth; this is also consistent with our findings, since the hybrid AGN feedback simulations use lower SN feedback heating temperatures by necessity.

\section{Black hole spin$-$mass relation}
\label{sec:BHspin}

\begin{figure*}
\includegraphics[width=1\textwidth, trim = 0 10 0 0]{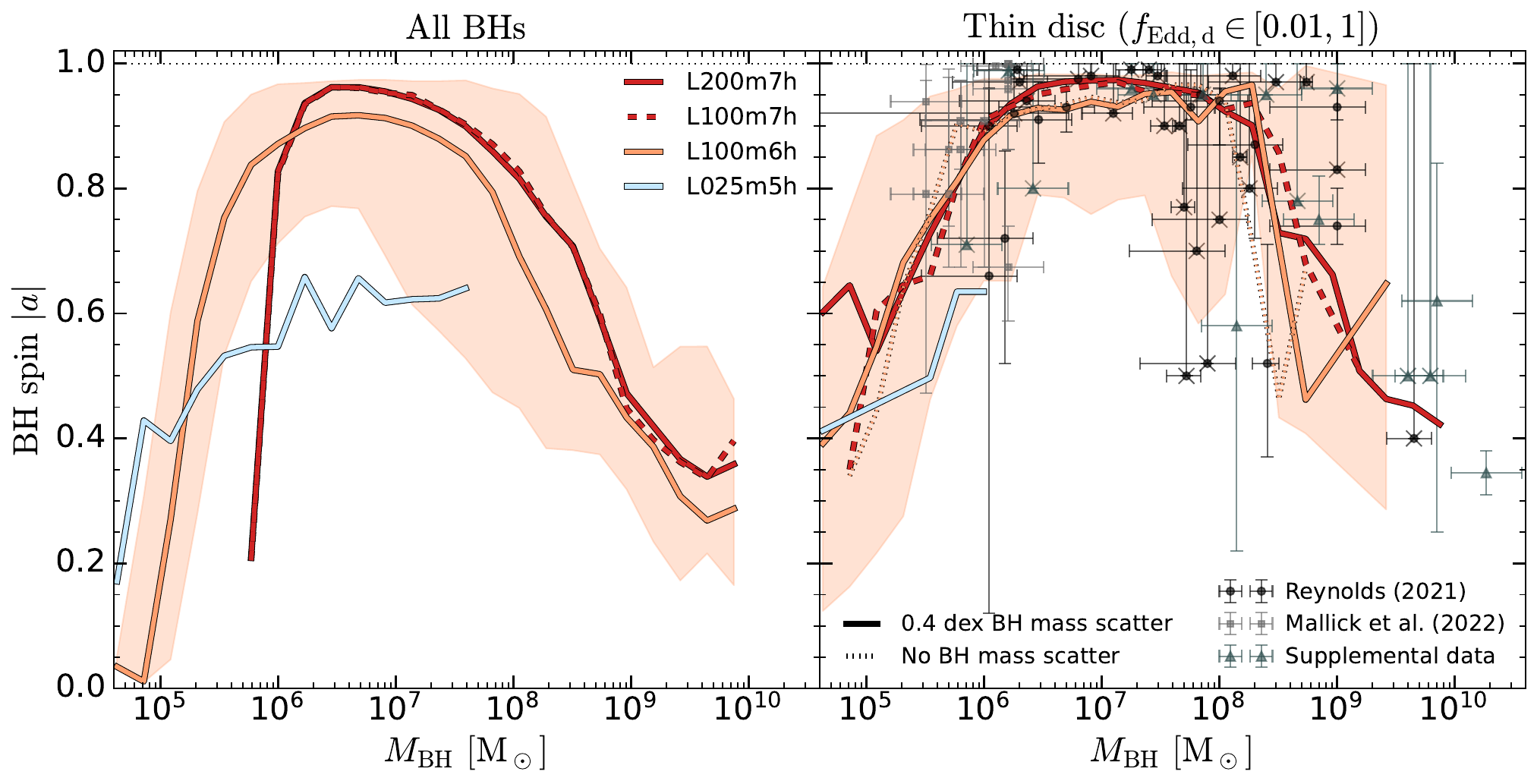}
\caption{The BH spin$-$mass relation for the L200m7h (red solid), L100m7h (red dashed), L100m6h (orange solid) and L025m5h (light blue solid) simulations. The orange shaded region shows the 16-84 percentile range for L100m6h. The left panel shows the relation for all BHs, while the right panel shows it only for BHs in the thin disc state, where we have also added 0.4 dex log-normal scatter to BH masses. The exception in this panel is the dotted line, which shows the case of no added scatter for L100m6h. The spins of BHs in the thin disc state are compared with observations of moderately-accreting BHs (\protect\citealt{Reynolds2021} and \protect\citealt{Mallick2022}, as well as our compilation of other studies; see the text for further details). For these observed BHs, many measured spins are lower limits. We mark these measurements with crosses, show no lower error bar and extend the upper one to $\vert a \vert=1$.}
\label{fig:BHspins}
\end{figure*}%

In this work we largely refrain from providing predictions on galaxy and BH properties from simulations run with the hybrid AGN feedback model. Some of these are presented in \cite{Schaye2025} and others will be explored in future studies. However, we present in Fig.~\ref{fig:BHspins} the predicted BH spin$-$BH mass relation at $z=0$ (note that this shows the spin magnitude) for the L200m7h, L100m7h, L100m6h and L025m5h simulations, which use the final calibrated hybrid AGN feedback model presented in \S~\ref{sec:calibration_first_results}. BH spin is a new and important BH property introduced in the hybrid AGN model, and Fig.~\ref{fig:BHspins} illustrates the typical magnitudes of BH spin for BHs of different masses. 

From the left panel, which shows the relation for all BHs, we see that median BH spins are generally high, $\vert a \vert>0.8$, for intermediate-mass BHs ($M_\mr{BH}\in[10^6,10^8]$ M$_\odot$) at m7 and m6, with peak median values of $\vert a \vert=0.96$ and $\vert a \vert=0.92$, reached at a BH mass of $M_\mr{BH}=10^{6.5}$ M$_\odot$ at both resolutions. For BH masses outside the range $M_\mr{BH}\in[10^6,10^8]$ M$_\odot$, the magnitude of BH spin decreases towards even lower or even higher BH masses on either side of this range. At m5, the BH spin grows monotonically up to $M_\mathrm{BH}\approx10^8$ M$_\odot$, the highest BH mass at which we can constrain BH spin using a L025 volume, where it reaches a value $\vert a \vert\approx0.6$, which is significantly lower than at m7 and m6. For a discussion of the reasons behind this lack of convergence, see below.

The shape of the BH spin$-$mass relation is a result of a balance between gas accretion, BH-BH mergers and jet-induced spindown. At intermediate masses ($M_\mr{BH}\in[10^6,10^8]$ M$_\odot$), BHs are most likely to be in the thin disc state (see Fig.~\ref{fig:accretion_regimes}, which shows that BHs with $M_\mr{BH}=10^{7}$ M$_\odot$ are the ones most likely to be in the thin disc state, and this is close to the BH mass at which the median spin magnitude is highest). Spinup in the thin disc state is very efficient due to high gas accretion rates, very quick spin redirections, and negligible jet-induced spindown. Mergers tend to result in BH spins significantly smaller than $\vert a \vert=1$ (e.g.~$\vert a\vert\approx0.7$, \citealt{Barausse2012}), which can explain lower BH spins for massive BHs ($M_\mr{BH}>10^{8}$ M$_\odot$). These BHs are hosted by quenched, gas-poor galaxies whose growth is mainly driven by mergers. Jet-induced spindown due to prolonged accretion in the thick disc state may also play a role in driving the median BH spin to values as low $\vert a \vert\approx0.3$ at $M_\mr{BH}>10^{9.5}$ M$_\odot$, which is hard to achieve through mergers alone. BH seeds in our simulations mostly grow through mergers. However, since we seed BHs with $\vert a \vert\approx0$, these initial mergers are unlikely to lead to significant spinup. The increase in BH spin with BH mass at the low-mass end ($M_\mr{BH}<10^{6}$ M$_\odot$), visible in Fig.~\ref{fig:BHspins}, is thus likely to be due to gas accretion. Since the BH seed mass is a calibration parameter that changes with resolution, our prediction for the BH spin$-$mass relation is likely resolution-dependent at these low BH masses, as is visible in the figure. However, the large difference between the seed mass at m6 and m7 is mostly due to the two resolutions using a different halo threshold mass for BH seeding. 

In the right panel of Fig.~\ref{fig:BHspins} we compare the predicted BH spin magnitudes with observational measurements of BH spin. These measurements rely on X-ray reflection spectroscopy and thermal continuum fitting (see review by \citealt{Reynolds2021}). They should be treated with care, since they rely on uncertain models (e.g.~\citealt{Hagen2023}). In addition to the compilation of BH spin measurements provided by \cite{Reynolds2021}, which span a wide range of BH masses, we also include measurements for low-mass BHs ($M_\mr{BH}\approx10^6$ M$_\odot$) from \cite{Mallick2022}, who used X-ray reflection spectroscopy. We supplement these datasets with our own compilation of individual measurements from the following studies, using largely but not exclusively X-ray reflection spectroscopy: \cite{Walton2013}, \cite{Valtonen2016}, \cite{Vasudevan2016}, \cite{EHT2019}, \cite{Jiang2019}, \cite{Laine2020}, \cite{Bambi2021}, \cite{Ghosh2021}, \cite{Walton2021}, and \cite{SiskReynes2022}. Many of these observational measurements are lower limits rather than true measurements of BH spin, so we do not bin the data.

All of the above spin measurements are for BHs accreting at moderate rates (\citealt{Reynolds2021}). For this reason, we restrict our comparison to BHs in the thin disc state ($0.01<f_\mr{Edd,d}<1$), where the AGN spectrum can be distinguished from that of the host galaxy. The inferred BH masses in the observed sample have a median uncertainty of $\approx0.4$ dex. We therefore add $0.4$ dex log-normal scatter to our simulated BH masses to replicate Eddington bias (for reference, we also show the case with no scatter added using a dotted line for L100m6h$-$the difference between the two cases is small). Comparing the simulation lines in the right panel with those in the left panel of Fig.~\ref{fig:BHspins}, the latter of which are for all BHs, we see that applying this selection results in higher BH spins at fixed BH mass, at least at intermediate BH masses ($M_\mr{BH}\in[10^6,10^8]$ M$_\odot$). This is indicative of those BHs spinning up more efficiently, which is expected for the thin disc state.

Our predictions for both m7 and m6 resolution are broadly consistent with the observed trend of the BH spin$-$mass relation, while m5 seems to predict too low BH spins for intermediate-mass BHs.. At a mass of $M_\mr{BH}\approx10^7$ M$_\odot$, where the relation peaks, the predicted spin magnitudes are in agreement with the observations at m7 and m6, all of which are close to $\vert a \vert=1$. At lower BH masses, our predictions are also close the observations. The scatter is reproduced there, while the low-spin percentile range of the scatter is possibly too large at $M_\mr{BH}\approx10^7$ M$_\odot$. We predict massive BHs ($M_\mr{BH}>10^{8.5}$ M$_\odot$) to have slightly lower BH spin than the dozen or so observed ones in this mass range, at least at m7. If this is not a result of observational bias, and is instead true disagreement, it could be indicative of overly strong jet spindown of massive BHs, since BHs cannot otherwise obtain low spins. However, at m6, we predict BH spin to be somewhat higher at the high-mass end ($M_\mr{BH}>10^8$ M$_\odot$) than at m7, at least up to where predictions are available at m6 from the L100m6h simulation ($M_\mr{BH}\approx10^{8.5}$ M$_\odot$). Larger volumes are needed to probe higher masses at m6, in order to determine whether the hybrid AGN feedback model predicts correct BH spins in massive galaxies. Finally, we do not compare in detail with other simulations, but our predictions for the relation are broadly similar to those of other recent cosmological hydrodynamical simulations that include BH spin modeling (\citealt{Bustamante2019}, \citealt{Sala2024}, \citealt{Beckmann2025}). 

We stress that BH spins were not used during the calibration of the hybrid AGN model, and are thus true predictions. However, during calibration we found that the prediction for BH spins is intimately tied to the radiative activity of BHs. In particular, variations of the accretion efficiency of the thick disc, which strongly influence the AGN LF (\S~\ref{sec:efficiency_variations}), also had a large impact on the predicted BH spin$-$mass relation. Larger values of the accretion efficiency result in lower BH spins. This is a consequence of BHs more frequently being in the thick disc rather than in the thin disc state near $z=0$, which results in less efficient spinup. Interestingly, we found that models that do not match the AGN LF also do not match the observed BH spin$-$mass relation. 

The good agreement of predicted BH spins at m7 and m6 resolutions, as well as that of \cite{Bustamante2019} and \cite{Sala2024}, all of which are relatively low-resolution ($m_\mr{g}>10^6$ M$_\odot$), is intriguing, and somewhat unexpected, especially given the too low BH spins at m5 resolution. This can be explained as follows. Efficient spinup of BHs requires: 1) that a sufficient fraction of the mass accretion should occur in the thin disc state, where spinup is easy, and 2) good alignment of BH spin with the surrounding gas on resolved scales, i.e.~the BH spin, $a$, needs to be positive often enough. We claim that the first condition is fulfilled by matching the observed AGN LF. The second condition is easy to fulfill at low resolutions. Low-resolution simulations, even with a multiphase ISM, do not resolve the turbulent nature of the gas as well as high-resolution simulations. Furthermore, BH kernels are larger at low resolution, encompassing a larger fraction of the galaxy and thus probing better the average angular momentum of the gas, which is well-defined for disc galaxies. This results in the angular momentum of the gas accreting onto a subgrid BH being aligned with the large-scale angular momentum of gas in the galaxy disc than it might be at higher resolution. These considerations imply that spinup should be less efficient at higher resolutions, which we can indeed see from Fig.~\ref{fig:BHspins}. We interpret this as a result of gas turbulence being resolved better and BH kernels being smaller.

These considerations show that predicting BH spins is a bigger challenge in high-resolution simulations than low-resolution ones, in a model that assumes that the subgrid accretion disc is aligned with the direction of the angular momentum in the BH kernel on resolved scales of the simulation. This strong assumption works in low-resolution simulations, where turbulence is not resolved well and where BH kernels encompass a substantial fraction of the galaxy. The assumption breaks down at higher resolutions, where the subgrid accretion disc is not resolved, nor angular momentum flows onto it, while the gas turbulence is resolved and the BH kernels are small. In high resolution simulations, the angular momentum of gas accreting onto the BH is stochastic and often misaligned with respect to a galaxy disc (\citealt{Levine2010}, \citealt{Hopkins2012}, \citealt{AnglesAlcazar2017}). This is also consistent with observations that show nuclear molecular gas discs that are misaligned with respect to their host galaxies (\citealt{Greene2013}, \citealt{Ruffa2020}). On the other hand, observations of radio jets indicate broad alignment with galaxy discs (\citealt{Kaviraj2015}, \citealt{Najar2019}, \citealt{Zheng2024}), which may imply alignment in a time-averaged sense.

\section{AGN feedback in the hybrid simulations}
\label{sec:feedback}

In this section we analyze in more detail how AGN feedback operates in practice in the COLIBRE simulations. We focus on jet feedback and we only consider basic properties; more detailed analyses will be presented in future studies. In \S~\ref{sec:large_scale_outflows} we highlight the qualitative differences between the thermal and hybrid AGN feedback simulations using visualizations of the IGM and large-scale AGN activity. We also quantify the far-reaching effects of jets. In \S~\ref{sec:feedback_energetics} we compare the feedback energies in the hybrid and thermal AGN feedback simulations, focusing on how much energy is injected as a function of cosmic time. For the hybrid AGN feedback simulations, we discuss the relative importance of the two feedback channels and the three accretion disc states.

\subsection{Large-scale outflows}
\label{sec:large_scale_outflows}

\begin{figure*}
\includegraphics[width=1\textwidth, trim = 0 10 0 0]{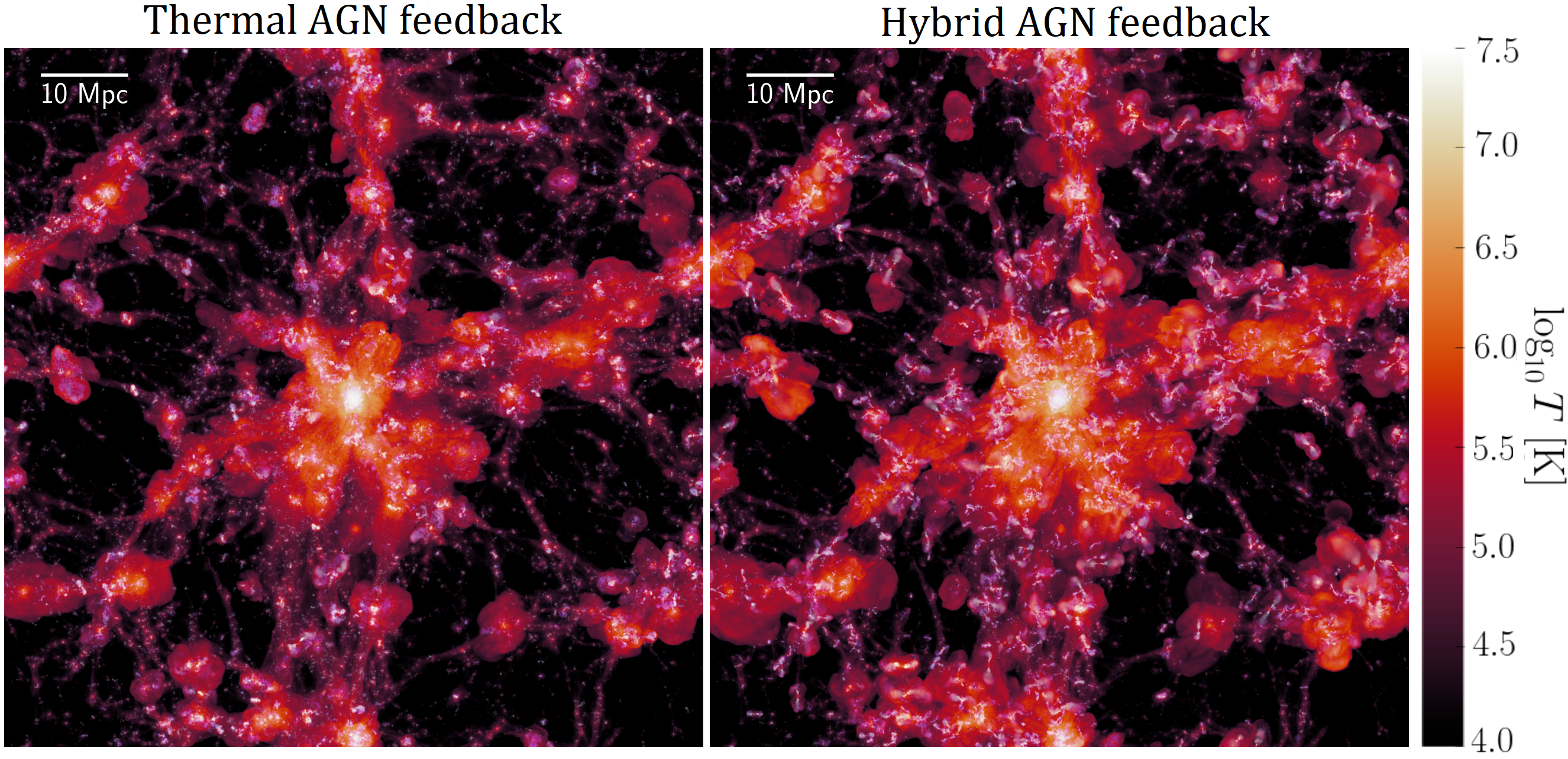}
\caption{Temperature in slices of $30$ cMpc at $z=0.2$ through the L100m6 (left) and L100m6h (right) simulations, using thermal and hybrid AGN feedback, respectively. The images are in proper coordinates. Red to yellow colours encode the mass-weighted projected temperature, while blue to white colours highlight particles affected by thermal or jet feedback up to 5 Gyr in lookback time. For these particles, the surface density is used to determine the opacity, while brighter (white) colours indicate more recent AGN events.}
\label{fig:IGM}
\end{figure*}%

In Fig.~\ref{fig:IGM} we show the temperature slices through the L100m6 and L100m6h simulations (run with the thermal and hybrid AGN feedback models, respectively) at $z=0.2$. We also overlay particles directly heated or kicked by AGN feedback (see caption for details). AGN feedback drives outflows and heats the IGM through shocks. These are frequently bipolar even in the thermal AGN feedback case (which is nominally isotropic, but bipolar outflows tend to form as gas escapes galaxies; \citealt{Mitchell2020}), but more often so in the hybrid AGN feedback case, due to the explicit two-sided jet activity. Individual jets may be discerned in the right panels as well as the bow shocks ahead of the jets. The jets appear to push out gas to larger radii compared to thermal feedback. This is consistent with large-scale jet activity in the SIMBA simulations (\citealt{Dave2019}, \citealt{Borrow2020}).


\begin{figure}
\includegraphics[width=1\columnwidth, trim = 10 10 0 0]{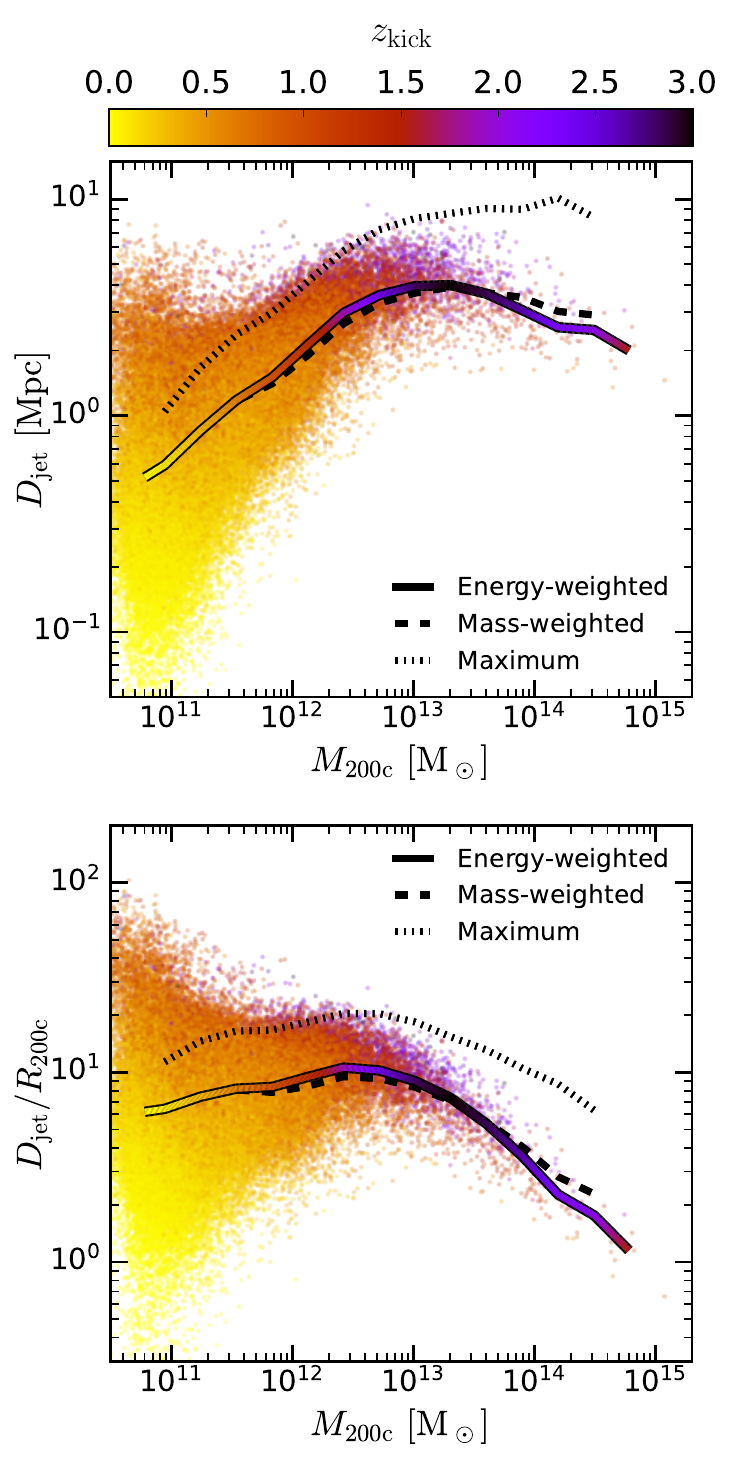}
\caption{The distances out to which a given halo can affect its surroundings through jet feedback in the L200m7h simulation. We show at $z=0$ and for central haloes, as a function of halo mass, three distance scales (top: raw scale in physical units, bottom: the same, normalized by the virial radius). The solid line shows the median distance to the centre of energy of jet-kicked particles, where the weight for every particle is $m_\mr{g}v_\mr{jet}^2/2$. The coloured dots show these median centres of energy for individual haloes. Colours represent the mean redshift at kick time, weighted in the same way as the distance. Dashed and dotted lines show median distances computed using the centre of mass and the farthest kicked jet particle, respectively. All distances are computed with respect to the halo centre (the most bound particle), and accounting for periodic boundary conditions. The BH used in the analysis is the most massive bound BH located in a 50 kpc exclusive aperture.}
\label{fig:jet_distances}
\end{figure}%

We now quantify some of our discussion from above. When kicking particles into jets, we tag them by the ID of the BH that kicked those particles. We do not do this for thermal feedback, so we restrict the following analysis to jet feedback only. In addition to recording the ID of the BH particle, we record how much energy was given to the gas particles, and the time when the kick occurred. This can be used to track individual jet particles and their positions at any given time. Using this information, for every central halo in L200m7h at $z=0$, we find all particles that were kicked as part of jet feedback by the current most massive BH (in a 50 kpc 3D aperture centered on the most bound particle) hosted by that halo. We then compute the distance to both the centre of mass and centre of energy of these particles, where for the latter, the weight is $m_\mr{g}v_\mr{jet}^2$, $v_\mr{jet}$ being the target jet velocity at kick time. We compute mean redshifts at kick time for the centre of energy case. For every BH, we also find the farthest kicked particle.

Fig.~\ref{fig:jet_distances} shows the results of this analysis, as a function of halo mass, defining the haloes here as objects whose mean density is 200 times larger than the critical density at $z=0$. In the top panel, we show the raw distances, while the bottom panel shows distances normalized by the virial radius. The median jet distances increase with halo mass from $\approx0.5$ Mpc ($\approx6$ virial radii) at $M_\mathrm{200,c}=10^{11}$ M$_\odot$ to $\approx4$ Mpc ($\approx10$ virial radii) at $M_\mathrm{200,c}\approx10^{13}$ M$_\odot$. They then drop off, both in raw terms and when normalized by virial radius. This is true for both the centre of mass and centre of energy cases. The median jet distance for the most massive haloes in L200m7h, with $M_\mathrm{200,c}\approx10^{15}$ M$_\odot$, is 2 Mpc ($\approx1$ virial radius). 

This decrease in the median jet distance is likely caused by two effects. First, AGN activity is most efficient at expelling gas around the $M_\mathrm{200,c}\approx10^{13}$ M$_\odot$ mass scale (see the scaling of gas fraction with halo mass; Davies et al.~in prep.). For these haloes, jet particles are therefore most likely to end up at many virial radii away. Second, more massive haloes have recurrent jet activity at lower redshifts, which can be seen in Fig.~\ref{fig:jet_distances} in terms of the mean kick redshift for the most massive haloes ($z\approx1.5$ for $M_\mathrm{200,c}\approx10^{15}$ M$_\odot$ compared to $z\approx3$ for $M_\mathrm{200,c}\approx10^{13}$ M$_\odot$). These jet particles kicked at lower redshifts are less likely to be located as far away as the ones kicked at high redshift.

In Fig.~\ref{fig:jet_distances} we also show the median distance to the farthest kicked jet particle. This saturates to $10$ Mpc, which corresponds to $\approx20$ virial radii for $M_\mathrm{200,c}=10^{13}$ M$_\odot$ and $\approx5$ virial radii for $M_\mathrm{200,c}=10^{15}$ M$_\odot$. While not shown here, in the case of individual haloes, the farthest jet particles can be located up to $20$ Mpc away (largely at $M_\mathrm{200,c}\leq10^{13}$), which is, incidentally, similar to SIMBA (\citealt{Borrow2020}, \citealt{Gebhardt2024}), but less extreme than IllustrisTNG (\citealt{Ayromlou2023}). Note that even the centre of energy can be as far away as $\approx100$ virial radii for low-mass haloes ($M_\mathrm{200,c}\approx10^{11}$ M$_\odot$).

We warn the reader that the above discussion does not necessarily pertain to the actual size of currently active jets, nor even jet remnants. This is for the following two reasons. First, BHs may kick particles that exit haloes hosting them and then effectively detach from the halo's motion. Even if we are not considering satellites here, the haloes may move considerably between the time at kick and $z=0$, and thus the farthest jet particles may be located far away simply due to this motion. Second, much of the jet feedback occurs at high redshift, especially for high-mass BHs, hosted by high-mass haloes, that do jet feedback in the thin disc state (see the next subsection). The peak redshift of this activity is at $z\approx3$. The fact that BHs hosted by more massive haloes kick more particles at higher redshifts, as seen in Fig.~\ref{fig:jet_distances}, may be an effect of jet feedback being stronger in the thin disc state, but it may also simply be a consequence of more massive haloes having formed earlier. In either case, these jet particles have departed their haloes at relatively early redshifts. Such particles join the Hubble flow and are therefore located even farther from their host haloes than they would otherwise be (compared to a simple scenario of ballistic ejection from a self-gravitating halo, without cosmic expansion).

\subsection{Feedback energetics}
\label{sec:feedback_energetics}

Here we focus on injected AGN energies, comparing the thermal AGN feedback simulations to those using the hybrid model, as well as comparing different channels and modes in the hybrid simulations. We also compare with injected SN energies. We study the energetics both globally as a function of redshift, and in terms of cumulative energies as a function of BH mass at $z=0$. We use the L200m7 and L200m7h simulations since they are converged (volume-wise) in terms of these energy measures, exhibit the smallest variation with time between snapshots, and are the largest simulations using the hybrid model that we have available, in terms of volume. Some of the quantitative analysis in the following discussion depends slightly on resolution, but the qualitative trends that we find also apply for the L100m6 simulations, which we do not show here.

\begin{figure*}
\includegraphics[width=1\textwidth, trim = 0 10 0 0]{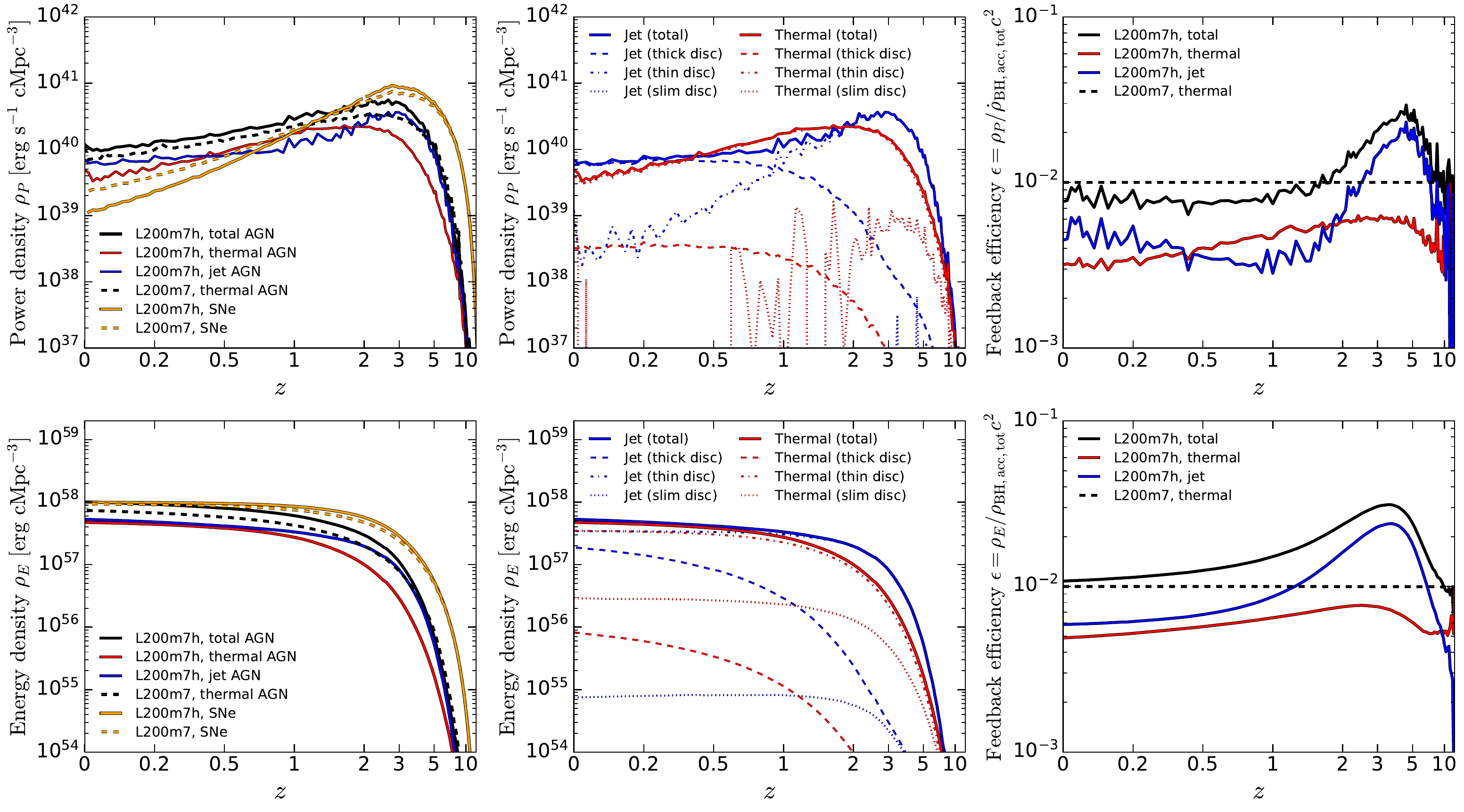}
\caption{Cosmic evolution of feedback energetics in the L200m7 (dashed) and L200m7h (solid) simulations, using thermal and hybrid AGN feedback, respectively. The top row shows the feedback power densities, while the bottom row shows the cumulative injected feedback energy densities. Yellow lines show energies from SNe, while black lines show total AGN energies. The thin red and blue lines show the thermal and kinetic jet components of AGN feedback in the hybrid simulation. The middle panels show the thermal and jet components in the hybrid AGN feedback simulation split onto individual contributions from the three accretion disc states, as given by the legend. In the right panels, the top (bottom) shows the effective feedback efficiency, computed as the total power (energy) density divided by the total rest mass-energy accretion rate density (accreted rest mass-energy density). }
\label{fig:power_densities}
\end{figure*}%

The left column of Fig.~\ref{fig:power_densities} shows the energetics in terms of the cosmic evolution of feedback power densities (top row) and cumulative injected feedback energy densities (bottom row). In both the thermal and hybrid AGN feedback simulations, SN feedback dominates above $z\approx1.5$, after which AGN take over (see top left). At $z=0$, AGN are injecting $3$ times as much energy as SN in the thermal AGN feedback simulation, and an order of magnitude more in the hybrid AGN feedback simulation. By $z=0$ the total injected AGN energy (bottom left) is nearly equal to the total injected SN energy in the thermal AGN feedback simulation, and equal for the hybrid AGN case. 

From the left panels we can also see that the AGN are injecting similar amounts of energy in the two AGN models up to $z\approx5$, with the thermal model showing slightly more AGN energy at these early redshifts. This is related to the thermal AGN feedback simulations exhibiting more efficient BH growth at the earliest redshifts (Huško et al.~in prep.). AGN start injecting more energy (by $\approx50$ per cent) in the hybrid model compared to the thermal one below $z=5$ and overall around $25$ per cent more energy is injected in the hybrid model by $z=0$.

The left panels also show the hybrid AGN power and energy density components due to thermal isotropic and kinetic jet feedback. Jet feedback is dominant at low redshift $z\leq0.5$. This is in agreement with the canonical picture of radiatively-driven `quasar' feedback dominating at high redshifts, and mechanical feedback dominating at low redshifts. Thermal feedback is, however, non-negligible even at $z=0$. Finally, jet feedback is again more important than thermal feedback at $z>2$. By $z=1$, both feedback modes have injected nearly equal amounts of energy, and this remains the case down to $z=0$. Our findings are in good agreement with the observationally motivated empirical model of \cite{Mocz2013}.

In the middle column of Fig.~\ref{fig:power_densities}, we show the individual contributions to the two feedback modes (thermal and jet) from the three different accretion disc states in the hybrid AGN simulations. In general, three feedback channels are important: 1) thick disc jets (from BHs with $f_\mr{Edd,d}<0.01$) at low redshift ($z<1$), 2) thin disc thermal feedback (from BHs with $0.01<f_\mr{Edd,d}<1$) at all redshifts, and 3) thin disc jets at $z>1$. The dominance of jets over thermal feedback at low ($z<0.5$) and high redshifts ($z>2$) is thus due to different types of discs. By $z=0$, thin disc jets have injected a very similar amount of energy as thin disc thermal feedback. The total thick disc jet energy injected by $z=0$ is a factor of $\approx2$ lower than the thin disc jet and thermal components. Winds from the thick and slim discs ($f_\mr{Edd,d}>1$ for the latter), implemented using thermal feedback in the hybrid model, are generally negligible in terms of global energetics with the exception of slim disc winds at $z>5$, when they are energetically comparable to thin disc jets and thermal feedback. Jets from the slim disc are always negligible, although we suspect that this may not be the case in a more realistic model\footnote{Low jet efficiencies are a consequence of our use of constant and low accretion efficiencies ($\epsilon_\mr{acc}=0.01$) in the slim disc. Given the fact that the disc-scale accretion rate cannot exceed $100$ times the Eddington rate, the horizon-scale Eddington ratio in the slim disc cannot be above the Eddington rate. This means that the dimensionless magnetic flux is always low in the slim disc, and thus also the jet efficiency (see Eqns.~\ref{eq:phi_a} and \ref{eq:epsilon_jet}). A more realistic choice for $\epsilon_\mr{acc}$, e.g.~one where it scales with $f_\mr{Edd,d}$ but such that the horizon-scale accretion rate does not drop below Eddington, would ensure that jets are at least as important in the slim disc state as they are in the thin disc state.}.

In the right column of Fig.~\ref{fig:power_densities}, we show the effective feedback efficiencies in the same simulations. These are computed by dividing the power (energy) densities by the rest mass-energy accretion rate densities (rest accreted mass-energy densities) onto all BHs. By construction, the thermal run has a total efficiency of $\epsilon=\epsilon_\mr{f}\epsilon_\mr{rad}=0.01$. Note that jet efficiencies here do not have the same meaning as the `true' jet efficiencies defined in \S~\ref{sec:hybrid_model}, since we are here dividing the energies that may be injected in any accretion state with masses that may have been accreted in other accretion states. Most BH accretion occurs in our model at moderate Eddington ratios, $f_\mathrm{Edd,d}=0.01-0.3$, while most jet energy is launched at Eddington ratios below (in the thick disc state) or above those values, with little growth occurring at such Eddington ratios, largely due to low values of the accretion efficiency.

The total instantaneous efficiency in the hybrid AGN feedback simulation is slightly lower (by $\approx0.1$ dex) than in the thermal AGN feedbck simulations at $z<1$, despite it having a larger power density at these redshifts. The total instantaneous efficiency is larger in the hybrid AGN feedback simulations at $z>2$, and reaches a maximum of $\approx3$ per cent at $z=5$. The jet channel has a mean instantaneous efficiency of $0.3-0.6$ per cent at $z<2$ (increasing to $\approx2$ per cent at $z=5$, likely due to thin disc jets becoming important). The thermal instantaneous efficiency increases monotonically with redshift from $0.3$ per cent at $z=0$ to $0.6$ per cent at $z=3$. This increase with redshift is probably due to the median BH spin also increasing with redshift (Huško et al.~in prep.), which also increases the radiative efficiencies. The cumulative total efficiency (in the bottom panel) is almost always higher in the hybrid AGN feedback model compared to the thermal one, with the exception of very high redshifts ($z>8$).

\begin{figure*}
\includegraphics[width=1\textwidth, trim = 0 10 0 0]{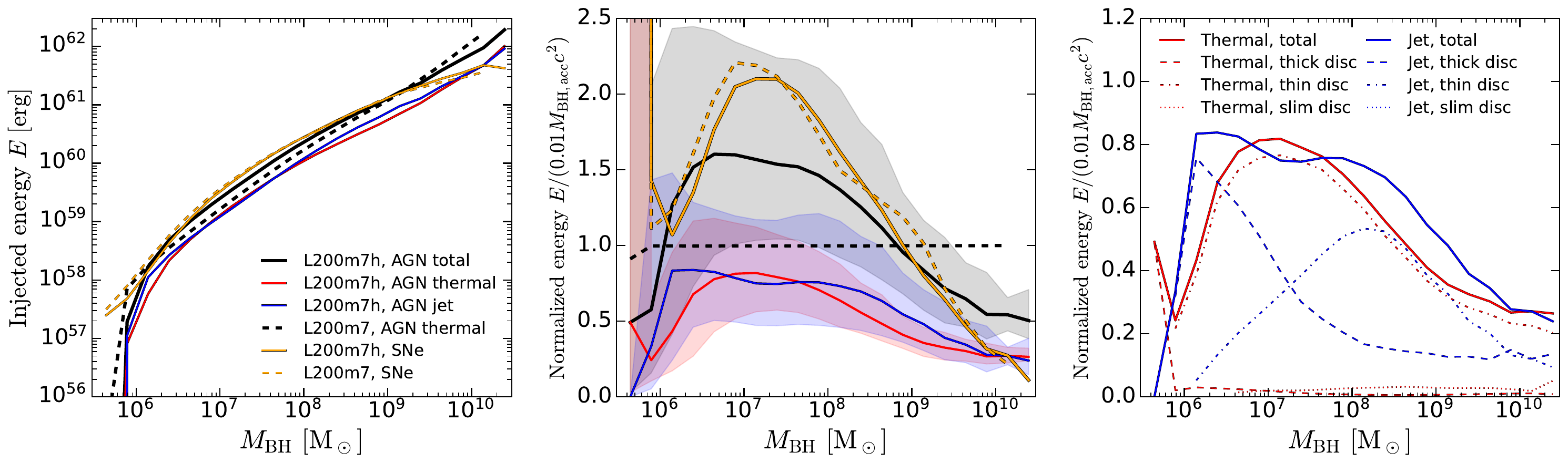}
\caption{\textit{Left:} BH mass dependence of the cumulative injected AGN and SN feedback energies at $z=0$ in the L200m7 (thermal) and L200m7h (hybrid) AGN feedback simulations. \textit{Middle:} the cumulative injected feedback energies in units of $0.01M_\mr{BH,acc}c^2$. \textit{Right:} as the middle panel but splitting the jet and thermal contributions in the hybrid AGN feedback simulation onto contributions from different accretion disc states.}
\label{fig:energies_BH_mass}
\end{figure*}%

The left panel Fig.~\ref{fig:energies_BH_mass} shows the BH mass dependence of the cumulative injected AGN feedback energies per BH by $z=0$ for the L200m7 thermal and hybrid AGN feedback simulations. In the middle and right panels we show the same quantity but divided by $0.01M_\mr{BH,acc}c^2$ in order to reduce the dynamic range, and since we expect the total AGN energy to roughly scale with the accreted BH mass. We also show energies injected by SNe of the host galaxies of these BHs. We find that the AGN-injected energies are comparable to SN-injected ones at all BH masses, and for both the thermal and hybrid AGN feedback models. This suggests that AGN feedback is relevant in both low-mass and high-mass galaxies, which is consistent with the effects we see when varying feedback parameters (\S~\ref{sec:param_variations}), since these variations impact galaxies of all masses. This is true for parameter variations that affect either of the two AGN feedback modes.

Cumulative AGN-injected energies clearly overtake SN-injected ones only above $M_\mr{BH}\approx10^9$ M$_\odot$ (see middle panel of Fig.~\ref{fig:energies_BH_mass}). The reason that this does not occur at lower mass, e.g.~at $M_\mr{BH}\approx10^7-10^8$ M$_\odot$, where AGN-induced quenching begins to occur, is due to the fact that some of the SN energy does not do useful work since it is injected at low heating temperatures ($\Delta T<10^{7.5}$ K) and is thus more prone to cooling losses. Furthermore, a clear transition at $M_\mr{BH}\approx10^7-10^8$ M$_\odot$ may be visible when comparing instantaneous feedback energies, but in Fig.~\ref{fig:energies_BH_mass} we are comparing cumulative energies injected over the entire history of a given BH (galaxy).

The most massive BHs appear to be injecting more feedback energy in the thermal AGN simulation than in the hybrid one, or at least to have injected it when integrated over time. In the hybrid AGN simulation, approximately half of the cumulative injected energy is in thermal and half in kinetic jet form by $z=0$, regardless of BH mass. As can be seen in the right panel, the thermal component is almost entirely due to thin disc feedback, while for the jet channel, the thick disc contribution is dominant below $M_\mr{BH}\approx10^7$ M$_\odot$ and the thin disc one above that mass, up to $M_\mr{BH}\approx10^{10}$ M$_\odot$. For the most massive BHs, thin and thick disc jets are approximately equally important in terms of total energy injected by $z=0$, although this energy was injected at very different times for the two feedback channels.

\section{Conclusions}
\label{sec:conclusions}

We introduce a new, hybrid AGN feedback model for cosmological, hydrodynamical galaxy formation simulations. The model includes two feedback modes: thermal isotropic and kinetic jets. The thermal mode represents the effects of AGN disc winds shocked on subgrid scales and/or radiative heating on subgrid or larger scales. The jet mode represents the effects of relativistic jets driven by the \cite{Blandford1977} mechanism. The model includes three accretion disc states (\S~\ref{sec:accretion_states}): the thick, thin and slim disc states. BHs occupy these states depending on their Eddington ratio, with the three states populated with BHs that have low ($f_\mr{Edd,d}<0.01$), moderate ($0.01<f_\mr{Edd,d}<1$) and super-Eddington ($f_\mr{Edd,d}>1$) accretion rates, respectively. Here, $f_\mr{Edd,d}$ is the Eddington ratio defined using the accretion rate onto a subgrid accretion disc. Both feedback modes are used in all three accretion states, but with different efficiencies. We also include indirect effects of mass loss due to accretion disc winds in the thick and slim disc states (\S~\ref{sec:accretion_eff}), through a new parameter: the accretion efficiency, $\epsilon_\mr{acc}$.

The model includes a subgrid prescription for BH spin (\S~\ref{sec:BH_spin}, \S~\ref{sec:BH_mergers}), which evolves due to: 1) gas accretion, 2) BH-BH mergers, 3) jet-induced spindown and 4) Lense-Thirring torques. This prescription is based on accretion disc theory, although it mostly uses numerical fitting functions obtained from general-relativistic (radiation) magneto-hydrodynamical simulations. The BH spin affects the feedback efficiencies of all feedback modes, as well as jet directions (\S~\ref{sec:feedback_effs}). 

The thermal isotropic and kinetic jet feedback modes are implemented through BH heating and kicking events, respectively (\S~\ref{sec:jet_scheme}), with the heating temperature, $\Delta T_\mathrm{AGN}$, and kick velocity, $v_\mr{jet}$, modulating the explosiveness and frequency of AGN feedback (\S~\ref{sec:AGN_implementation}). The energy of each feedback event scales in proportion with the BH mass. In addition to $\Delta T_\mathrm{AGN}$ and $v_\mr{jet}$, our model has one free parameter that is directly related to AGN feedback: the coupling efficiency, $\epsilon_\mr{f}$, of radiation to thermal feedback in the thin disc state. This same parameter is present in single-mode AGN feedback models often used in galaxy formation simulations. However, parameters related to the accretion efficiency, introduced in our model, also have a strong indirect impact on feedback.

The hybrid AGN feedback model is implemented in the \textsc{swift} code (\citealt{Schaller2024}) used for galaxy formation simulations, and has been coupled to the COLIBRE galaxy formation simulation suite (\citealt{Schaye2025}). While the flagship simulations of the COLIBRE suite used a thermal-only AGN feedback model (\citealt{Booth2009}), a set of runs use the hybrid AGN feedback model presented here. These runs were performed at all three resolutions included in the COLIBRE suite (m7, m6, m5), but in boxes up to 8 times smaller in volume than the flagship runs, at each resolution.

In \S~\ref{sec:calibration_first_results} we demonstrated the impact of varying three model parameters: the jet velocity normalization (Fig.~\ref{fig:velocity_variations}, the thick disc accretion efficiency (Fig.~\ref{fig:efficiency_variations}), and the thermal feedback coupling efficiency (Fig.~\ref{fig:coupling_efficiency_variations}). We also discussed how the final model was calibrated. As shown there, AGN feedback has important effects in intermediate-mass and low-mass galaxies, all the way down to $M_*\approx10^8$ M$_\odot$, and not just in massive galaxies as typically assumed. We confirmed this by considering the energies injected by AGN and comparing them to those injected by SNe (Fig.~\ref{fig:power_densities}, Fig.~\ref{fig:energies_BH_mass}).

New parameters introduced in the hybrid AGN feedback model require calibration, especially those that are more numerical in nature (e.g.~jet velocities). Parameters that are also present in the thermal AGN feedback model, as well as SN-related parameters, require recalibration when used with the new hybrid AGN feedback model. Instead of an \textit{ab initio} calibration of the entire set of stellar and AGN feedback parameters, as an initial guess we used the fiducial set of parameters from the thermal AGN COLIBRE simulations (\citealt{Chaikin2025}). We then varied these parameters as necessary to reproduce the same calibration galaxy and BH properties as the thermal AGN model (the GSMF, the stellar mass$-$size relation, and the stellar mass$-$BH mass relation, all at $z=0$), but we also calibrate on the bolometric AGN luminosity function (at $z=0.2$) since it holds useful information on how many BHs are radiatively-efficient, corresponding to our thin disc state ($0.01<f_\mr{Edd,d}<1$). The calibration was done first for m7, and the parameter values were then slightly adjusted for m6 and m5. The parameters we (re)calibrate for the hybrid AGN feedback model are: 1) the jet velocity normalization, $v_\mr{jet,0}$; 2) the accretion efficiency-related parameter, $r_\mr{tr,0}$; 3) the coupling efficiency $\epsilon_\mr{f}$; 4) the BH seed mass, $M_\mr{BH,seed}$; and 5) the stellar feedback pivot density $n_\mr{H,pivot}$. The final, calibrated values of the parameters for the three resolutions are given in Table \ref{tab:tab2}.

The calibration of the hybrid AGN feedback model presented in this study likely did not yield the best-fitting set of parameter values, particularly in relation to the slim accretion disc state (used for super-Eddington accretion). In the future, it will be useful to perform a calibration wider in scope (utilizing more parameters) and more efficient (utilizing emulators). Despite the more limited approach to the calibration, the hybrid AGN feedback simulations generally reproduce the observational data well (Fig.~\ref{fig:convergence_hybrid}).

The BH spin$-$mass relation (Fig.~\ref{fig:BHspins}), arguably the most important new prediction from the hybrid AGN model in relation to BHs, is in broad agreement with observations. The predicted relation has a peak for intermediate-mass BHs ($M_\mr{BH}\in[10^6,10^8]$ M$_\odot$). These BHs typically have near-maximal spin, $\vert a \vert\approx1$. Lower- and higher-mass BHs have lower spins . This shape of the BH spin$-$mass relation is linked to the relevance of: 1) accretion in the thin disc state, which tends to spin up BHs towards $\vert a \vert\approx1$, 2) BH-BH mergers, which tend to result in spins significantly less than $\vert a \vert\approx1$, and 3) jet-induced spindown, which can almost completely spin down BHs.

Finally, in \S~\ref{sec:feedback} we presented the first results on the operation of AGN feedback in the hybrid AGN feedback model, as found in the larger, production runs included in the COLIBRE suite. By $z=0$, jet particles are able to reach large distances (of order a dozen Mpc, in some cases 100 virial radii) from their host galaxies and impact the IGM on larger scales than in the purely thermal AGN model (Figs.~\ref{fig:IGM}, \ref{fig:jet_distances}). This has implications for cosmological studies, where the impact of baryonic effects on matter power spectra are of great interest. The total amount of injected AGN energy (Fig.~\ref{fig:power_densities}) is larger in the hybrid model than in the thermal one at all but the highest ($z>5$) redshifts, by nearly 50 per cent. Jets provide roughly half of the total AGN energy (independent of BH mass; Fig.~\ref{fig:energies_BH_mass}), with the thick disc state ($f_\mr{Edd,d}<0.01$) dominating the jet channel at lower redshift ($z<1$) and the thin disc state ($0.01<f_\mr{Edd,d}<1$) dominant at higher redshifts ($z>1$, peaking at $z\approx3$). Thermal isotropic feedback from the thin disc is important at all redshifts and dominates over jet feedback at $z\approx0.5-1.5$, but only slightly. Other feedback channels contribute negligibly to cosmic energy averages.

The hybrid AGN simulations use a different AGN feedback model compared to the fiducial COLIBRE simulations, but they are calibrated to the same observational data. This provides a useful estimate of the impact of different implementations of AGN feedback on: 1) galaxy properties that were not calibrated, 2) CGM, IGM and ICM properties, and 3) baryonic effects on cosmology, among others. These simulations can also be used to study various aspects of AGN feedback, including AGN jets. 

The hybrid AGN feedback simulations will be a powerful tool facilitating studies focusing on BH and accretion physics in a fully cosmological context, especially since our model includes BH spin evolution and three distinct accretion disc states. In particular, the simulations can be used to study the statistics of spin for the BH population, as well as BH spin evolution. The inclusion of BH spin will also facilitate more accurate predictions of gravitational wave emission due to BH-BH mergers. While not discussed here, the outputs of the simulations include quantities necessary for the calculation of the radio emission of jets and lobes in post-processing. Predictions using this synthetic radio emission will be discussed in future studies. 


\section*{Acknowledgements}
FH thanks Nina Karačić for her work on Fig.~\ref{fig:accretion_schematic}, Bert Vandenbroucke for help with implementing the hybrid AGN feedback model into \textsc{SWIFT} and COLIBRE, and Rob McGibbon for help with the processing and analysis of the hybrid AGN feedback simulations. The research in this paper made use of the \textsc{swift} open-source simulation code (\url{http://www.swiftsim.com}, \citealt{Schaller2024}) version 1.0.0. The \textsc{swift}simio Python library was used to analyze and visualize the data from the simulations (\citealt{Borrow2020_swiftsimio}, \citealt{Borrow_2021_swiftsimio}). 
FH acknowledges support from the Science Technology Facilities Council through a CDT studentship (ST/P006744/1). 
CGL acknowledges support from STFC consolidated grants ST/T000244/1 and ST/X001075/1. 
JT acknowledges support of a STFC Early Stage Research and Development grant (ST/X004651/1). ABL acknowledges support by the Italian Ministry for Universities (MUR) program “Dipartimenti di Eccellenza 2023-2027” within the Centro Bicocca di Cosmologia Quantitativa (BiCoQ), and support by UNIMIB’s Fondo Di Ateneo Quota Competitiva (project 2024-ATEQC-0050).
SP acknowledges support from the Austrian Science Fund (FWF) through grant number V 982-N.
This project has received funding from the Netherlands Organization for Scientific Research (NWO) through research programme Athena 184.034.002.
This work used the DiRAC@Durham facility managed by the Institute for Computational Cosmology on behalf of the STFC DiRAC HPC Facility (www.dirac.ac.uk). The equipment was funded by BEIS capital funding via STFC capital grants ST/K00042X/1, ST/P002293/1, ST/R002371/1 and ST/S002502/1, Durham University and STFC operations grant ST/R000832/1. DiRAC is part of the National e-Infrastructure.


\section*{Data availability}

The data underlying the figures shown in this article will be provided upon reasonable request to the corresponding author. The SWIFT code is public (\url{https://gitlab.cosma.dur.ac.uk/swift/swiftsim}), while the COLIBRE modules will be made available after the public data release. People interested in using the hybrid AGN feedback simulations before that point are encouraged to contact the corresponding author.

\bibliographystyle{mnras}
\bibliography{jet_bibliography} 

\appendix

\section{Simplified versions of the model}
\label{sec:appendix_model}

In \citet{Husko2022_spin_driven, Husko_winds, Husko_2025_SE} we introduced the aspects of the model that concern the thick, thin and slim disc states, respectively. We studied AGN feedback in these individual accretion disc states separately, but also in hybrid variants. These studies were performed using idealized galaxy clusters in the first two cases, and high-redshift cosmological simulations of a protocluster in the latter case. The first study modeled only the thick disc and only kinetic jet feedback, using BH spin evolution. The second study also considered the thin disc, comparing models that apply either the thick disc state and its jets or the thin disc and its thermal feedback at all Eddington ratios. There, a hybrid model where the two modes switched at $f_\mr{Edd,d}=0.01$ was also considered. In the final study, the super-Eddington slim disc was introduced, with its jet and wind (thermal) feedback, considering only hybrid models where the thin disc transitions to the slim disc state at $f_\mr{Edd,d}=1$.

In the hybrid model that we introduce here, we combine these three accretion disc states in a similar way as in the above studies. Similar to the last of the above three studies, we include jets in the thin disc, which was not done in the second study. Here we also include wind (thermal) feedback in the thick disc, which was not included in either of the first two studies. Another major modification to the model presented here is that we include accretion efficiencies in the thick and slim disc, both on account of their mass loss induced by strong winds. Finally, the previous versions of the hybrid AGN feedback model were coupled with the \textsc{swift}-EAGLE galaxy formation model, whereas here we couple it with COLIBRE.

\section{The transition between accretion disc states}

The state of the accretion flow (thick, thin or slim disc) is assumed to depend on the dimensionless accretion rate, i.e.~the Eddington ratio. However, it is not immediately clear if one should use the disc-scale Eddington ratio, $f_\mr{Edd,d}$, or the horizon-scale one, $f_\mr{Edd,H}$, to determine whether the accretion disc should transition to another state. This choice is discussed in \S~\ref{sec:appendix_transition_large_vs_net}, whereas in \S~\ref{sec:appendix_transition_crit} we discuss the choice of critical Eddington ratios at which these transitions occur.

\subsection{Disc-scale versus horizon-scale Eddington ratios}
\label{sec:appendix_transition_large_vs_net}

From a modeling perspective, it is not clear whether accretion discs should transition between different states based on the value of the horizon-scale accretion rate or the inflow rate onto the disc. This is a question that concerns only simulations and not reality. The reason is that the transition between the thick and thin disc is, in reality, probably smooth rather than abrupt. In an ideal scenario, as a BH that is deep in the thick disc (i.e.~both $f_\mathrm{Edd,d}\ll0.01$ and $f_\mathrm{Edd,H}\ll0.01$) progressively increases its accretion rate, its accretion efficiency should also progressively increase until it reaches $\epsilon_\mathrm{acc}\approx1$ at the critical Eddington ratio $f_\mathrm{Edd}\approx0.01$. At this point, it does not matter whether one uses $f_\mathrm{Edd,d}$ or $f_\mathrm{Edd,H}=\epsilon_\mathrm{acc}f_\mathrm{Edd,d}$ to decide the transition between the states, since both will give the same outcome. The same is true if the disc is transitioning from the thin to the thick disc state.

Let us now consider the above scenario in the context of our model. If the disc is currently thin, $\epsilon_\mathrm{acc}=1$ and $f_\mathrm{Edd,H}>0.01$ (therefore also $f_\mathrm{Edd,d}>0.01$), the disc will stay thin. Let us imagine then slightly reducing $f_\mathrm{Edd,H}$ to $f_\mathrm{Edd,H}=0.009$. The disc will become thick, with $\epsilon_\mathrm{acc}\ll1$ and $f_\mathrm{Edd,d}=\epsilon_\mathrm{acc}f_\mathrm{Edd,H}\ll0.009$. If the accretion rate again slightly increases (so that $f_\mathrm{Edd,H}>0.01$), the disc may or may not transition to the thin state depending on whether we consider $f_\mathrm{Edd,H}$ or $f_\mathrm{Edd,d}$. If we consider $f_\mathrm{Edd,H}$, the accretion rate would need to increase by nearly a factor of 100 (assuming a per-cent level accretion efficiency), i.e. $f_\mathrm{Edd,d}$ would need to be of order unity, for the disc to transition back to the thin state. We consider this to be unrealistic, and instead use $f_\mathrm{Edd,d}$ to decide the transition. This is supported by our numerical experiments (not shown here) where we attempted to use $f_\mathrm{Edd,H}$ to decide the transition, but found it impossible to calibrate realistic models with that choice.

The above discussion may point to our accretion efficiency prescription being somewhat unrealistic, more specifically, the fact that accretion efficiencies suddenly drop to very low values even if the disc is only slightly in the thick state. This is an outcome of our calibration rather than an \textit{a priori} assumption we made, with the $r_\mathrm{tr,0}$ parameter having a large value (see \S~\ref{sec:efficiency_variations}). We considered models that have accretion efficiencies of order unity for accretion discs that are only slightly in the thick state, but our attempt at calibrating such models was unsuccessful. We believe that this is related to the jet feedback prescription for the thick disc. We use jet efficiency formulas that depend only on BH spin, i.e.~jets are very strong for discs that are only slightly in the thick disc state. In reality, however, jets should be relatively weak at such high Eddington ratios, and radiative feedback stronger (\citealt{YuanNarayan2014}). Our need to use very low accretion efficiencies therefore probably compensates for the overly strong jets at the point of transition between the thick and thin disc states. Future general-relativistic radiative magnetohydrodynamical (GRRMHD) simulations of the transition between the thick and thin disc state will likely resolve this issue from a theoretical perspective.

\subsection{Critical Eddington ratios}
\label{sec:appendix_transition_crit}

The transition between the thin disc at moderate Eddington ratios and the slim disc is simply set to occur at $f_\mr{Edd,H}=1$, in other words, we assume that the slim disc coincides with the super-Eddington accretion disc state. The transition between the thick (\citealt{NarayanYi1994}) and thin (\citealt{ShakuraSunyaev1973}) discs is more subtle. It should occur around $f_\mr{Edd,d}\approx\alpha^2$ (\citealt{Narayan1995}), where $\alpha$ is the accretion disc viscosity parameter. More recent calculations suggest that the properties of the thick disc already begin to change at $f_\mr{Edd,H}=0.2\alpha^2$, and the transition appears to be complete by $f_\mr{Edd,d}=0.7\alpha$ (\citealt{YuanNarayan2014}). Between these two values, the disc takes on a transition state whose properties are not well understood. For conceivable values of $\alpha$, which may be as low as $0.05$ and as high as $0.1-0.4$ (\citealt{King2007}, \citealt{YuanNarayan2014}), the transition state may occupy the range $0.001-0.3$ in $f_\mr{Edd,d}$. Observations of both X-ray binary (\citealt{Done2007}) and AGN spectra (\citealt{Noda2018}) find this transition to occupy a narrower range of $f_\mr{Edd,d}=0.01-0.03$. An analysis of the radiative and mechanical powers of AGN shows the transition to span the same range (\citealt{Russell2013}). We assume the lower end of this range to be the critical transition rate at which the two accretion disc states (the thin and thick disc) interchange; $f_\mr{Edd,d,crit}=0.01$.

Given this choice, we can set a value for the viscosity parameter $\alpha$, which appears in many of the equations describing accretion disc structure that we will discuss. For this purpose we use the finding of numerical calculations that the transition spans the range between $0.2\alpha^2$ and $0.7\alpha$ in $f_\mathrm{Edd,d}$ (\citealt{YuanNarayan2014}). We assume that the geometric mean of these two boundaries corresponds to $f_\mathrm{Edd,crit,0}=0.01$, which is true for $\alpha\approx0.1$, so we set $\alpha=0.1$ in this model.

\section{Innermost stable circular orbit (ISCO)}
\label{sec:appendix_ISCO}

The radius of the innermost stable circular orbit (ISCO, \citealt{Kerr}) is given by $R_\mr{ISCO}=r_\mathrm{ISCO} R_\mr{G}$, where $r_\mathrm{ISCO}$ is the dimensionless radius of the ISCO. $r_\mathrm{ISCO}$ depends on BH spin as
\begin{equation}
    r_\mathrm{ISCO}=3+Z_2\mp\sqrt{(3-Z_1)(3+Z_1+2Z_2)},
\label{eq:Lambda}
\end{equation}
where the minus and plus sign are for prograde and retrograde accretion, respectively, and $Z_1$ and $Z_2$ are functions of BH spin given by
\begin{equation}
    Z_1(a)=1+\Big(1-a^2\Big)^{1/3}\Big[\big(1+\vert a \vert\big)^{1/3}+\big(1-\vert a\vert\big)^{1/3}\Big]
\label{eq:Z1}
\end{equation}
and
\begin{equation}
    Z_2(a)=\sqrt{3a^2+Z_1^2}.
\label{eq:Z2}
\end{equation}
The specific angular momentum at the ISCO is given by $L_\mr{ISCO}=M_\mr{BH}G\ell_\mr{ISCO}/c$, where $\ell_\mr{ISCO}$ is a dimensionless function of BH spin:
\begin{equation}
    \ell_\mr{ISCO}(a)=\frac{2}{3\sqrt{3}}\Big(1+2\sqrt{3r_\mathrm{ISCO}-2}\Big).
\label{eq:l_ISCO}
\end{equation}

\section{Radiative efficiency in the thick and slim discs}
\label{sec:appendix_rad_eff}

In the thick disc state, most of the thermal energy is advected inwards towards the black hole before it has time to radiate away. This results in a low, but non-zero radiative efficiency. This efficiency is mostly governed by radiative cooling rates of electrons through synchrotron, bremsstrahlung, and Compton processes. Numerical simulations have not focused on radiative efficiencies of thick discs, so we take the results from an early numerical study (\citealt{Mahadevan}), in which cooling processes were studied in the context of the original thick disc solution of \cite{NarayanYi1994} (note that more recent results have been obtained by \citealt{Xie2012}, although their formulae are not as convenient for general numerical implementation). 

The numerical study in question (\citealt{Mahadevan}) found two different regimes: for $f_\mathrm{Edd,H}<f_\mathrm{Edd,crit,visc}$, viscous heating dominates the heating of electrons, whereas for $f_\mr{Edd,crit,visc}<f_\mathrm{Edd,H}<f_\mr{Edd,crit,thick}$, it is dominated by ion-electron heating. Here, $f_\mr{Edd,crit,visc}$ is the transitional value between the two thick disc state, and $f_\mathrm{Edd,crit,thick}=0.01$ is the transitional accretion rate which separates thin and thick discs. The radiative efficiency in the viscous heating regime is given by
\begin{equation}
\epsilon_\mr{rad,thick}=0.04\epsilon_\mr{rad,thin}\bigg(\frac{\delta}{0.1}\bigg)\bigg(\frac{1-\beta}{0.5}\bigg)\bigg(\frac{6}{r_\mr{ISCO}}\bigg),
\label{eq:eq5}
\end{equation}
and in the ion-electron heating regime by
\begin{equation}
\epsilon_\mr{rad,thick}=0.2\epsilon_\mr{rad,thin}\bigg(\frac{f_\mr{Edd,H}}{\alpha^2}\bigg)\bigg(\frac{\beta}{0.5}\bigg)\bigg(\frac{6}{r_\mr{ISCO}}\bigg).
\label{eq:eq6}
\end{equation}
Here, $\beta$ is the ratio of gas pressure and total pressure (which includes the magnetic pressure). A somewhat different parameter, $\beta_\mr{m}$, can be defined as the ratio of gas pressure and magnetic pressure (\citealt{YuanNarayan2014}). The two parameters are related by $\beta=\beta_\mr{m}/(1+\beta_\mr{m})$. $\beta_\mr{m}$ is not an independent parameter; many simulations have found that $\alpha\beta_\mr{m}\approx0.5$ (e.g.~\citealt{Begelman2021}, \citealt{YuanNarayan2014}), which we adopt. $\delta$ represents the fraction of viscous energy dissipation directly transferred to the electrons, and is constrained in theoretical studies between $0.1$ and $0.5$ (\citealt{Sharma2007}, \citealt{YuanNarayan2014}). Observations imply a value close to $0.2$ (\citealt{Yuan2003}, \citealt{Liu2013}), which we adopt. The critical Eddington ratio between the two thick disc states can be found by ensuring that both formulae (Eqns.~\ref{eq:eq5} and \ref{eq:eq6}) yield the same radiative efficiency. This gives an accretion rate equal to
\begin{equation}
f_\mr{Edd,crit,visc}=0.2\bigg(\frac{\delta}{0.1}\bigg)\bigg(\frac{1-\beta}{\beta}\bigg)\alpha^2,
\label{eq:eq7}
\end{equation}
which is equal to $1.28\times10^{-3}$ for our assumed values of the different parameters.

For the slim disc, we take results based on numerical works (\citealt{Sadowski2009}). We use the following fitting formula adopted for their results by (\citealt{Madau2014}):
\begin{equation}
\epsilon_\mr{rad,SD}=\frac{0.1}{f_\mr{Edd,H}}A(a)\bigg[ \frac{0.985}{1.6/f_\mr{Edd,H}+B(a)}+\frac{0.015}{1.6/f_\mr{Edd,H}+C(a)}\bigg],
\label{eq:rad_SD}
\end{equation}
where the three BH spin-dependent functions are given by $A(a)={(0.9663-0.9292a)}^{-0.5639}$, 
$B(a)={(4.627-4.445a)}^{-0.5524}$ and $C(a)={(827.3-718.1a)}^{-0.7060}$. For accretion that is sufficiently sub-Eddington, this formula yields the radiative efficiency of a thin disc (Eqn.~\ref{eq:eps_rad_NT}). For super-Eddington accretion, the radiative efficiency drops with the Eddington ratio.

\section{Warp radii and disc surface densities}
\label{sec:appendix_warp}

The radius $R_\mr{warp}$, which separates the inner and outer accretion disc, can be calculated by equating the Lense-Thirring precession time-scale ($t_\mr{p}=2\pi/\Omega_\mr{p}$, with $\Omega_\mr{p}=2GJ_\mr{BH}/c^2R^3$ the precession rate) and the vertical warp propagation time-scale ($t_\mr{warp}=R^2/\nu_2$, with $\nu_2$ the kinematic viscosity in the vertical direction) (\citealt{Pringle1992}, \citealt{Martin2007}, \citealt{Cielo2014}). We use the relation $\dot{M}=3\pi\nu_1 \Sigma$ (for $R\gg R_\mr{ISCO}$, \citealt{Fiacconi2018}) to calculate $\nu_1$, and therefore $\nu_2$. For the thick and slim disc, we assume $\nu_2=\nu_1$. For the thin disc, $\nu_2$ can be related to $\nu_1$ by $\nu_2=\xi\nu_1$, with $\xi$ a numerical factor (\citealt{Lodato2010}). The constant parameter $\xi$ is often also expressed in the form $\alpha_2/\alpha$. Early theoretical calculations predicted $\alpha_2=1/2\alpha$ for small $\alpha$ (\citealt{Papaloizou1983}), which has also been confirmed by simulations (\citealt{Lodato2007}). Later simulations found that higher-order corrections to this prediction may need to be included for realistic values of $
\alpha$ (\citealt{Lodato2010}), such as $\alpha=0.2$. These numerical results agree with an early theoretical prediction (\citealt{Ogilvie1999}) which we assume here:
\begin{equation}
    \xi=\frac{\nu_2}{\nu_1}=\frac{\alpha_2}{\alpha}=\frac{2}{\alpha^2}\frac{1+7\alpha^2}{4+\alpha^2},
\label{eq:alpha_2}
\end{equation}
which reduces to $1/2\alpha^2$ for small $\alpha$.

The warp radius, $R_\mr{warp}$, can be derived for the advection-dominated thick and slim discs as (see \citealt{Ogilvie1999}):
\begin{equation}
R_\mr{warp,adv}=R_\mr{G}\bigg(\frac{384\vert a\vert}{25(H/R)^2}\bigg)^{2/5}.
\label{eq:r_warp_adaf}
\end{equation}
Thus, it depends only on the magnitude of the current BH spin, the aspect ratio of the disc (and the BH mass through $R_\mr{G}$). For both the thick and slim disc we assume $H/R=0.3$, based on general-relativistic magnetohydrodynamical (GRMHD) simulations (\citealt{Narayan2021}, \citealt{Ricarte}). This value is somewhat lower than the aspect ratio in the self-similar analytical solutions for these discs ($H/R\approx0.4-0.5$ as in \citealt{NarayanYi1994}, \citealt{Wang1999}, respectively). Regardless of these small differences, Eqn.~(\ref{eq:r_warp_adaf}) yields small warp radii (no larger than $\approx10R_\mr{G}$), much smaller than for the thin disc (see below). While we take the aspect ratio $H/R$ from the GRMHD simulations, we assume surface densities based on the analytical solutions since they are more wieldy and easily accessible. These take the same form for the two types of accretion disc:
\begin{equation}
\Sigma_\mr{adv}=\frac{\dot{M}_\mathrm{acc,H}}{2\pi R\vert v_\mr{r} \vert},
\label{eq:Sigma_adv}
\end{equation}
where $v_\mr{r}$ is the radial inflow velocity of the gas, and the only source of difference between the expression for $\Sigma_\mr{adv}$ between the two discs. For the slim disc, $v_\mr{r}=-\alpha v_\mr{K}/\sqrt{5}$ (\citealt{Wang1999}), where $v_\mr{K}=\sqrt{M_\mr{BH}G/R}$ is the Keplerian velocity. For the thick disc we take $v_\mr{r}=-\alpha v_0 v_\mr{K}$ (\citealt{NarayanYi1994}), where the parameter $v_0$ is not $1/\sqrt{5}\approx0.45$ but is instead dependent on $\alpha$ through $v_0=3/(5+2\varepsilon)$, where $\varepsilon=(5/3-\gamma)/(\gamma-1)$, $\gamma$ is the adiabatic index of the gas, related to the gas-to-total pressure ratio $\beta$ by $\gamma = (8-3\beta)/(6-3\beta)$. Finally, $\beta$ can be connected to $\alpha$ through the findings of GRMHD simulations (\citealt{YuanNarayan2014}) as $\beta=1/(1+2\alpha)$, which is equivalent to the relation $\alpha\beta_\mathrm{m}=0.5$ mentioned in Appendix \ref{sec:appendix_rad_eff}. $v_0$ depends on $\alpha$ only weakly; for $\alpha=0.2$, we obtain $v_0=0.52$, very close to the value for the slim disc. Having introduced these dependencies, we are also in a position to state the specific angular momentum $L(R)$ for the thick and slim discs. For the slim disc, it is a factor $1/\sqrt{5}\approx0.45$ of the Keplerian value $L_\mr{K}=\sqrt{M_\mr{BH}GR}$, while for the thick disc, it is $\Omega_0=\sqrt{2\varepsilon/(5+2\varepsilon)}$ times the Keplerian value. For $\alpha=0.2$, $\Omega_0\approx0.37$, again close to the slim disc value.

For the thin disc, $R_\mr{warp}$ also depends on the surface density $\Sigma$, unlike for the thick and slim discs. The solution for the thin disc (\citealt{ShakuraSunyaev1973}), assumed in this model, describes three regions: a) an inner region where radiation pressure dominates the pressure and electron scattering the opacity, which is often unstable and usually does not extend far out, b) a middle region where gas dominates the pressure and electron scattering dominates the opacity and c) an outer region where gas also dominates the pressure, but the opacity is dominated by free-free absorption. Our fiducial choice is to describe the disc using the region b) from \cite{ShakuraSunyaev1973} (but we found that using region c) makes little difference). In region b), the surface density of a thin disc can be expressed as
\begin{equation}
\Sigma_\mr{TD}=6.84 \times 10^{5} \mathrm{~g} \mathrm{~cm}^{-2} \alpha^{-4 / 5} f_\mathrm{Edd,H}^{3 / 5}\bigg(\frac{\epsilon_\mr{rad}}{0.1} \bigg)^{3/5}\left(\frac{M_{\mathrm{BH}}}{10^{8} \mathrm{M}_{\odot}}\right)^{1 / 8}\left(\frac{R}{R_{\mathrm{S}}}\right)^{-3 / 5}.
\label{eq:sigma_2}
\end{equation}
Using the surface density, the warp radius can be calculated as
\begin{equation}
R_{\text {warp,TD}}=3410 R_{S} a^{5 / 8} \xi^{-5/8}\alpha^{-1 / 2} f_\mathrm{Edd,H}^{-1 / 4}\bigg(\frac{\epsilon_\mr{rad}}{0.1} \bigg)^{-1/4}\left(\frac{M_{\mathrm{BH}}}{10^{8} \mathrm{M}_{\odot}}\right)^{1 / 8}.
\label{eq:Rwarp_2}
\end{equation}
for region b) (\citealt{Griffin2019a}). 

At large radii, thin accretion discs can be prone to the effects of self-gravity (see \citealt{Lodato2008} for a review). The gravity there due to the disc locally becomes comparable to that due to the BH. In order to estimate the radius at which this becomes true, the disc self-gravity radius $R_\mr{sg}$, we use the Toomre instability parameter, $Q=\Omega c_{\mathrm{s}} /\pi G \Sigma$, and set its value equal to the critical value of $1$. This equation, the Toomre instability criterion, be solved for the self-gravity radius, giving
\begin{equation}
R_{\text {sg,TD}}=6460 R_{S} \alpha^{28/51} f_\mathrm{Edd,H}^{-18/51}\bigg(\frac{\epsilon_\mr{rad}}{0.1} \bigg)^{-18/51}\left(\frac{M_{\mathrm{BH}}}{10^{8} \mathrm{M}_{\odot}}\right)^{-49/51}.
\label{eq:Rsg_2}
\end{equation}
If $R_\mr{sg}<R_\mr{warp}$, we assume that the entire accretion disc is (counter-) aligned and use $R_\mr{sg}$ instead of $R_\mr{warp}$ in all equations where $R_\mr{warp}$ appears.


\section{Kinetic jets as a representation of relativistic jet feedback}
\label{sec:kinetic_jets}

For use as a feedback mechanism in cosmological simulations of galaxy formation, which have relatively poor resolution, it is not clear how best to represent the effects of relativistic AGN jets. Real jets and the lobes they inflate are affected by many physical processes, including basic (relativistic) hydrodynamics, magnetic fields, instabilities and cosmic ray physics. Many idealized jet simulations have been performed, varying the different physical aspects at play (see the review by \citealt{Bourne2023}). While the details change, it appears that the main characteristic of jet physics is that, as the jet material is shocked at the jet head, a hot lobe of material is inflated. The lobe then expands and drives a bow shock that propagates through the ambient medium and heats it anisotropically. This is the main way jets deposit their energy; most or all simulations find that of order half of the injected jet energy is deposited into the medium through the bow shock while the jet is still active. This fraction varies somewhat (in the range 30-70 per cent) with how the jets are simulated. The behaviour of lobes may also differ somewhat depending on whether they constitute nonrelativistic particles or relativistic ones, on the magnetic fields of the ambient medium, and on how well instabilities are captured in the simulations. However, as at least half of the energy is deposited in the ambient medium independently of the exact contents of the lobes, it appears that for capturing the dominant effects, the precise details of jet and lobe modeling do not matter. The main effects of relativistic AGN jet feedback can thus be represented with a mechanism as simple as non-relativistic kinetic kicks/injections. 

The caveat to the above statement is that one needs to resolve the presence of a lobe to capture the driving of a bow shock. With of order dozens or hundreds of particles, we begin to resolve a lobe, so the energy injection should broadly be correct (\citealt{Husko2022_self_similar}). With only a dozen or fewer particles representing an entire jet event, this is not true, and instead all of the energy injection into the ambient medium is done through shock heating of individual kicked particles. While this scenario is very common in our simulations, more common than well-resolved jets, this is a resolution limitation that cannot be circumvented by more physical modeling of jets. It instead requires numerical advancements such as adaptive particle splitting/merging (see e.g.~\citealt{Bourne2020}, \citealt{Su2021}, \citealt{Talbot2020}).

\section{The observed median black hole mass$-$stellar mass relation}
\label{app:BH_data}

We compute the weighted median BH mass as a function of the total stellar mass using the stellar masses and dynamical BH mass measurements from \cite{Graham2023}, with the stellar masses corrected as per the erratum \cite{Graham2024erratum} (a decrease of $0.15$ dex relative to \citealt{Graham2023}). The stellar masses from \cite{Graham2023} were inferred from deep infrared observations with \textit{Spitzer} using colour-dependent mass-to-light ratios. We confirmed that these are consistent with our own mass-to-light ratios, after converting from their assumed \cite{Kroupa2001} to our adopted \cite{Chabrier2003} IMF. The dynamical BH mass measurements collected in \cite{Graham2023} may be biased towards galaxies of a particular morphological type, at least in the case of some of the methods of BH mass measurement. For this reason, we decided to compute weighted medians that take into account the expected and realized numbers of galaxies of different morphological types in different stellar mass bins. 

Our binning procedure consists of the following steps. We compute weighted median BH masses in four bins that each contain $\approx30-40$ galaxies, with the limits between bins chosen by hand so that no bin has extreme outliers. The weights used for the medians for a given type of galaxy (spheroidal or disky) are equal to $f_\mr{exp}/f_\mr{obs}$, where $f_\mr{obs}$ is the actual fraction found for a given galaxy type in a given stellar mass bin in the \cite{Graham2023} data, while $f_\mr{exp}$ is the expected fraction in that same stellar mass bin. For this purpose, ellipticals and lenticulars are both considered spheroidal, since they appear to follow the same $M_\mr{BH}-M_*$ relation in the \cite{Graham2023} data. For $f_\mr{exp}$, we use the morphological fractions (and their dependence on stellar mass) from \cite{Moffett2016}. The binned stellar and BH masses that we obtain as a result of this procedure are provided in Table \ref{tab:tab4}.

\captionsetup[table]{skip=0pt} 
\begin{table}
\begin{center}
\caption{The binned median galaxy stellar and BH masses (and their associated errors) from the \protect\cite{Graham2023} dataset, corrected for morphological sampling effects. These represent the median BH mass$-$stellar mass relation for the overall galaxy population, although one that includes Eddington bias, rather than the intrinsic median relation. See the text for a description of the binning procedure and the morphological sampling correction.}
\label{tab:tab4}
\centering
\begin{tabular*}{0.6\columnwidth}{@{\extracolsep{\fill}}lr}
   $\log_{10}M_\mr{*}$ $[\mr{M}_\odot]$ & $\log_{10}M_\mr{BH}$ $[\mr{M}_\odot]$  \\
  \hline 
  $10.50\pm0.03$ & $7.44\pm0.14$ \\
  $10.85\pm0.02$ & $7.95\pm0.10$  \\
  $11.37\pm0.02$ & $8.85\pm0.10$  \\
  $11.70\pm0.03$ & $9.40\pm0.13$  \\
\end{tabular*}
\end{center}
\end{table}

\section{Simpler recalibrated models}
\label{app:recal_models}

\begin{figure*}
\includegraphics[width=1\textwidth, trim = 0 5 0 0]{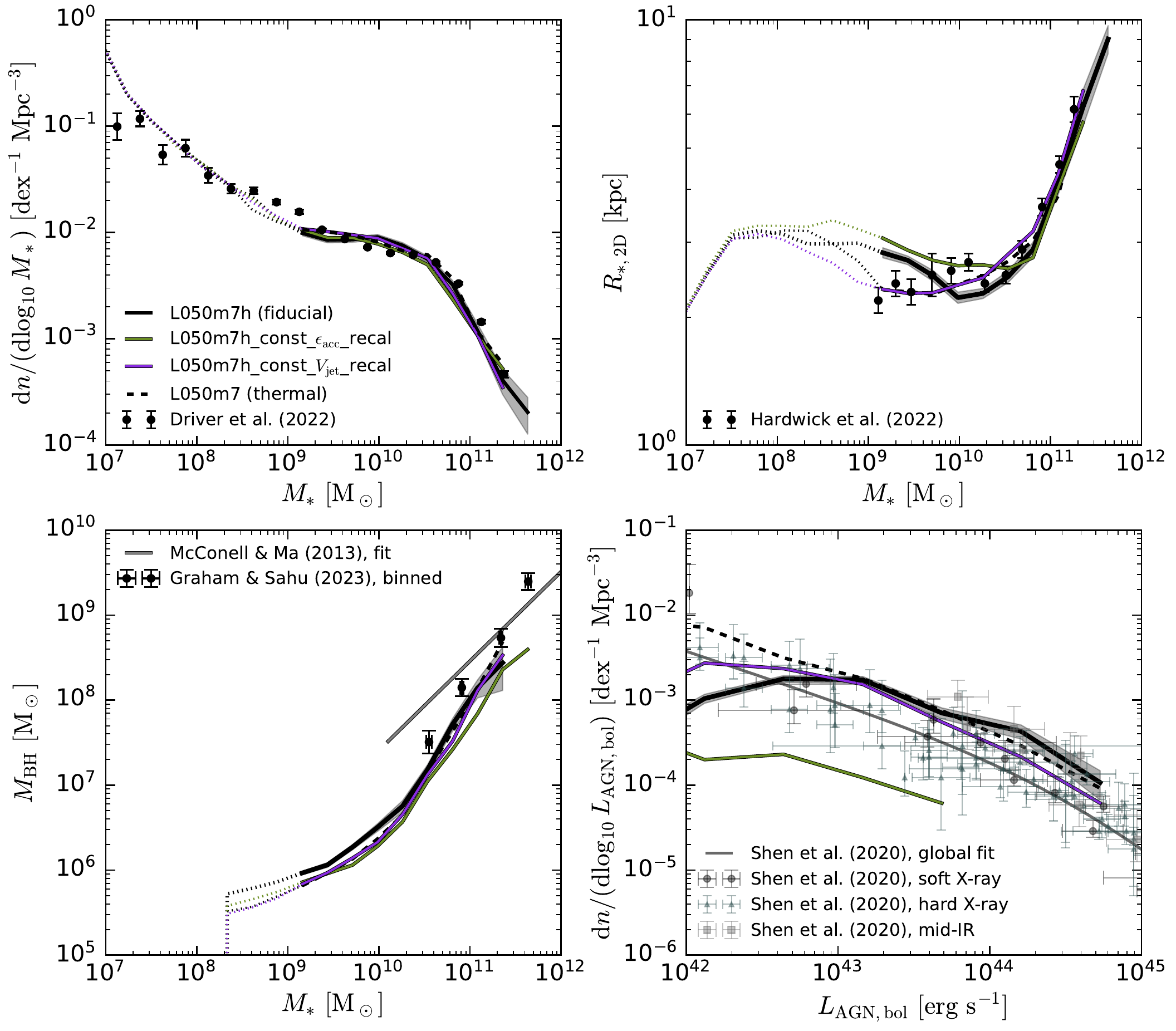}
\caption{As Figs.~\ref{fig:velocity_variations}-\ref{fig:coupling_efficiency_variations}, but showing the comparison (in L050 boxes at m7 resolution) between the fiducial hybrid AGN feedback model and two simpler models that, separately, used constant accretion efficiencies in the thick disc (of 100 per cent) or constant jet velocities (of $10^4$ km~s$^{-1}$). Both of the simpler models have been recalibrated to reproduce the data shown here as best they can.}
\label{fig:recalibrated_models}
\end{figure*}%

In Figs.~\ref{fig:velocity_variations} and \ref{fig:efficiency_variations} we showed the effects of varying the jet velocities and the thick disc accretion efficiencies, respectively. There we also showed models that had constant values of the associated parameters, $v_\mr{jet}=10^4$ km~s$^{-1}$ and $\epsilon_\mr{acc}=1$, respectively. However, showing that such models do not perform as well as our fiducial one when compared against calibration data is not sufficient to rule them out. This is because such parameter variations were not coupled to simultaneous changes of other parameter values. In Fig.~\ref{fig:recalibrated_models} we show our best attempts at recalibrating such models. Both of the models were recalibrated by reducing the BH seed masses (since AGN feedback is more effective in lower-mass galaxies/earlier in both runs, this had to be offset by delaying BH growth). The constant-velocity model has a decreased SN pivot density (making SN feedback more effective), while the opposite is true for the constant-efficiency model. That model also has a lower thermal AGN feedback coupling efficiency.

Both recalibrated models perform well when compared against the $z=0$ GSMF and galaxy size$-$mass relation (see top row in Fig.~\ref{fig:recalibrated_models}). The constant-velocity model better reproduces galaxy sizes at $M_*<10^{9.5}$ M$_\odot$, and reproduces BH masses and the AGN LF equally well (or even slightly better for the latter) than the fiducial model. Despite these small benefits, the variable-velocity model was chosen for reasons explained in \S~\ref{sec:thermal_feedback}. The fact that these models all produce the main observables to a similar degree of accuracy means that these observables cannot be used to discriminate between them. Other galaxy properties, which remain true predictions, must be used.

The model with a constant accretion efficiency falls slightly short on the BH$-$mass relation at the massive end (see bottom left in Fig.~\ref{fig:recalibrated_models}). This could have been remedied by further reducing its coupling efficiency, but note that it already has a very low value of $\epsilon_\mr{f}=0.01$. As a result, jet AGN feedback is almost completely dominant over thermal AGN feedback at all masses, and even over SN feedback. The AGN LF (see bottom right in Fig.~\ref{fig:recalibrated_models}) is too low, and we were not able to reproduce it with any parameter variations for the model in question. This is indicative of AGN jet feedback from the thick disc accretion state being overly dominant in such a model. Our remedy was the introduction of thick disc accretion efficiencies $\epsilon_\mr{acc}\ll1$.

\section{Impact of parameter variations on the stellar mass$-$halo mass relation}
\label{app:SHMR}

\begin{figure*}
\includegraphics[width=1\textwidth, trim = 0 5 0 0]{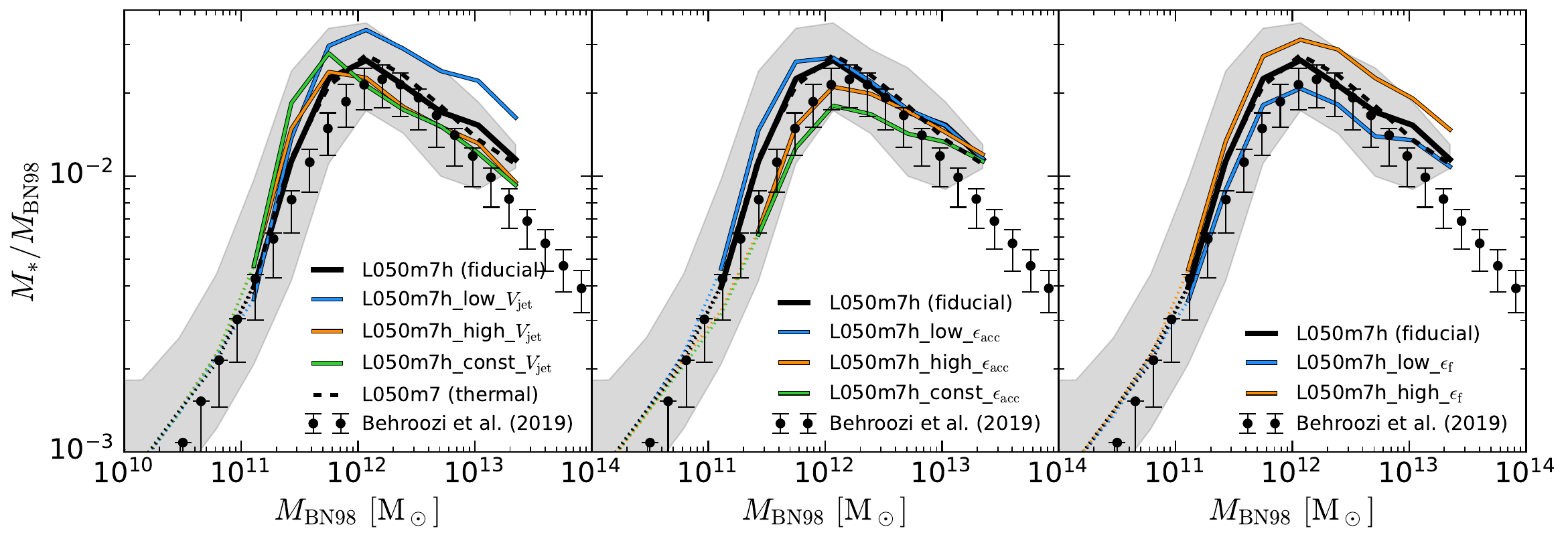}
\caption{The impact of variations of the jet velocity normalization (left), the thick disc accretion efficiency (middle) and the thin disc thermal coupling efficiency (right) on the stellar mass$-$halo mass relation for central subhaloes at $z=0$. The lines show the median from different simulations as given by the legends, with the black shaded region corresponding to the 16$-$84 per-centile range for the fiducial hybrid simulation. The lines become dotted at a halo mass where the median stellar mass is $\leq10^9$ M$_\odot$, corresponding to $\approx$100 star particles. The variations shown here are the same as presented in Figs.~\ref{fig:velocity_variations}-\ref{fig:coupling_efficiency_variations}. For reference, we show the prediction from the \protect\cite{Behroozi2019} semi-empirical model. Halo masses follow the \protect\cite{BryanNorman1998} definition.}
\label{fig:SHMR}
\end{figure*}%

In \S~\ref{sec:param_variations} we showed the effects of varying different parameters in terms of the impact on two main galaxy observables: the GSMF and galaxy sizes as a function of stellar mass. Differences in the GSMF can be hard to interpret, since changes in stellar mass at fixed halo mass produce horizontal shifts in the GSMF.

To make the impact of these parameter variations on galaxy stellar masses easier to interpret, in Fig.~\ref{fig:SHMR} we show the effects of the three parameter variations included in \S~\ref{sec:param_variations} (the jet velocity normalization, the thick disc accretion efficiency, and the thin disc thermal feedback coupling efficiency) on the stellar mass$-$halo mass relation for central subhaloes at $z=0$. We use bins of width 0.3 dex in halo mass. For reference, we also show the \cite{Behroozi2019} semi-empirical model, which reproduces the observed GSMF). For consistency with \cite{Behroozi2019}, we define the halo mass as the virial mass using the definition from \cite{BryanNorman1998}, which is a fitting function to the results of \cite{Eke1996}. 

The left panel of Fig.~\ref{fig:SHMR} shows that higher jet velocities lead to a decrease in stellar masses in haloes more massive than $M_\mr{BN98}\approx10^{12}$ M$_\odot$, likely due to more explosive feedback. In lower-mass haloes, the effect is opposite, likely due to differences in the sampling of AGN feedback. Lower jet velocities, however, also lead to higher stellar masses in low-mass haloes. It thus appears that the fiducial velocity model is close to optimal in terms of the balance between sampling effects and the how efficient (explosive) the feedback is. The middle panel shows that higher accretion efficiencies in the thick disc state lead to lower stellar masses in all haloes down to $M_\mr{BN98}\approx10^{11}$ M$_\odot$, likely due to more effective jet feedback from the thick disc. The right panel shows that higher thin disc thermal coupling efficiencies lead to higher stellar masses in all haloes down to $M_\mr{BN98}\approx10^{11}$ M$_\odot$. This is likely a consequence of the coupling efficiency mainly changing BH masses due to self-regulation, which in turn changes the AGN heating temperatures in those galaxies due to different BH masses (Eqn.~\ref{eq:delta_T_AGN}).

\bsp	
\label{lastpage}
\end{document}